\documentclass[reqno,11pt]{amsart}
\usepackage[utf8]{inputenc}
\usepackage{graphicx}
\usepackage{slashed}
\usepackage{amscd}
\usepackage{amssymb}
\usepackage{esint}
\usepackage{enumitem}
\usepackage{comment}
\usepackage{amsmath}
\usepackage{dsfont}
\usepackage[mathscr]{eucal}
\usepackage[linktocpage]{hyperref}
\usepackage{pstricks}
\usepackage{pst-all}

\numberwithin{equation}{section}
\allowdisplaybreaks[1]
\definecolor{labelkey}{gray}{.65}

\title[Causal Fermion Systems as an Effective Collapse Theory]{Causal Fermion Systems as an \\ Effective Collapse Theory}

\author[F.\ Finster]{Felix Finster}
\address{Fakult\"at f\"ur Mathematik \\ Universit\"at Regensburg \\ D-93040 Regensburg \\ Germany}%
\email{finster@ur.de}

\author[J.\ Kleiner]{Johannes Kleiner}
\address{Munich Center for Mathematical Philosophy \\ Ludwig-Maximilians-Universität \\ D-80539 München \\ Germany}
\email{contact@jkleiner.de}

\author[C.F.\ Paganini]{Claudio F. Paganini \\ \\ May 2024}
\address{Fakult\"at f\"ur Mathematik \\ Universit\"at Regensburg \\ D-93040 Regensburg \\ Germany}%
\email{claudio.paganini@ur.de}

\newcommand{\bl}{\textcolor{blue}}

\newtheorem{Def}{Definition}[section]
\newtheorem{Thm}[Def]{Theorem}
\newtheorem{Prp}[Def]{Proposition}
\newtheorem{Lemma}[Def]{Lemma}

\newcommand{\Thanks}{\vspace*{.5em} \noindent \thanks}
\newcommand{\beq}{\begin{equation}}
\newcommand{\eeq}{\end{equation}}
\newcommand{\Proof}{\begin{proof}}
	\newcommand{\QED}{\end{proof} \noindent}

\newcommand{\la}{\langle}
\newcommand{\ra}{\rangle}

\newcommand{\ket}{\mathclose{>}}
\newcommand{\bbra}{\mathopen{\ll}}
\newcommand{\kket}{\mathclose{\gg}}
\newcommand{\Sl}{\mbox{$\prec \!\!$ \nolinebreak}}
\newcommand{\Sr}{\mbox{\nolinebreak $\succ$}}
\newcommand{\C}{\mathbb{C}}
\newcommand{\R}{\mathbb{R}}
\newcommand{\1}{\mathds{1}}

\newcommand{\N}{\mathbb{N}}
\newcommand{\Pdd}{\mbox{$\partial$ \hspace{-1.2 em} $/$}}

\renewcommand{\Tr}{\text{\rm{Tr}}}
\DeclareMathOperator{\tr}{tr}
\renewcommand{\O}{{\mathscr{O}}}
\renewcommand{\L}{{\mathcal{L}}}
\newcommand{\Sact}{{\mathcal{S}}}

\newcommand{\B}{{\mathscr{B}}}

\renewcommand{\H}{\mathscr{H}}

\newcommand{\Lin}{\text{\rm{L}}}
\newcommand{\F}{{\mathscr{F}}}

\DeclareMathOperator{\Symm}{\mbox{\rm{Symm}}}

\DeclareMathOperator{\re}{Re}

\DeclareMathOperator{\supp}{supp}
\newcommand{\Sig}{\mathscr{S}}

\newcommand{\scrN}{\mycal N}
\newcommand{\itemD}{\item[{\raisebox{0.125em}{\tiny $\blacktriangleright$}}]}
\newcommand{\scrU}{{\mathscr{U}}}

\newcommand{\s}{\mathfrak{s}}

\newcommand{\bitem}{\begin{itemize}[leftmargin=2.5em]}
\newcommand{\eitem}{\end{itemize}}

\newcommand{\sint}{\sum  \!\!\!\!\!\!\! \int }
\newcommand{\res}{\text{\rm{\tiny{res}}}}

\newcommand{\bM}{\mathbf{M}}
\newcommand{\bW}{\mathbf{W}}

\DeclareFontFamily{OT1}{rsfso}{}
\DeclareFontShape{OT1}{rsfso}{m}{n}{ <-7> rsfso5 <7-10> rsfso7 <10-> rsfso10}{}
\DeclareMathAlphabet{\mycal}{OT1}{rsfso}{m}{n}

\newcommand\Felix[1]{}
\newcommand\Claudio[1]{}
\newcommand\Johannes[1]{}

\begin{document}
\maketitle

\begin{abstract}
It is shown that, in the non-relativistic limit, causal fermion systems give rise to an effective collapse theory. The nonlinear and stochastic correction terms to the Schr\"odinger equation are derived from the causal action principle. The dynamics of the statistical operator is described by a deterministic equation of Kossakowski-Lindblad form. Moreover, the quantum state undergoes a dynamical collapse compatible with the Born rule. The effective model has similarities with the continuous spontaneous localization model, but differs from it by a conservation law for the probability integral as well as a non-locality in time on a microscopic length scale~$\ell_{\min}$.
\end{abstract}

\tableofcontents

\section{Introduction} \label{secintro} 
It is one of the most outstanding open problems of contemporary physics to reconcile the linear dynamics
of quantum theory with the reduction of the state vector in a measurement process.
Different solutions to this so-called {\em{measurement problem}} have been proposed,
among them Bohmian mechanics~\cite{duerr+teufel}, the many-worlds interpretation~\cite{dewitt-manyworlds},
decoherence~\cite{joos} and collapse models~\cite{gao}.
Moreover, there are attempts to derive the state reduction from a pre-quantum theory
(see for example~\cite{thooft1, thooft2, khrennikov, khrennikov2}).
Despite all these different proposals, a derivation of the state reduction from first principles is still lacking.
In the present paper, such a derivation is given, taking the theory of causal fermion systems as the fundamental
physical theory. In this theory, the dynamics of the physical system is described by
a novel action principle, the {\em{causal action principle}}. The resulting dynamical equations
are {\em{nonlinear}}. Moreover, the corresponding field equations allow for a plethora of solutions,
which can best be described {\em{stochastically}}.
Therefore, all the basic ingredients for an effective collapse
model are already present. Here we shall work out the resulting effects in detail
in the non-relativistic limit to obtain an effective collapse model.
In this way, the measurement problem is solved from first principles
starting from a fundamental physical theory.

{\em{Collapse theories}} were introduced more than thirty years ago
in order to explain the collapse of the wave function in the measurement process~\cite{karolyhazy,
karolyhazy2, ghirardi1, gisin, pearle0, ghirardi-pearle-rimini, ghirardi2}.
The general concept is to regard the wave function as the fundamental physical object of the theory,
whereas the particle character is a consequence of the dynamical collapse.
The reduction of the wave function is realized by modifying the Schr\"odinger dynamics by nonlinear
and stochastic terms. The most prominent example is the
{\em{continuous spontaneous localization (CSL) model}}.
In the collapse models, the modifications of the Schr\"odinger dynamics are introduced
ad hoc. One idea to explain why the collapse comes about is to consider gravity as
being fundamentally classical, and to associate the nonlinear correction term to
the nonlinear coupling of the wave function in Newton's law of gravity
(see~\cite{diosi1, diosi2, penrose-collapse, penrose-collapse2} or the more recent approach
in~\cite{lulli-marciano-psciccia}). Detailed introductions to and surveys on
collapse models can be found for example~\cite{ bassi-ghirardi, pearle, bassi, bassi-duerr-hinrichs}.

The theory of {\em{causal fermion systems}} is a recent approach to fundamental physics
(see the basics in Section~\ref{secprelim}, the reviews~\cite{dice2014, review}, the textbooks~\cite{cfs, intro}
or the website~\cite{cfsweblink}).
In this approach, spacetime and all objects therein are described by a measure~$\rho$
on a set~$\F$ of linear operators on a Hilbert space~$(\H, \la .|. \ra_\H)$. 
The physical equations are formulated by means of the so-called {\em{causal action principle}},
a nonlinear variational principle where an action~$\Sact$ is minimized under variations of the measure~$\rho$.
Causal fermion systems allow for the description of generalized ``quantum'' spacetimes,
which macroscopically look like Minkowski space, but on a microscopic length scale~$\varepsilon$
(which can be identified with the Planck scale) may have a different, possibly discrete structure.

The general idea that the causal action principle should incorporate collapse phenomena
was first proposed in~\cite[Chapter~5]{kleinerdr}. However, working out the resulting effects 
quantitatively and in detail became possible only more recently, after having gained a more complete
understanding of the structure of the solutions of the corresponding Euler-Lagrange equations
and its linearized version, the so-called linearized field equations.
Our starting point is the observation in~\cite{nonlocal} that the causal action principle
does not only allow for classical fields (like electromagnetic or gravitational waves),
but that it gives rise to a multitude of additional fields. The number~$N$ of these fields scales like
\beq \label{Nscaleintro}
N \simeq \frac{\ell_{\min}}{\varepsilon} \:,
\eeq
where~$\ell_{\min}$ is a physical parameter which lies between the regularization scale~$\varepsilon$ and the
length scale~$\ell_\text{macro}$ of macroscopic physics,
\beq \label{ellmindef}
\varepsilon \ll \ell_{\min} \ll \ell_\text{macro} \:.
\eeq
In particular, the number of fields tends to infinity if the regularization length~$\varepsilon$ tends to zero.
{Each of these fields propagates with the speed of light.
Intuitively speaking, these fields can be thought of as being formed of a plethora of
electromagnetic fields coupling to different wave packets and propagating in different directions,
each of them localized on the scale~$\ell_{\min}$.
This multitude of fields has far-reaching consequences for the dynamics of the system.
Namely, as worked out in~\cite{fockdynamics}, it gives rise to an effective dynamics described by
second-quantized fields. Moreover, in the non-relativistic limit, it gives rise to the stochastic term of
collapse models as follows. Suppose we consider a wave function in a laboratory.
Then the laboratory is penetrated by all the fields which all couple to the wave function
(see Figure~\ref{figcoll1}).
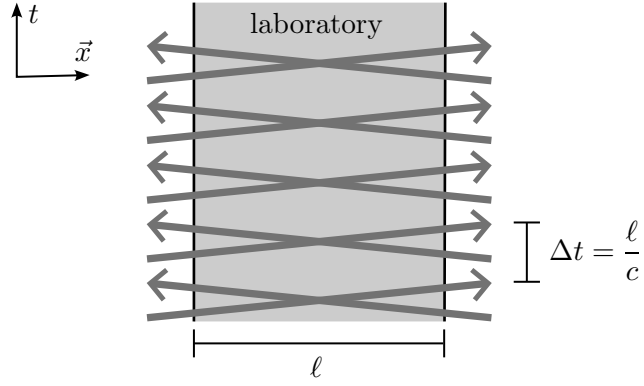
\begin{figure}
\psset{xunit=.5pt,yunit=.5pt,runit=.5pt}
\begin{pspicture}(398.05444913,267.05817275)
{
\newrgbcolor{curcolor}{0.80000001 0.80000001 0.80000001}
\pscustom[linestyle=none,fillstyle=solid,fillcolor=curcolor]
{
\newpath
\moveto(138.05960817,266.55476172)
\lineto(325.7080101,266.55476172)
\lineto(325.7080101,27.53117524)
\lineto(138.05960817,27.53117524)
\closepath
}
}
{
\newrgbcolor{curcolor}{0.80000001 0.80000001 0.80000001}
\pscustom[linewidth=1.00157103,linecolor=curcolor]
{
\newpath
\moveto(138.05960817,266.55476172)
\lineto(325.7080101,266.55476172)
\lineto(325.7080101,27.53117524)
\lineto(138.05960817,27.53117524)
\closepath
}
}
{
\newrgbcolor{curcolor}{0 0 0}
\pscustom[linewidth=2.00000125,linecolor=curcolor]
{
\newpath
\moveto(137.55882331,267.05554976)
\lineto(137.55882331,27.03039322)
}
}
{
\newrgbcolor{curcolor}{0 0 0}
\pscustom[linewidth=2.00000125,linecolor=curcolor]
{
\newpath
\moveto(326.83915465,267.05554976)
\lineto(326.2087937,27.03039322)
}
}
{
\newrgbcolor{curcolor}{0.44705883 0.44705883 0.44705883}
\pscustom[linewidth=5.50000619,linecolor=curcolor]
{
\newpath
\moveto(102.26135811,29.81339543)
\lineto(358.1827389,55.60161464)
}
}
{
\newrgbcolor{curcolor}{0.44705883 0.44705883 0.44705883}
\pscustom[linewidth=3.77952756,linecolor=curcolor]
{
\newpath
\moveto(348.97819636,43.27810581)
\lineto(359.12285993,55.69634704)
\lineto(346.7046187,65.84101061)
}
}
{
\newrgbcolor{curcolor}{0.44705883 0.44705883 0.44705883}
\pscustom[linewidth=5.50000619,linecolor=curcolor]
{
\newpath
\moveto(361.50622866,29.81339543)
\lineto(105.58482898,55.60161464)
}
}
{
\newrgbcolor{curcolor}{0.44705883 0.44705883 0.44705883}
\pscustom[linewidth=3.77952756,linecolor=curcolor]
{
\newpath
\moveto(117.0629491,65.8410107)
\lineto(104.64470794,55.69634704)
\lineto(114.7893716,43.27810588)
}
}
{
\newrgbcolor{curcolor}{0.44705883 0.44705883 0.44705883}
\pscustom[linewidth=5.50000619,linecolor=curcolor]
{
\newpath
\moveto(102.26137323,74.09495637)
\lineto(358.1827389,99.88317559)
}
}
{
\newrgbcolor{curcolor}{0.44705883 0.44705883 0.44705883}
\pscustom[linewidth=3.77952756,linecolor=curcolor]
{
\newpath
\moveto(348.97819643,87.5596667)
\lineto(359.12285993,99.97790799)
\lineto(346.70461864,110.12257149)
}
}
{
\newrgbcolor{curcolor}{0.44705883 0.44705883 0.44705883}
\pscustom[linewidth=5.50000619,linecolor=curcolor]
{
\newpath
\moveto(361.50622866,74.09495637)
\lineto(105.58484409,99.88317559)
}
}
{
\newrgbcolor{curcolor}{0.44705883 0.44705883 0.44705883}
\pscustom[linewidth=3.77952756,linecolor=curcolor]
{
\newpath
\moveto(117.06296428,110.12257157)
\lineto(104.64472306,99.97790799)
\lineto(114.78938665,87.55966677)
}
}
{
\newrgbcolor{curcolor}{0.44705883 0.44705883 0.44705883}
\pscustom[linewidth=5.50000619,linecolor=curcolor]
{
\newpath
\moveto(102.26138835,118.37651732)
\lineto(358.1827389,144.16473653)
}
}
{
\newrgbcolor{curcolor}{0.44705883 0.44705883 0.44705883}
\pscustom[linewidth=3.77952756,linecolor=curcolor]
{
\newpath
\moveto(348.97819651,131.84122759)
\lineto(359.12285993,144.25946894)
\lineto(346.70461858,154.40413237)
}
}
{
\newrgbcolor{curcolor}{0.44705883 0.44705883 0.44705883}
\pscustom[linewidth=5.50000619,linecolor=curcolor]
{
\newpath
\moveto(361.50622866,118.37651732)
\lineto(105.58485921,144.16473653)
}
}
{
\newrgbcolor{curcolor}{0.44705883 0.44705883 0.44705883}
\pscustom[linewidth=3.77952756,linecolor=curcolor]
{
\newpath
\moveto(117.06297946,154.40413245)
\lineto(104.64473818,144.25946894)
\lineto(114.78940169,131.84122766)
}
}
{
\newrgbcolor{curcolor}{0.44705883 0.44705883 0.44705883}
\pscustom[linewidth=5.50000619,linecolor=curcolor]
{
\newpath
\moveto(102.26140346,163.30927952)
\lineto(358.1827389,189.09749874)
}
}
{
\newrgbcolor{curcolor}{0.44705883 0.44705883 0.44705883}
\pscustom[linewidth=3.77952756,linecolor=curcolor]
{
\newpath
\moveto(348.97819658,176.77398974)
\lineto(359.12285993,189.19223116)
\lineto(346.70461852,199.3368945)
}
}
{
\newrgbcolor{curcolor}{0.44705883 0.44705883 0.44705883}
\pscustom[linewidth=5.50000619,linecolor=curcolor]
{
\newpath
\moveto(361.50622866,163.30927952)
\lineto(105.58487433,189.09749874)
}
}
{
\newrgbcolor{curcolor}{0.44705883 0.44705883 0.44705883}
\pscustom[linewidth=3.77952756,linecolor=curcolor]
{
\newpath
\moveto(117.06299464,199.33689459)
\lineto(104.6447533,189.19223115)
\lineto(114.78941674,176.77398981)
}
}
{
\newrgbcolor{curcolor}{0.44705883 0.44705883 0.44705883}
\pscustom[linewidth=5.50000619,linecolor=curcolor]
{
\newpath
\moveto(102.26140346,208.24204173)
\lineto(358.1827389,234.03026094)
}
}
{
\newrgbcolor{curcolor}{0.44705883 0.44705883 0.44705883}
\pscustom[linewidth=3.77952756,linecolor=curcolor]
{
\newpath
\moveto(348.97819658,221.70675195)
\lineto(359.12285993,234.12499336)
\lineto(346.70461852,244.26965671)
}
}
{
\newrgbcolor{curcolor}{0.44705883 0.44705883 0.44705883}
\pscustom[linewidth=5.50000619,linecolor=curcolor]
{
\newpath
\moveto(361.50622866,208.24204173)
\lineto(105.58487433,234.03026094)
}
}
{
\newrgbcolor{curcolor}{0.44705883 0.44705883 0.44705883}
\pscustom[linewidth=3.77952756,linecolor=curcolor]
{
\newpath
\moveto(117.06299464,244.26965679)
\lineto(104.6447533,234.12499335)
\lineto(114.78941674,221.70675202)
}
}
{
\newrgbcolor{curcolor}{0 0 0}
\pscustom[linewidth=1.49999998,linecolor=curcolor]
{
\newpath
\moveto(4.20000378,257.60555165)
\lineto(4.20000378,211.16055133)
\lineto(49.60548661,212.29569086)
}
}
{
\newrgbcolor{curcolor}{0 0 0}
\pscustom[linestyle=none,fillstyle=solid,fillcolor=curcolor]
{
\newpath
\moveto(0.00000384,255.50555168)
\lineto(4.20000378,267.05555151)
\lineto(8.40000372,255.50555168)
\curveto(5.95350375,257.60555165)(2.45700381,257.60555165)(0.00000384,255.50555168)
\closepath
}
}
{
\newrgbcolor{curcolor}{0 0 0}
\pscustom[linestyle=none,fillstyle=solid,fillcolor=curcolor]
{
\newpath
\moveto(47.40117516,216.44189519)
\lineto(59.05253472,212.53186758)
\lineto(47.61111002,208.0445191)
\curveto(49.64931052,210.5427386)(49.56192513,214.0381464)(47.40117516,216.44189519)
\closepath
}
}
{
\newrgbcolor{curcolor}{0 0 0}
\pscustom[linewidth=1.49999998,linecolor=curcolor]
{
\newpath
\moveto(378.3066822,101.53772425)
\lineto(398.05444913,101.53772425)
}
}
{
\newrgbcolor{curcolor}{0 0 0}
\pscustom[linewidth=1.49999998,linecolor=curcolor]
{
\newpath
\moveto(388.18054677,101.53772425)
\lineto(388.18054677,56.89061496)
}
}
{
\newrgbcolor{curcolor}{0 0 0}
\pscustom[linewidth=1.49999998,linecolor=curcolor]
{
\newpath
\moveto(378.3066822,56.89061496)
\lineto(398.05444913,56.89061496)
}
}
{
\newrgbcolor{curcolor}{0 0 0}
\pscustom[linewidth=1.49999998,linecolor=curcolor]
{
\newpath
\moveto(137.55882709,21.03566944)
\lineto(137.55882709,0.00001338)
}
}
{
\newrgbcolor{curcolor}{0 0 0}
\pscustom[linewidth=1.49999998,linecolor=curcolor]
{
\newpath
\moveto(137.55882709,10.51784141)
\lineto(325.20879496,10.51784141)
}
}
{
\newrgbcolor{curcolor}{0 0 0}
\pscustom[linewidth=1.49999998,linecolor=curcolor]
{
\newpath
\moveto(325.94323276,21.03566944)
\lineto(325.94323276,0.00002094)
}
\rput[bl](180,240){laboratory}
\rput[bl](405,58){$\displaystyle \Delta t = \frac{\ell}{c}$}
\rput[bl](13,252){$t$}
\rput[bl](48,223){$\vec{x}$}
\rput[bl](225,-15){$\ell$}
}
\end{pspicture}
\caption{Stochastic fields coupling to the wave function.}
\label{figcoll1}
\end{figure}%
We drew the speed of light with an angle smaller than~$45$ degrees in order to point out that
measurements and the collapse of the wave function typically occur on time scales which are
much larger than the size of the laboratory divided by the speed of light.
Describing all these fields stochastically with Gaussian distributions
will give rise to a stochastic correction term to the
evolution equation for the quantum state.
In the non-relativistic limit} is obtained by taking the limiting case~$\Delta t \searrow 0$,
the stochastic fields at different times are uncorrelated, giving rise to Markovian stochastic fields.

In our derivation of the collapse model we also make essential use of the specific features
of the causal action principle. Most importantly, as already mentioned, it is nonlinear and
the resulting fields can be described stochastically.
Moreover, in causal fermion systems there are Noether-like conservation laws.
In particular, current conservation holds in the sense that the so-called commutator inner product
is time independent (for details see Section~\ref{secnonlinear}).
This means that, despite the complicated stochastic and nonlinear nature of the interaction, the
causal action principle ensures that the wave function is properly normalized and that
its probabilistic interpretation remains meaningful. This gives a compelling explanation for the
fact that, in our effective collapse model, the nonlinear and the stochastic terms is
of the Kossakowski-Lindblad form.
Another essential ingredient to the derivation of collapse is that the
dynamical equations as well as the conserved commutator inner product are {\em{nonlocal}} in space and time
on the scale~$\ell_{\min}$.

We now describe the structure of our effective collapse model in some more detail.
In the non-relativistic limit, the dynamics is described by the Schr\"odinger-type equation
\beq \label{schintro}
i \partial_t \psi = H \psi
\eeq
where the Hamiltonian is the sum of the free Hamiltonian~$H_0$
(being the Dirac Hamiltonian~$H_0 := -i \gamma^0 \vec{\gamma} \vec{\nabla}$ or the
Schr\"odinger Hamiltonian~$H_0 = -\Delta/2m$) and a potential~$V$,
\[ H = H_0 + V \:. \]
This potential is nonlocal in space and time, meaning that it can be written as the integral operator with
nonlocal kernel
\beq \label{Vnonloc}
(V \psi)(t, \vec{x}) = \int_{-\infty}^\infty dt' \int_{\R^3} d^3y \:V(t, \vec{x}; t', \vec{y})\: \psi(t', \vec{y}) \:.
\eeq
The nonlocal potential is expressed stochastically as
\beq \label{Mdefintro}
V(t, \vec{x}; t', \vec{y}) = \sum_{k,a} \: M_{k,a}(t, \vec{x};t', \vec{y})\: W_{k,a}\Big( \frac{t+t'}{2} \Big) \:,
\eeq
where~$M_{k,a}$ are given kernels, and~$W_{k,a}$ are normalized Markovian Gaussian fields, i.e.\
\beq \label{Wintro}
\bbra W_{k,\alpha}(\tau) \kket = 0 \qquad \text{and} \qquad
\bbra W_{k,a}(\tau)\, W_{l,b}(\tau') \kket = \delta(\tau-\tau')\, \delta_{ab} \, \delta_{kl} \:,
\eeq
where~$\bbra \cdots \kket$ denotes the statistical mean.
Here~$k \in \N$ is a quantum number which arises when diagonalizing the covariance operator
(for details see Section~\ref{secmarkov} and the proof of Lemma~\ref{lemmadiag}).
An important feature of this setup is that the time~$t$ in the Schr\"odinger equation~\eqref{schintro}
is in general different from the time~$\tau$ at which the Gaussian field in~\eqref{Wintro} is evaluated.
Indeed, the nonlocal potential~$V$ in time~\eqref{Vnonloc} involves two times~$t$
and~$t'$ which are in general different. According to~\eqref{Mdefintro}, the Gaussian field is evaluated
at time~$(t+t')/2$.
Denoting the difference of these two times by~$\zeta$, we get a new parameter which can be treated
on the same footing as the labels~$k$ and~$a$ of the Gaussian field in~\eqref{Mdefintro}.
We thus combine all these parameters to a new parameter~$\varkappa := (k, a, \zeta)$.
After doing so, our model is similar to the CSL model, but still has a different structure.
In particular, due to the nonlocality in time, the evolution for the state cannot be described
as a stochastic partial differential equation (for example in the It{\^o} calculus).
Instead, the time evolution is described by a nonlocal Dyson series. The statistical mean
of composite expressions and their dynamics can be computed using the Wick rules for
Markovian Gaussian fields~\eqref{Wintro}.

At this stage, the subtle question arises which composite expressions are physically meaningful.
Here we must make use of the fact that the scalar product which is conserved in time and gives rise
to the probabilistic interpretation of the wave function is {\em{not}} the standard $L^2(\R^3)$-scalar product,
but instead the so-called {\em{commutator inner product}}~$\la .|. \ra_t$ which is again nonlocal both
in space and time
\begin{align}
\la \psi | \phi \ra_t &= \int \psi^\dagger(t,\vec{x}) \,\phi(t,\vec{x}) \: d^3x \label{commintro1} \\
&\quad\:+i \int_{-\infty}^t d\tau \int_{\R^3} d^3x \int_t^\infty d\tau' \int_{\R^3} d^3y \:
\psi(\tau, \vec{x})^\dagger\, V(\tau,\vec{x}; \tau', \vec{y}) \:\phi(\tau', \vec{y}) \label{commintro2} \\
&\quad\:-i \int_t^\infty d\tau \int_{\R^3} d^3x \int_{-\infty}^t d\tau' \int_{\R^3} d^3y \:
\psi(\tau, \vec{x})^\dagger\, V(\tau,\vec{x}; \tau', \vec{y}) \:\phi(\tau', \vec{y}) \:, \label{commintro3}
\end{align}
where the dagger is the complex wave function (thus~\eqref{commintro1} is the standard $L^2$-scalar product,
whereas~\eqref{commintro2} and~\eqref{commintro3} are the nonlocal contributions).
Consequently, the {\em{statistical operator}}~$\sigma_t$ must be formed with respect to this scalar product, i.e.\
\[ \big( \sigma_t \phi \big)(t) := \bbra \:\psi(t) \,\la \psi | \phi \ra_t \:\kket \:, \]
(where~$\phi$ is any wave function chosen independent of the stochastic fields, and $\bbra \cdots \kket$
again denotes the statistical average). In bra/ket notation, the statistical operator can be written shorter as
\[ \sigma_t = \bbra \: |\psi \ra_t \la \psi | \:\kket \:. \]
To leading order in the length scale~$\ell_{\min}$, its
time evolution is shown to be of the {\em{Kossa\-kow\-ski-Lindblad form}}
\[ \frac{d \sigma_t}{dt} = -i [H, \sigma_t] - \frac{1}{2} \sint_\varkappa 
\big[ K_\varkappa, [K_\varkappa, \sigma_t] \big] \; \Big( 1 + \O \big( \ell_{\min}\, \|V\| \big) \Big)\:, \]
where~$K_\varkappa$ are spatial operators formed of the kernels~$M_{k,a}$ in~\eqref{Mdefintro}
(for details see Theorem~\ref{thmlindblad}).
In order to derive collapse, one must take into account that the
commutator inner product involves the stochastic potential (as is obvious from~\eqref{commintro2} and~\eqref{commintro3}).
In order to get a closer connection to the usual Schr\"odinger dynamics, we
transform to the standard $L^2(\R^3)$-scalar product by a nonlocal spatial
transformation~$\psi \mapsto \tilde{\psi}$.
The fact that the nonlocal fluctuating terms in the resulting
Hamiltonian do not commute with the operators describing local measurements is what
triggers the collapse (the detailed mechanism will be worked out in Section~\ref{seccollapse}).

Our effective collapse model has similarities with the continuous spontaneous localization (CSL) model.
A major difference is the non-locality of the dynamical equations and the conserved current
in time on the scale~$\ell_{\min}$. Due to this non-locality in time, it does not seem possible to
relate our model directly to the CSL model. Our model gives rise to a deterministic evolution equation
for the statistical operator of Kossakowski-Lindblad form. It also has the no-signalling property.
Similar to the CSL model, our effective model involves two parameters:
\bitem
\itemD The length scale~$\ell_{\min}$ in~\eqref{ellmindef}.
\itemD The strength of the stochastic background field as described by the
operators~$M_{k,a}$ in~\eqref{Mdefintro}.
\eitem
In contrast to the concept behind the Diosi-Penrose model,
in our model it is {\em{not}} the gravitational field which triggers the collapse.
Instead, the collapse is a collective effect of the multitude of bosonic fields which all
couple to the wave functions. We remark that, as worked out in~\cite{fockdynamics}, this multitude of
bosonic fields can also be described in the standard formalism of quantum field theory by a second-quantized electromagnetic field. With this in mind, the collapse is closely related to the
electromagnetic interaction in quantum field theory.

We finally remark that our physical picture is reminiscent and has similarities with J\"urg Fr\"ohlich's
ETH formulation of quantum theory~\cite{froehlich2019review, frohlich2015math, F, froehlich-pizzo2,
froehlich-gang-pizzo}. In simple terms, the connection can be made as follows.
The fields which couple to the collapsing wave function (as shown in Figure~\ref{figcoll1})
all leave the laboratory after a short time~$\Delta t$. They carry information on the physical system
inside the laboratory. This information is lost when the fields leave the system.
This is very similar to Fr\"ohlich's {\em{principle of diminishing potentialities}}.
The loss of information is the reason why the effective dynamics is irreversible and not unitary.
However, our model also has the major differences to the ETH formulation that the collapse is
caused by a nonlocal term in the evolution equations, rather than by a branching process taking place
at the ``events'' of the ETH formulation. At present it is unclear how these two mechanisms are related
to each other. A general comparison between
structures of the ETH formulation and causal fermion systems can be found in~\cite{eth-cfs}.

The paper is structured as follows. Section~\ref{secprelim} provides the necessary background on
causal fermion systems and the causal action principle. After introducing the basic
structure (Section~\ref{seccfs}), we explain how the dynamics can be described with
a Dirac equation involving a nonlocal potential (Section~\ref{secdir}). Moreover, we explain
how the nonlinearity comes about (Section~\ref{secnonlinear}).
In Section~\ref{secnonrelativ} we consider a simplified model of
a purely spatial nonlocality. This section also introduces the needed concepts from stochastic analysis
and makes the contact to collapse models. As we shall see, this model is too simple for describing
collapse phenomena. In Section~\ref{secmodel} we proceed with a more realistic model which incorporates
all the relevant features of the dynamics as described by the causal action principle.
In the one-particle picture, we consider a Dirac equation involving a potential which is nonlocal both in space
and time (Section~\ref{secnonloc}). After writing the dynamics in the Hamiltonian form (Section~\ref{sechamilton})
and transforming to the standard $L^2$-scalar product in space (Section~\ref{sectransform}),
we describe the fields stochastically (Section~\ref{secmarkov}) to obtain an effective dynamics of
Kossakowski-Lindblad form (Section~\ref{seclindblad}).
We then derive that the wave function collapses (Section~\ref{seccollapse}).
Our results carry over to the Fock space description (Section~\ref{secfock}).
We conclude with a discussion of the effective parameters of the model (Section~\ref{secparameters})
and of the nature of the nonlinear term (Section~\ref{secnonlin}).
In Section~\ref{secrelativity} we give an outlook on the corresponding relativistic collapse model.
Finally, the appendix provides detailed computations needed for taking the non-relativistic limit.

\section{Preliminaries} \label{secprelim}
This section provides the necessary background on causal fermion systems
and introduces all the basic structures needed later on.

\subsection{A Few Basics on Causal Fermion Systems} \label{seccfs}
\subsubsection{Causal Fermion Systems and the Reduced Causal Action Principle}
We now recall the basic setup and introduce the main objects to be used later on.
\begin{Def} \label{defcfs} (causal fermion systems) {\em{ 
Given a separable complex Hilbert space~$\H$ with scalar product~$\la .|. \ra_\H$
and a parameter~$n \in \N$ (the {\em{``spin dimension''}}), we let~$\F \subset \Lin(\H)$ be the set of all
symmetric operators on~$\H$ of finite rank, which (counting multiplicities) have
at most~$n$ positive and at most~$n$ negative eigenvalues. On~$\F$ we are given
a positive measure~$\rho$ (defined on a $\sigma$-algebra of subsets of~$\F$).
We refer to~$(\H, \F, \rho)$ as a {\em{causal fermion system}}.
}}
\end{Def} \noindent

A causal fermion system describes a spacetime together
with all structures and objects therein.
In order to single out the physically admissible
causal fermion systems, one must formulate physical equations. To this end, we impose that
the measure~$\rho$ should be a minimizer of the causal action principle,
which we now introduce. For brevity of the presentation, we only consider the
{\em{reduced causal action principle}} where the so-called boundedness constraint has been
incorporated by a Lagrange multiplier term. This simplification is no loss of generality, because
the resulting EL equations are the same as for the non-reduced action principle
as introduced for example in~\cite[Section~\S1.1.1]{cfs}.

For any~$x, y \in \F$, the product~$x y$ is an operator of rank at most~$2n$. 
However, in general it is no longer a symmetric operator because~$(xy)^* = yx$,
and this is different from~$xy$ unless~$x$ and~$y$ commute.
As a consequence, the eigenvalues of the operator~$xy$ are in general complex.
We denote these eigenvalues counting algebraic multiplicities
by~$\lambda^{xy}_1, \ldots, \lambda^{xy}_{2n} \in \C$
(more specifically,
denoting the rank of~$xy$ by~$k \leq 2n$, we choose~$\lambda^{xy}_1, \ldots, \lambda^{xy}_{k}$ as all
the non-zero eigenvalues and set~$\lambda^{xy}_{k+1}, \ldots, \lambda^{xy}_{2n}=0$).
Given a parameter~$\kappa>0$ (which will be kept fixed throughout this paper),
we introduce the $\kappa$-Lagrangian and the causal action by
\begin{align*}
\text{\em{$\kappa$-Lagrangian:}} && \L(x,y) &= 
\frac{1}{4n} \sum_{i,j=1}^{2n} \Big( \big|\lambda^{xy}_i \big|
- \big|\lambda^{xy}_j \big| \Big)^2 + \kappa\: \bigg( \sum_{j=1}^{2n} \big|\lambda^{xy}_j \big| \bigg)^2 \\
\text{\em{causal action:}} && \Sact(\rho) &= \iint_{\F \times \F} \L(x,y)\: d\rho(x)\, d\rho(y) \:. 
\end{align*}
The {\em{reduced causal action principle}} is to minimize~$\Sact$ by varying the measure~$\rho$
under the following constraints,
\begin{align*}
\text{\em{volume constraint:}} && \rho(\F) = 1 \quad\;\; \\
\text{\em{trace constraint:}} && \int_\F \tr(x)\: d\rho(x) = 1 \:. 
\end{align*}
This variational principle is mathematically well-posed if~$\H$ is finite-dimensional.
For a review of the existence theory and the analysis of general properties of minimizing measures
we refer to~\cite[Chapter~12]{intro}.

\subsubsection{Spacetime, Causal Structure and the Physical Wave Functions} \label{secweo}
Let~$\rho$ be a {\em{minimizing}} measure. {\em{Spacetime}}~$M$ is defined as the support
of this measure,
\[ 
M := \supp \rho \subset \F \:. \]
the spacetimes points are symmetric linear operators on~$\H$.
These operators contain a lot of information which, if interpreted correctly,
gives rise to spacetime structures like causal and metric structures, spinors
and interacting fields (for details see~\cite[Chapter~1]{cfs}).
Here we restrict attention to those structures needed in what follows.
We begin with a basic notion of causality.

\begin{Def} (causal structure) \label{def2}
{\em{For any~$x, y \in \F$, the product~$x y$ is an operator
of rank at most~$2n$. We denote its non-trivial eigenvalues (counting algebraic multiplicities)
by~$\lambda^{xy}_1, \ldots, \lambda^{xy}_{2n}$.
The points~$x$ and~$y$ are
called {\em{spacelike}} separated if all the~$\lambda^{xy}_j$ have the same absolute value.
They are said to be {\em{timelike}} separated if the~$\lambda^{xy}_j$ are all real and do not all 
have the same absolute value.
In all other cases (i.e.\ if the~$\lambda^{xy}_j$ are not all real and do not all 
have the same absolute value),
the points~$x$ and~$y$ are said to be {\em{lightlike}} separated. }}
\end{Def} \noindent
Restricting the causal structure of~$\F$ to~$M$, we get causal relations in spacetime.

Next, for every~$x \in M$ we define the {\em{spin space}}~$S_xM$ by~$S_xM = x(\H)$;
it is a subspace of~$\H$ of dimension at most~$2n$.
It is endowed with the {\em{spin inner product}} $\Sl .|. \Sr_x$ defined by
\beq \label{ssp}
\Sl u | v \Sr_x = -\la u | x v \ra_\H \qquad \text{(for all $u,v \in S_xM$)}\:.
\eeq
A {\em{wave function}}~$\psi$ is defined as a function
which to every~$x \in M$ associates a vector of the corresponding spin space,
\beq \label{psirep}
\psi \::\: M \rightarrow \H \qquad \text{with} \qquad \psi(x) \in S_xM \quad \text{for all~$x \in M$}\:.
\eeq

It is an important observation that every vector~$u \in \H$ of the Hilbert space gives rise to a distinguished
wave function. In order to obtain this wave function, denoted by~$\psi^u$, we simply project the vector~$u$
to the corresponding spin spaces,
\beq \label{psiudef}
\psi^u \::\: M \rightarrow \H\:,\qquad \psi^u(x) = \pi_x u \in S_xM \:.
\eeq
We refer to~$\psi^u$ as the {\em{physical wave function}} of~$u \in \H$.

\subsubsection{The Euler-Lagrange Equations and the Dynamical Wave Equation}
We now state the Euler-Lagrange equations.
\begin{Prp} Let~$\rho$ be a minimizer of the reduced causal action principle.
Then the local trace is constant in spacetime, meaning that
\[ 
\tr(x) = 1 \qquad \text{for all~$x \in M$} \:. \]
Moreover, there are parameters~$\mathfrak{r}, \s>0$ such that
the function~$\ell$ defined by
\[ 
\ell \::\: \F \rightarrow \R\:,\qquad \ell(x) := \int_M \L(x,y)\: d\rho(y) - \mathfrak{r}\, \big( \tr(x) -1 \big) - \s \]
is minimal and vanishes in spacetime, i.e.\
\beq \label{EL}
\ell|_M \equiv \inf_\F \ell = 0 \:.
\eeq
\end{Prp} \noindent
For the proof of the EL equations and more details we refer for example to~\cite{nonlocal}.
The parameter~$\mathfrak{r}$ can be viewed as the Lagrange parameter corresponding to the
trace constraint. Likewise, $\s$ is the Lagrange parameter of the volume constraint.

For our purposes, it is most convenient to formulate the EL equations in terms of the operator~$Q$
which we now introduce (for more details see~\cite[Sections~2 and~3]{nonlocal}).
We first note that the Lagrangian~$\L(x,y)$ can be expressed in terms of the 
{\em{kernel of the fermionic projector}} $P(x,y)$ defined by
\[ 
P(x,y) := -\pi_x y \::\: S_yM \rightarrow S_xM \:. \]
Consequently, the first variation of the Lagrangian can be expressed in terms
of the first variation of this kernel. Being real-valued and real-linear in~$\delta P(x,y)$,
it can be written as
\[ 
\delta \L(x,y) = 2 \re \Tr_{S_xM} \!\big( Q(x,y)\, \delta P(x,y)^* \big) \]
(where~$\Tr_{S_xM}$ denotes the trace on the spin space~$S_xM$)
with a kernel~$Q(x,y)$ which is again symmetric (with respect to the spin inner product), i.e.
\beq \label{Qxydef}
Q(x,y) \::\: S_yM \rightarrow S_xM \qquad \text{and} \qquad Q(x,y)^* = Q(y,x) \:.
\eeq
More details on this method and many computations can be found in~\cite[Sections~1.4 and~2.6
as well as Chapters~3-5]{cfs}.
Using these formulas, the EL equation~\eqref{EL} imply that the physical wave functions
must all satisfy the equation
\[ \int_M Q(x,y)\, \psi^u(y) \:d\rho(y) = \mathfrak{r}\, \psi^u(x) \qquad \text{for all~$x \in M$ and~$u \in \H$}\:, \]
where~$\mathfrak{r} \in \R$ is the Lagrange parameter of the trace constraint.
This equation is referred to as the {\em{dynamical wave equation}}.
Denoting the integral operator with kernel~$Q(x,y)$ by~$Q$, the dynamical wave equation
can be written in the shorter form
\beq \label{ELQ}
Q \psi^u = \mathfrak{r}\, \psi^u \qquad \text{for all~$u \in \H$}\:.
\eeq
Similar to the Dirac equation, the dynamical wave equation can be regarded as a linear equation
for the physical wave functions~$\psi^u$. But as~$Q$ is also formed of the physical wave function,
the dynamical wave equation is indeed {\em{nonlinear}}. The nonlinear corrections will be discussed in
more detail in Section~\ref{secnonlinear}; they will be crucial for getting a connection to collapse phenomena.

\subsubsection{The Conservation Law for the Commutator Inner Product} \label{secconserve}
The connection between symmetries and conservation laws made by Noether's theorem
extends to causal fermion systems~\cite{noether}. However, the conserved quantities of a causal fermion system
have a rather different structure, being formulated in terms of so-called surface layer integrals.
A {\em{surface layer integral}} is a double integral of the form
\beq \label{intdouble}
\int_\Omega \bigg( \int_{M \setminus \Omega} (\cdots)\: \L(x,y)\: d\rho(y) \bigg)\, d\rho(x)
\eeq
where the two variables~$x$ and~$y$ are integrated over~$\Omega$ and its complement,
and~$(\cdots)$ stands for variational derivatives acting on the Lagrangian.
Since in typical applications,
the Lagrangian is small if~$x$ and~$y$ are far apart, the main contribution to the
surface layer integral is attained when both~$x$ and~$y$ are near the boundary~$\partial \Omega$.
With this in mind, a surface layer integral can be thought of as a ``thickened'' surface integral,
where we integrate over a spacetime strip of a certain width. For systems in Minkowski space as considered
here, the length scale of this strip is the Compton scale~$m^{-1}$.
For more details on the concept of a surface layer integral we refer to~\cite[Section~9.1]{intro}.

There are various Noether-like theorems for causal fermion systems, which relate
symmetries to conservation laws (for an overview 
see~\cite{osi} or~\cite[Chapter~9]{intro}). The conserved quantity of relevance here is the
commutator inner product (for more details see~\cite[Section~9.4]{intro}
or~\cite[Section~5]{noether} and~\cite[Section~3]{dirac}):
The causal action principle is invariant under unitary transformations of the measure~$\rho$, i.e.\ under
transformations
\[ \rho \rightarrow \scrU \rho \qquad \text{with} \qquad
(\scrU \rho)(\Omega) := \rho \big( \scrU^{-1} \,\Omega\, \scrU \big) \:, \]
where~$\scrU$ is a unitary operator on~$\H$ and~$\Omega \subset \F$ is any measurable subset.
The conserved quantity corresponding to this symmetry is the so-called {\em{commutator inner product}}
\beq \label{cip}
\la \psi | \phi \ra^t := -2i \,\bigg( \int_{\Omega} \!d\rho(x) \int_{M \setminus \Omega} \!\!\!\!\!\!\!\!d\rho(y) 
- \int_{M \setminus \Omega} \!\!\!\!\!\!\!\!d\rho(x) \int_{\Omega} \!d\rho(y) \bigg)\:
\Sl \psi(x) \:|\: Q(x,y)\, \phi(y) \Sr_x \:,
\eeq
where~$\psi, \phi$ are wave functions~\eqref{psirep}, and~$\Omega \subset M$ describes a
spacetime region (and the kernel~$Q(x,y)$ as in~\eqref{Qxydef}).
The set~$\Omega$ should be thought of as the past of a Cauchy surface
at time~$t$, so that the surface layer integral describes a ``thickened'' integral over the Cauchy surface.\Claudio{Bildchen dazu?}%
Then the conservation law states that the commutator inner product of any two physical wave
functions~$\psi$ and~$\phi$ (as defined by~\eqref{psiudef}) does not depend on the choice of
the Cauchy surface.

\subsection{Effective Description by a Nonlocal Dirac Equation} \label{secdir}
In~\cite{nonlocal} the EL equations and linearizations thereof (the so-called linearized field equations)
were studied in detail for causal fermion systems describing Minkowski space.
In this setting, the abstract structures of a causal fermion system become more concrete,
opening the door for a detailed analysis.
We now explain what these findings mean for the structure of the dynamical wave equation.
For causal fermion systems describing Minkowski space, the physical wave functions~\eqref{psiudef} can be represented
by usual spinorial wave functions in spacetime. Moreover, the spin inner product~\eqref{ssp}
goes over to the usual pointwise inner product on Dirac spinors, i.e.\
\beq \label{sip}
\Sl \psi | \phi \Sr(x) = \overline{\psi(x)} \phi(x) = \psi(x)^\dagger \gamma^0 \phi(x) \:,
\eeq
where~$\overline{\psi(x)}$ is sometimes referred to as the adjoint spinor
(for more details on the correspondence of the abstract objects with objects in Minkowski space
see~\cite[Section~1.2]{cfs}).
Moreover, in the Minkowski vacuum, the dynamical wave equation~\eqref{ELQ} reduces to the Dirac equation
in Minkowski space
\[ \big( i \Pdd - m \big) \psi(x) = 0 \:, \]
where, for ease in notation, the superscript~$u$ of the wave function was omitted.
In order to describe an interacting situation, one must insert potentials into the Dirac equation. It turns out that
the EL equations do not allow only for homogeneous classical fields
(like plane electromagnetic waves), but instead for a plethora of fields coupling to different wave
packets propagating in different directions. All these fields are described by a {\em{nonlocal potential}}~$\B$.
More precisely, the Dirac equation becomes
\beq \label{dirnonloc}
\big( i \Pdd + \B - m \big) \psi = 0 \:,
\eeq
where~$\B$ is an integral operator in spacetime with kernel~$\B(x,y)$, i.e.\
\beq \label{Bnonlocal}
\big( \B \psi \big)(x) = \int_M \B(x,y)\, \psi(y)\: d^4y \:.
\eeq
This integral operator has the structure
\beq \label{Bstochastic}
\B(x,y) = \sum_{a=1}^N B_a \Big( \frac{x+y}{2} \Big) \:L_a(y-x)\:,
\eeq
where~$B_a()$ are multiplication operators acting on the spinors and the~$L_a$ are 
complex-valued functions. The factors in the nonlocal potential are symmetric in the sense that
\beq \label{symmpot}
B_a(z)^* = B_a(z) \qquad \text{and} \qquad \overline{L_a(\xi)} = L_a(-\xi)
\eeq
(where the star is the adjoint with respect to the spin inner product).
The number~$N$ of these potentials is very large and scales like~\eqref{Nscaleintro}
with~$\ell_{\min}$ in the range~\eqref{ellmindef}.
The scale~$\ell_{\min}$ also determines the scale of the non-locality of the potential.
It is an important consequence of~\eqref{symmpot} that the nonlocal potential
is {\em{symmetric}}, meaning that
\beq \label{Bsymm}
\B(x,y)^* = \B(y,x) \:.
\eeq

The conserved commutator inner product~\eqref{cip} takes the following form
\begin{align}
\la \psi | \phi \ra_t &:= \int \Sl \psi \,|\, \gamma^0\, \phi \Sr_{(t,\vec{x})} \: d^3x \label{c11} \\
&\quad\;\; -i \int_{x^0<t} d^4x \int_{y^0>t} d^4y\;
\Sl \psi(x) \,|\, \B(x,y)\, \phi(y) \Sr_x \label{c2} \\
&\quad\;\; +i \int_{x^0>t} d^4x \int_{y^0<t} d^4y\;
\Sl \psi(x) \,|\, \B(x,y)\, \phi(y) \Sr_x \:. \label{c3}
\end{align}
Note that~\eqref{c11} is the usual scalar product on Dirac wave functions.
The additional summands~\eqref{c2} and~\eqref{c3} can be understood as correction terms
which take into account the nonlocality of the potential~$\B$ in~\eqref{sip}.
The mathematical structure of these additional terms is that of a surface layer integral,
similar as explained for abstract causal fermion systems after~\eqref{intdouble}.
Making use of the symmetry of the nonlocal potential~\eqref{Bsymm}, a straightforward computation
shows that {\em{current conservation}} holds in the sense that the quantity
is independent of the time~$t$ (for the derivation see~\cite[Proposition~B.1]{baryogenesis}).

\begin{figure}
\psset{xunit=.5pt,yunit=.5pt,runit=.5pt}
\begin{pspicture}(654.39143624,196.84522734)
{
\newrgbcolor{curcolor}{0.74901962 0.74901962 0.74901962}
\pscustom[linestyle=none,fillstyle=solid,fillcolor=curcolor]
{
\newpath
\moveto(484.9008189,79.87636687)
\curveto(484.9008189,79.87636687)(486.38605984,70.55285065)(486.97048819,56.03074167)
\curveto(487.55491654,41.50862891)(471.66034016,21.46118702)(471.66034016,21.46118702)
\lineto(489.96292913,2.4394749)
\lineto(509.0304189,20.71977222)
\curveto(509.0304189,20.71977222)(498.10365354,31.97315301)(498.81669921,49.4397823)
\curveto(499.52974488,66.90641159)(502.60170709,79.42466419)(502.60170709,79.42466419)
\closepath
}
}
{
\newrgbcolor{curcolor}{0.74901962 0.74901962 0.74901962}
\pscustom[linewidth=1.20188972,linecolor=curcolor]
{
\newpath
\moveto(484.9008189,79.87636687)
\curveto(484.9008189,79.87636687)(486.38605984,70.55285065)(486.97048819,56.03074167)
\curveto(487.55491654,41.50862891)(471.66034016,21.46118702)(471.66034016,21.46118702)
\lineto(489.96292913,2.4394749)
\lineto(509.0304189,20.71977222)
\curveto(509.0304189,20.71977222)(498.10365354,31.97315301)(498.81669921,49.4397823)
\curveto(499.52974488,66.90641159)(502.60170709,79.42466419)(502.60170709,79.42466419)
\closepath
}
}
{
\newrgbcolor{curcolor}{0.90196079 0.90196079 0.90196079}
\pscustom[linestyle=none,fillstyle=solid,fillcolor=curcolor]
{
\newpath
\moveto(159.31893543,114.03735836)
\lineto(153.27110173,113.2111612)
\lineto(140.69603906,113.45769978)
\lineto(139.67797039,124.07638828)
\lineto(138.32090835,134.82361096)
\lineto(136.62054803,143.91922923)
\lineto(135.23355591,150.17405632)
\lineto(133.03922646,156.89477821)
\lineto(130.66084157,162.6976645)
\lineto(128.58723024,166.81147994)
\lineto(152.86016126,191.86696214)
\lineto(177.39190299,166.97203805)
\lineto(173.56362331,162.60630198)
\lineto(170.18796472,156.24387931)
\lineto(166.74466394,146.9633212)
\lineto(165.11775874,141.62339301)
\lineto(163.0972422,132.71962734)
\lineto(161.61350551,126.21741128)
\lineto(160.80185575,121.14235238)
\closepath
}
}
{
\newrgbcolor{curcolor}{0 0 0}
\pscustom[linewidth=0,linecolor=curcolor]
{
\newpath
\moveto(159.31893543,114.03735836)
\lineto(153.27110173,113.2111612)
\lineto(140.69603906,113.45769978)
\lineto(139.67797039,124.07638828)
\lineto(138.32090835,134.82361096)
\lineto(136.62054803,143.91922923)
\lineto(135.23355591,150.17405632)
\lineto(133.03922646,156.89477821)
\lineto(130.66084157,162.6976645)
\lineto(128.58723024,166.81147994)
\lineto(152.86016126,191.86696214)
\lineto(177.39190299,166.97203805)
\lineto(173.56362331,162.60630198)
\lineto(170.18796472,156.24387931)
\lineto(166.74466394,146.9633212)
\lineto(165.11775874,141.62339301)
\lineto(163.0972422,132.71962734)
\lineto(161.61350551,126.21741128)
\lineto(160.80185575,121.14235238)
\closepath
}
}
{
\newrgbcolor{curcolor}{0.74901962 0.74901962 0.74901962}
\pscustom[linestyle=none,fillstyle=solid,fillcolor=curcolor]
{
\newpath
\moveto(140.67810217,113.66265751)
\lineto(149.29360176,113.66265751)
\lineto(158.56370494,113.14926917)
\lineto(153.27341254,62.30635238)
\lineto(149.0039554,4.79805884)
\lineto(145.09012611,4.88377852)
\closepath
}
}
{
\newrgbcolor{curcolor}{0.74901962 0.74901962 0.74901962}
\pscustom[linewidth=0,linecolor=curcolor]
{
\newpath
\moveto(140.67810217,113.66265751)
\lineto(149.29360176,113.66265751)
\lineto(158.56370494,113.14926917)
\lineto(153.27341254,62.30635238)
\lineto(149.0039554,4.79805884)
\lineto(145.09012611,4.88377852)
\closepath
}
}
{
\newrgbcolor{curcolor}{0.74901962 0.74901962 0.74901962}
\pscustom[linestyle=none,fillstyle=solid,fillcolor=curcolor]
{
\newpath
\moveto(502.47403465,80.07801222)
\lineto(493.85852598,80.07801222)
\lineto(484.6872,80.07801222)
\lineto(473.45380157,196.33007017)
\lineto(522.83408504,193.87349821)
\closepath
}
}
{
\newrgbcolor{curcolor}{0.74901962 0.74901962 0.74901962}
\pscustom[linewidth=0,linecolor=curcolor]
{
\newpath
\moveto(502.47403465,80.07801222)
\lineto(493.85852598,80.07801222)
\lineto(484.6872,80.07801222)
\lineto(473.45380157,196.33007017)
\lineto(522.83408504,193.87349821)
\closepath
}
}
{
\newrgbcolor{curcolor}{0 0 0}
\pscustom[linewidth=2.26771663,linecolor=curcolor]
{
\newpath
\moveto(10.34325165,47.94309317)
\lineto(153.1038652,191.91919899)
\lineto(297.71344252,45.49890293)
}
}
{
\newrgbcolor{curcolor}{0 0 0}
\pscustom[linewidth=2.26771663,linecolor=curcolor]
{
\newpath
\moveto(331.9672252,159.04237758)
\lineto(490.33555276,1.60155364)
\lineto(647.6992063,159.11102513)
}
}
{
\newrgbcolor{curcolor}{0 0 0}
\pscustom[linewidth=1.19999845,linecolor=curcolor]
{
\newpath
\moveto(86.90170356,125.60923767)
\curveto(94.96008416,121.46687698)(103.01686375,117.32534173)(120.92631836,115.25100699)
\curveto(138.83577298,113.17667225)(166.59617461,113.16896202)(184.4675452,115.31554621)
\curveto(202.33891124,117.46213039)(210.32011691,121.76262803)(218.30102324,126.06296239)
}
}
{
\newrgbcolor{curcolor}{0.74901962 0.74901962 0.74901962}
\pscustom[linestyle=none,fillstyle=solid,fillcolor=curcolor]
{
\newpath
\moveto(523.00435276,194.6981949)
\lineto(502.72586457,79.60664466)
}
}
{
\newrgbcolor{curcolor}{0 0 0}
\pscustom[linewidth=1.36062998,linecolor=curcolor]
{
\newpath
\moveto(523.00435276,194.6981949)
\lineto(502.72586457,79.60664466)
}
}
{
\newrgbcolor{curcolor}{0 0 0}
\pscustom[linewidth=2.04094496,linecolor=curcolor,linestyle=dashed,dash=3.24000001 3.24000001]
{
\newpath
\moveto(152.98944378,191.67302702)
\lineto(122.76131906,5.61345411)
}
}
{
\newrgbcolor{curcolor}{0 0 0}
\pscustom[linewidth=2.04094496,linecolor=curcolor,linestyle=dashed,dash=3.24000001 3.24000001]
{
\newpath
\moveto(153.00550677,191.56175773)
\lineto(168.95405102,5.52815773)
}
}
{
\newrgbcolor{curcolor}{0 0 0}
\pscustom[linewidth=1.13385827,linecolor=curcolor]
{
\newpath
\moveto(145.10768277,5.35355868)
\curveto(144.49650142,31.91405254)(143.88522028,58.47876435)(142.7336829,81.85747049)
\curveto(141.58215005,105.23613128)(139.89120907,125.41557921)(137.47319282,138.95628246)
\curveto(135.05517657,152.4969857)(131.91245254,159.38719295)(128.76746532,166.28235742)
}
}
{
\newrgbcolor{curcolor}{0 0 0}
\pscustom[linewidth=1.13385827,linecolor=curcolor]
{
\newpath
\moveto(149.28273487,5.02610041)
\curveto(151.14253002,34.02751899)(153.0027515,63.03560466)(155.87324523,87.89789632)
\curveto(158.74373896,112.76016983)(162.62190161,133.45837606)(166.3419908,146.01600151)
\curveto(170.06208454,158.57362243)(173.6214788,162.98048958)(177.18165317,167.38832731)
}
}
{
\newrgbcolor{curcolor}{0 0 0}
\pscustom[linewidth=1.19999845,linecolor=curcolor]
{
\newpath
\moveto(424.36583433,67.92078387)
\curveto(432.42416504,72.06313096)(440.4809537,76.20466167)(458.39042646,78.27894198)
\curveto(476.29989921,80.3532223)(504.06028724,80.36075112)(521.93166236,78.21408529)
\curveto(539.80303748,76.06741947)(547.78422047,71.76692183)(555.76513134,67.46660561)
}
}
{
\newrgbcolor{curcolor}{0 0 0}
\pscustom[linewidth=1.58740153,linecolor=curcolor,linestyle=dashed,dash=2.51999998 2.51999998]
{
\newpath
\moveto(502.70065512,79.59489789)
\lineto(490.46545512,1.43038639)
}
}
{
\newrgbcolor{curcolor}{0 0 0}
\pscustom[linewidth=1.58740153,linecolor=curcolor,linestyle=dashed,dash=2.51999998 2.51999998]
{
\newpath
\moveto(484.76203465,80.24292057)
\lineto(490.49482205,2.00343836)
}
}
{
\newrgbcolor{curcolor}{0 0 0}
\pscustom[linewidth=1.36062998,linecolor=curcolor]
{
\newpath
\moveto(473.38996535,195.84198576)
\lineto(484.5672,79.752096)
}
}
{
\newrgbcolor{curcolor}{0 0 0}
\pscustom[linewidth=1.36062998,linecolor=curcolor]
{
\newpath
\moveto(6.16537323,166.96453191)
\lineto(6.51195591,129.89537537)
\lineto(42.71458394,129.8087108)
}
}
{
\newrgbcolor{curcolor}{0 0 0}
\pscustom[linewidth=1.36062998,linecolor=curcolor]
{
\newpath
\moveto(366.11990551,54.04755175)
\lineto(366.46648819,16.97841411)
\lineto(402.66914646,16.89178734)
}
}
{
\newrgbcolor{curcolor}{0 0 0}
\pscustom[linewidth=1.36062998,linecolor=curcolor]
{
\newpath
\moveto(0.52695307,159.9660601)
\lineto(6.24604346,166.97014072)
\lineto(12.48505323,159.9660601)
}
}
{
\newrgbcolor{curcolor}{0 0 0}
\pscustom[linewidth=1.36062998,linecolor=curcolor]
{
\newpath
\moveto(360.32301732,47.86997443)
\lineto(366.04208504,54.87404372)
\lineto(372.28114016,47.86997443)
}
}
{
\newrgbcolor{curcolor}{0 0 0}
\pscustom[linewidth=1.36062998,linecolor=curcolor]
{
\newpath
\moveto(36.48791055,136.13650923)
\lineto(43.66922457,129.8108538)
\lineto(36.57456378,123.91844561)
}
}
{
\newrgbcolor{curcolor}{0 0 0}
\pscustom[linewidth=1.36062998,linecolor=curcolor]
{
\newpath
\moveto(395.67456378,23.21980498)
\lineto(402.85589291,16.89414576)
\lineto(395.76122835,11.00171112)
}
}
{
\newrgbcolor{curcolor}{0 0 0}
\pscustom[linewidth=2.26771654,linecolor=curcolor]
{
\newpath
\moveto(10.59259767,48.19371742)
\curveto(6.02296471,42.59200403)(1.45496451,36.99228624)(8.78267414,31.15822261)
\curveto(16.1103874,25.32415899)(35.333952,19.25701947)(59.02701354,14.31276246)
\curveto(82.72007509,9.36850545)(110.88159043,5.54735773)(141.66337058,5.00133695)
\curveto(172.44515074,4.45531616)(205.84694173,7.18451301)(234.74607723,11.80103962)
\curveto(263.64520819,16.41761159)(288.04117569,22.92151332)(297.74259855,28.90710576)
\curveto(307.44401688,34.89274356)(302.45112189,40.35916498)(297.45691162,45.82703773)
}
}
{
\newrgbcolor{curcolor}{0 0 0}
\pscustom[linewidth=2.26771654,linecolor=curcolor]
{
\newpath
\moveto(332.44140472,158.44238589)
\curveto(327.35646312,164.91079332)(322.27255106,171.37789455)(329.13500145,176.98484183)
\curveto(335.99745638,182.59179364)(354.80612409,187.33758463)(385.07642079,190.83749669)
\curveto(415.34671748,194.33740422)(457.07737323,196.59124233)(496.28315339,195.38116158)
\curveto(535.48893354,194.17108535)(572.16938457,189.49708573)(599.25957165,185.27706028)
\curveto(626.34980409,181.05703484)(643.8492737,177.29106973)(650.20477606,172.86595654)
\curveto(656.56027843,168.44084787)(651.77126929,163.35654576)(646.98262299,158.27263824)
}
}
{
\newrgbcolor{curcolor}{0 0 0}
\pscustom[linewidth=1.65543311,linecolor=curcolor]
{
\newpath
\moveto(128.56686614,40.77888529)
\curveto(134.5310332,39.26790576)(140.49505512,37.75697159)(146.88093128,37.39078072)
\curveto(153.26681197,37.02477128)(160.07400718,37.80305159)(166.88133392,38.5817401)
}
}
{
\newrgbcolor{curcolor}{0 0 0}
\pscustom[linewidth=0.74834649,linecolor=curcolor]
{
\newpath
\moveto(159.44250557,31.56406828)
\curveto(161.69267906,28.55825537)(163.94283439,25.55244246)(169.68108699,24.66431395)
\curveto(175.41933959,23.77618545)(184.64524951,25.00583206)(193.87132724,26.23543332)
}
}
{
\newrgbcolor{curcolor}{0 0 0}
\pscustom[linestyle=none,fillstyle=solid,fillcolor=curcolor]
{
\newpath
\moveto(161.74778562,31.98108634)
\lineto(156.61712103,35.33825611)
\lineto(158.392952,29.46963341)
\curveto(158.74218406,31.03980248)(160.13863355,32.08519477)(161.74778562,31.98108634)
\closepath
}
}
{
\newrgbcolor{curcolor}{0 0 0}
\pscustom[linewidth=1.65543311,linecolor=curcolor]
{
\newpath
\moveto(476.82078614,156.59390891)
\curveto(483.59405178,157.7677697)(490.36714961,158.94158513)(496.88788687,158.8695171)
\curveto(503.40862413,158.79740372)(509.67642935,157.47949758)(515.9443525,156.16150072)
}
\rput[bl](20,140){$p$}
\rput[bl](380,30){$\xi$}
\rput[bl](165,90){$-\omega_{\min}$}
\rput[bl](438,85){$\ell_{\min}$}
\rput[bl](200,23){$\vartheta$}
\rput[bl](490,135){$\vartheta$}
\rput[bl](-40,-5){$\displaystyle -\frac{1}{\varepsilon}$}
}
\end{pspicture}
\caption{A homogeneous solution in momentum and position space.}
\label{fighom} 
\end{figure}
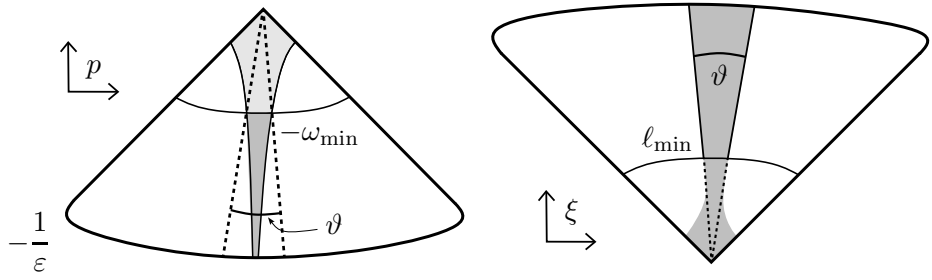

For the derivation of the collapse model, we will have to specify the form of the potentials in~\eqref{Bstochastic}.
Therefore, we now recall what the analysis in~\cite{nonlocal} revealed about these potentials.
Considering the kernel~$L_a(y-x)$ as a convolution operator, its Fourier transform~$\hat{L}_a$
is a multiplication operator in momentum space. The symmetry property in~\eqref{symmpot} means
that the function~$\hat{L}_a$ is real-valued. In Figure~\ref{fighom} the functions~$L_a$ and~$\hat{L}_a$
are depicted in a typical example.
On the left side, the support of~$\hat{L}_a$ is shown on the lower mass shell.
Given a parameter~$\omega_{\min}$ is in the range
\[ 
\frac{\pi^2}{\ell_{\min}} \lesssim \omega_{\min} \leq \frac{1}{\varepsilon} \:, \]
for frequencies smaller than~$-\omega_{\min}$, the function~$\hat{L}_a$ is supported inside
a cone of opening angle
\[ 
\vartheta = \frac{1}{\sqrt{\ell_{\min} \,\omega_{\min}}} \:. \]
This function is smooth in the sense that its derivatives have the scaling behavior
\[ \big| D^p \hat{L}(p) \big| \lesssim \frac{\big| \hat{L}(p)\big|}{\big( \vartheta\,(\omega_{\min} + |\omega|) \big)^p} \:. \]
For frequencies larger than~$-\omega_{\min}$, the support of the function~$\hat{L}_a$ is a bit more
spread out, as is indicated by the light gray region on the left of Figure~\ref{fighom}.
Taking the Fourier transform, the corresponding function~$L_a$
can be thought of as a wave packet propagating almost with the speed of light
localized in space on the scale~$\ell_{\min}$, as is shown on the
right of Figure~\ref{fighom}.

Next, the potential~$B_a(z)$ in~\eqref{symmpot} is found to be vectorial. Thus it can be written as
\beq \label{Bavec}
B_a = (A_a)_j\: \gamma^j
\eeq
with potentials~$A_a(z)$ which can be thought of as nonlocal generalizations of an electromagnetic potential.
In a suitable gauge, they satisfy the homogeneous wave equation with an error term,
\beq \label{Aerr}
\Box A_a = \O \big( \ell_{\min} \:D^3 A_a \big) \:,
\eeq
where~$\|D^3A_a\|$ denotes the third derivatives of~$A_a$. Thus the error term is of the 
multiplicative order~$\ell_{\min}/\ell_\text{macro}$, where~$\ell_\text{macro}$ is the length scale
on which the potential~$A_a$ varies. These error terms and their scaling behavior are worked out
in detail in~\cite{nonlocal}. In this paper it is also shown that the functions~$L_a$ and~$A_a$ can
be computed iteratively in an expansion in powers of~$\ell_{\min}/\ell_\text{macro}$.
Here we do not need the details, but it suffices to work with the error term~\eqref{Aerr}.

\subsection{Incorporating the Nonlinear Coupling} \label{secnonlinear}
As already pointed out, the EL equations~\eqref{EL} and the dynamical wave equation~\eqref{ELQ}
are nonlinear. This nonlinearity was studied systematically in~\cite{perturb} in an abstract setting.
More concretely, for causal fermion systems describing Minkowski space, this nonlinearity
becomes manifest in the classical field equations as derived in the continuum limit~\cite{cfs}.
One should keep in mind that, in the continuum limit analysis, one mainly restricts attention
to local potentials like a classical Maxwell or Yang-Mills potential or a classical gravitational field.
For the nonlocal potentials in~\eqref{dirnonloc} and~\eqref{Bnonlocal}, however, the nonlinear
coupling has not yet been worked out in detail. For this reason, we here merely describe 
a simple effective model, which seems to capture the essence of the
nonlinear coupling.
This model is obtained naturally by demanding the nonlinear system to be of variational form.
We begin by noting that the nonlocal Dirac equation~\eqref{dirnonloc} can be obtained
by varying the wave function~$\psi$ in the nonlocal Dirac action
\begin{align*}
S_\psi &:= \int_M \Sl \psi(x)\:|\: \big((i \Pdd + \B - m) \psi(x) \big) \Sr\: d^4x \\
&\,= \int_M \Sl \psi(x)\:|\: (i \Pdd - m) \psi(x) \big) \Sr\: d^4x 
+ \int_M d^4x \int_M d^4y \:\Sl \psi(x) \:|\: \B(x,y)\: \psi(y) \Sr \:.
\end{align*}
In order to allow for homogeneous solutions of the form~\eqref{Aerr}, we choose the
corresponding action for each potential~$A_a$ as the Maxwell action in the Lorenz gauge, i.e.
\beq \label{SAdef}
S_A := \frac{1}{e^2} \int_M \sum_{a=1}^N \partial_j (A_a)_k\: \partial^j (A_a)^k\: d^4x \:,
\eeq
where~$e$ is a corresponding coupling constant.
Varying the potentials, we obtain the nonlocal Maxwell equations
\beq \label{maxnonloc}
\Box (A_a)^k = e^2 J_a^k 
\eeq
where the~$J_a$ are the nonlocal Dirac currents
\beq \label{Jnonloc}
J_a^k(z) := \int_M \Sl \psi(x) | \gamma^k\, L_a(y-x)\, \psi(y)\Sr
\big|_{x=z-\xi/2, \;y = z+\xi/2}\: d^4 \xi \:.
\eeq
Together with the nonlocal Dirac equation~\eqref{dirnonloc}, we have a system of nonlinear
equations describing the dynamics of the system.
Clearly, this method has the shortcoming that the error terms in~\eqref{Aerr} are not specified.
Nevertheless, in view of the fact that these error terms will not be relevant for the analysis in this paper,
the nonlinear system~\eqref{dirnonloc} and~\eqref{maxnonloc}, \eqref{Jnonloc}
will be sufficient for our purposes.

Finally, it is convenient to write the resulting potentials~$B_a$ as
\beq \label{Ahominret}
B_a = B_a^\text{hom} + B_a^\text{ret} \:,
\eeq
where the potentials~$B_a^\text{hom}$ satisfy the homogeneous wave equation
\[ \Box B_a^\text{hom} = 0 \:, \]
whereas the potentials~$B_a^\text{ret}$ are retarded solutions of the inhomogeneous equation, i.e.\
\[ B_a = \gamma^k (A_a)_k \qquad \text{with} \qquad (A_a)^k = e^2 S^\wedge(J_a^k) \:, \]
where~$S^\wedge$ denotes the retarded Green's operator of the wave equation.

\section{A Model with a Purely Spatial Nonlocality} \label{secnonrelativ}
Our starting point is the causal fermion system describing Minkowski space modelled effectively
by the nonlocal Dirac equation~\eqref{dirnonloc}--\eqref{Bstochastic}, where the
bosonic potentials~$B_a$ satisfy the nonlocal Maxwell equation~\eqref{maxnonloc}, \eqref{Jnonloc}.
Our first task is to take the non-relativistic limit. Moreover, we need to describe the
bosonic background fields stochastically.
In order to explain the concepts step by step, in this section we begin
with the simplest possible setting where we take the non-relativistic limit in a naive way by
replacing the nonlocal potential~$\B$ by an operator which is nonlocal only in space.
As we shall see, the resulting model is too simple to describe collapse phenomena.
Nevertheless, it is a good starting point for developing our methods and getting a connection to
the CSL model. The shortcomings of this model will also serve as the motivation for the
refined model in Section~\ref{secmodel} which takes into account the nonlocality in time and
indeed describes collapse and the reduction of the state vector.

\subsection{The Nonrelativistic Limit with a Purely Spatial Nonlocality} \label{secpurespatial}
In order to get a direct connection to the standard setting of collapse models,
we now consider the non-relativistic limit.
The simplest approach is to assume that the nonlocal potential~$\B(x,y)$ is purely spatial, i.e.\ that
it can be written as
\beq \label{Bnonrelat}
(\B \psi)(t,\vec{x}) = \int_{\R^3} \B_t \big(\vec{x}, \vec{y} \big) \: \psi \big(t, \vec{y} \big)\: d^3y
\eeq
with new kernels~$\B_t$.
Under this simplifying assumption, the nonlocal Dirac equation~\eqref{dirnonloc} can be written
in the Hamiltonian form
\beq \label{dynHamilton}
i \partial_t \psi = H \psi \:,
\eeq
where the Hamiltonian is of the form
\beq \label{H0V}
H = H_0 + V_t
\eeq
with the standard Dirac Hamiltonian~$H_0 := -i \gamma^0 \vec{\gamma} \vec{\nabla}$ 
and the spatial operator~$V_t$ given by
\beq \label{Vdef}
\big(V_t\, \psi)(\vec{x}) := \int_{\R^3} V_t \big(\vec{x}, \vec{y} \big) \: \psi \big(\vec{y} \big)\: d^3y \qquad \text{with} \qquad
V_t(\vec{x}, \vec{y}) := -\gamma^0 \, \B_t \big( \vec{x}, \vec{y} \big) \:.
\eeq
The symmetry of the nonlocal potential~\eqref{Bsymm} means that the kernels~$V_t(\vec{x}, \vec{y})$
are symmetric in the sense that
\[ V_t(\vec{x}, \vec{y})^\dagger = V_t(\vec{y}, \vec{x}) \qquad \text{for all~$t \in \R$}\:. \]
(where the dagger denotes again transposition and complex conjugation).
Note that the operator~$V_t$ may be time dependent, but it is nonlocal only in the spatial variables.

For a purely spatial nonlocal potential the additional terms~\eqref{c2} and~\eqref{c3} vanish.
Therefore, the commutator inner product reduces to the usual scalar product on the
wave functions at time~$t$, i.e.\
\beq \label{print}
\la \psi |  \phi \ra_t = \int_{\R^3} \Sl \psi \,|\, \gamma^0 \,\phi \Sr(t,\vec{x})\: d^3x \:.
\eeq
We denote the corresponding Hilbert space by~$(\H_t, \la.|.\ra_t)$.

We remark that one could simplify the setting further by taking the non-relativistic limit of the
Dirac equation and disregarding the spin, in which case the Dirac Hamiltonian would be replaced by the Schr\"odinger operator~$H_0=-\Delta/2m$, whereas~\eqref{print} would go over to the standard scalar
product of quantum mechanics
\[ \la \psi |  \phi\ra_t = \int_{\R^3} \overline{\psi(t,\vec{x})} \phi (t,\vec{x})\: d^3x \:. \]
For the following considerations, this simplification is not needed. The reader can work either with
the Schr\"odinger or with the Dirac equation.

Finally, it is convenient to simplify the setting by working in the {\em{interaction picture}}. To this end, we set
\[ \check{\psi}(t) := e^{-i t H_0} \psi(t) \qquad \text{and} \qquad
\check{V}_t := e^{-i t H_0}\: V_t\: e^{i t H_0} \:. \]
Then the Dirac equation in the Hamiltonian form~\eqref{H0V} takes the simple form
\beq \label{HOVint}
i \partial_t \check{\psi} = \check{V}_t \,\check{\psi} \:.
\eeq
From now on, we shall always work in the interaction picture. With this in mind, the check will be omitted.

\subsection{Markovian Gaussian Background Fields}
The potentials~$V_t$ involve the ple\-tho\-ra of potentials~$B_a$.
More specifically, using~\eqref{Bstochastic} and taking the non-relativistic limit according
to~\eqref{Bnonrelat} and~\eqref{Vdef}, we find
\beq \label{Vtfirst}
V_t = -\sum_a \gamma^0 B_{t,a}\Big(\frac{\vec{x}+\vec{y}}{2} \Big)\: L_a \big(\vec{y}-\vec{x} \big) \:.
\eeq
According to~\eqref{Ahominret}, the potentials~$B_a$ can be decomposed into the homogeneous
and the retarded potentials. We now focus on the homogeneous potentials
(the retarded potentials will be considered in Section~\ref{secnocollapse} below).
These homogeneous potentials describe background fields which were created
at some time in the past and are penetrating our measurement device (as shown in Figure~\ref{figcoll1}).
Since these background fields are unknown, we describe each of them stochastically.
For simplicity and in order to get a connection to the familiar setting as
considered in~\cite{gisin, bassi-ghirardi}, we
assume that these stochastic potentials have mean zero and are Gaussian and Markovian, i.e.\ 
\begin{align}
\bbra \gamma^0 B_{t,a} (\vec{x}) \kket &= 0 \label{meanC} \\
\bbra \:\big(\gamma^0 B_{t,a}(\vec{x}) \big)^\alpha_\beta\: 
\big( \gamma^0 B_{t',b}(\vec{y}) \big)^\gamma_\delta \:\kket &= \delta(t-t')\: \delta_{ab}\:
C^{\;\:\,\alpha \gamma}_{a, \beta \delta}(t, \vec{x}, \vec{y}) \label{covC}
\end{align}
with covariances~$C_a$.
Here the Greek indices are spinorial indices, corresponding to the fact that the potentials~$B_{t,a}$
are matrix-valued (more precisely, they are vectorial according to~\eqref{Bavec}).
It is useful to simplify the form of the covariance by a change of basis.
Although this procedure is standard, we give the construction in detail.
\begin{Lemma} \label{lemmadiag}
There are real-valued Markovian Gaussian fields~$W_{k,a}(t)$ which are independent and normalized, i.e.\
\beq \label{gaussW}
\bbra W_{k,a}(t) \kket = 0 \qquad \text{and} \qquad
\bbra W_{k,a}(t)\, W_{l,b}(t') \kket = \delta(t-t')\, \delta_{ab} \, \delta_{kl} \:,
\eeq
as well as kernels~$M_{k,a}(\vec{x}, \vec{y})$ which are symmetric in the sense that
\beq \label{Masymm}
M_{k,a}(\vec{x}, \vec{y})^\dagger = M_{k,a}(\vec{y}, \vec{x})\:,
\eeq
such that the homogeneous stochastic potential can be written as
\beq \label{Vrep}
V_t^\text{\rm{hom}}(\vec{x}, \vec{y}) = \sum_{k,a} \: M_{k,a}(\vec{x}, \vec{y})\: W_{k,a}(t) \:.
\eeq
\end{Lemma}
\Proof Denoting the Hermitian $4\times4$-matrices by~$\Symm(\C^4)$, we consider the
Hilbert space~$L^2(\R^3, \Symm(\C^4))$ of square integrable functions taking values in the
Hermitian matrices. Given a fixed time~$t$ and~$a \in \{1,\ldots, N\}$,
we consider the covariance as an integral kernel of an operator on this Hilbert space,
\begin{align*}
&C_a : L^2(\R^3, \Symm(\C^4)) \rightarrow L^2(\R^3, \Symm(\C^4)) \\
&(C_a E)^\alpha_\beta(\vec{x}) := \int_{\R^3} \sum_{\gamma, \delta=1}^4
C^{\;\:\,\alpha \gamma}_{a, \beta \delta}(t, \vec{x}, \vec{y})\: E^\delta_\gamma(\vec{y})\: d^3y \:.
\end{align*}
This operator is positive, because for any matrix-valued function~$E \in L^2(\R^3, \Symm(\C^4))$
and any test function~$\eta \in C^\infty_0(\R)$,
\begin{align*}
0 &\leq \bbra\; \bigg( \int_{-\infty}^\infty dt \: \eta(t)\: \int_{\R^3} d^3x\:\Tr_{\C^4} \big(E(\vec{x})\: \gamma^0 B_{t,a}(\vec{x}) \big)  \bigg)^2\; \kket \\
&= \int_{-\infty}^\infty dt \: \eta(t)^2 \int_{\R^3} d^3x  \int_{\R^3} d^3y\:
\sum_{\alpha, \beta, \gamma, \delta=1}^4 \overline{E^\alpha_\beta(\vec{x})}\:
C^{\;\:\,\alpha \gamma}_{a, \beta \delta}(t, \vec{x}, \vec{y})\: E^\delta_\gamma(\vec{y}) \\
&= \int_{-\infty}^\infty \eta(t)^2 \:\la E, C_a E \ra_{L^2(\R^3, \text{\tiny{Symm}}(\C^4))} \:dt \:.
\end{align*}
For technical simplicity, we assume that this operator is compact, making it possible to
choose an orthonormal eigenvector basis denoted by~$(\phi_k)_{k \in \N}$
(more generally, bounded or unbounded selfadjoint operators
could be treated similarly if one replaced the sum over~$k$ by the integral with respect to the
spectral measure; for brevity we here omit the details). Moreover, the eigenvalues are
all non-negative. Leaving out the zero eigenvalues, we thus obtain a decomposition of the covariance
\beq \label{eigen}
C^{\;\:\,\alpha \gamma}_{a, \beta \delta}(t, \vec{x}, \vec{y})
= \sum_k \lambda_k \:\phi^{\alpha}_{k, \beta}(\vec{x})\: 
\phi^{\gamma}_{k, \delta}(\vec{y}) \qquad \text{with~$\lambda_k>0$}\:.
\eeq
Introducing the kernels~$M_{k,a}$ by
\[ (M_{k,a})^\alpha_\beta(\vec{x}, \vec{y}) := \sqrt{\lambda_k}\: 
\phi^{\alpha}_{k, \beta}\bigg( \frac{\vec{x}+\vec{y}}{2} \bigg) \: L_a(\vec{y}-\vec{x}) \:, \]
a direct computation using~\eqref{gaussW} and~\eqref{eigen} shows that
\begin{align*}
&\bbra\; \sum_{k,l,a,b} (M_{k,a})^\alpha_\beta(\vec{x}, \vec{y})\: W_{k,a}(t) \:
(M_{l,b})^\gamma_\delta(\vec{x}', \vec{y}')\: W_{l,b}(t') \;\kket \\
&= \bbra\; (V^\text{hom}_t)^\alpha_\beta(\vec{x}, \vec{y})\: 
(V^\text{hom}_t)^\gamma_\delta(\vec{x}', \vec{y}')\; \kket \:,
\end{align*}
as desired.
\QED
We note for clarity that, in the physical applications, the covariances~$C_a$ in~\eqref{covC}
typically give localizations in space. A simple example which
gives a direct connection to non-relativistic collapse models is to choose
\[ C^{\;\:\,\alpha \gamma}_{a, \beta \delta}(t, \vec{x}, \vec{y})
= \delta^\alpha_\delta \:\delta^\gamma_\beta
\sum_k \chi_{U_{k, a}}(\vec{x})\: \chi_{U_{k, a}}(\vec{y}) \:, \]
where the~$U_{k, a}$ are disjoint subsets of~$\R^3$ describing different spatial regions.
In this and other similar examples, the covariance is {\em{not}} translation invariant.
As a consequence, also the new kernels~$M_{k,a}$ in~\eqref{Vrep} are no longer
homogeneous. Thus, in contrast to the kernels~$L_a$ in~\eqref{Vtfirst}, they
do {\em{not}} depend only on the difference vector~$\vec{y}-\vec{x}$.

\subsection{The Kossakowski-Lindbad Dynamics}
We next rewrite the stochastic dynamics in the It\^o calculus. For the reader not familiar with
stochastic calculus, we recall a few basics (for more details see for example~\cite{arnold-l}).
We take a particular perspective which will also be a suitable starting for our later generalization to equations
which are nonlocal in time (see Section~\ref{secmodel}). The evolution equation~\eqref{HOVint} can be
solved iteratively with the familiar {\em{Dyson series}},
\Felix{Hier auch schon ein Bildchen?}%
\begin{align}
\psi(t) &= \psi(t_0) + \int_{t_0}^t \dot{\psi}(\tau)\: d\tau = \psi(t_0) -i  \int_{t_0}^t V_\tau\,  \psi(\tau)\: d\tau
= \cdots = \notag \\
&=  \psi(t_0) -i \int_{t_0}^t d\tau \,V_\tau\: \psi(t_0)
- \int_{t_0}^t d\tau_1 \,V_{\tau_1} \int_{t_0}^{\tau_1} d\tau_2 \,V_{\tau_2}\: \psi(t_0) + \cdots \:. \label{dyson}
\end{align}
Using this formula, one can also write down composite expressions like the projection
operator~$|\psi \ra \la \psi|$ onto the state. Taking the statistical mean, one gets Gaussian pairings
which involve the covariance of the stochastic field
(basics on Gaussian fields, the covariance, the Wick rule and resulting Gaussian pairings
can be found for example in~\cite[Section~6.2]{glimm+jaffe}).
Generally speaking, the statistical mean must be taken at the very end for the
expectation value of an observable describing a measurement. However, doing so is often not convenient,
because it would be necessary to express the expectation value in terms of the Dyson series.
The It\^o calculus make it possible to partly take the statistical mean of intermediate expressions
by carrying out those Gaussian pairings inside the expressions, while leaving the stochastic field
for those potentials which will be paired later outside.
In order to explain how this works in our setting, for notational simplicity, we combine the summation indices~$k$ and~$a$ to one index~$\varkappa = (k, a)$. Then the Schr\"odinger equation in the
interaction picture~\eqref{HOVint} with a homogeneous stochastic potential given by~\eqref{Vrep} can be written as
\beq \label{schstoch}
 \partial_t \psi(t) = -i V_t\, \psi(t) \qquad \text{with} \qquad V_t =  \sum_\varkappa \: M_\varkappa\: W_\varkappa(t) \:,
\eeq
where~$M_\varkappa$ denotes the spatial operator with integral as in~\eqref{Vrep}.
Similar to the notation in~\cite{bassi-ghirardi} we set
\[ 
\bM \cdot \bW_t = \sum_\varkappa M_\varkappa\: W_\varkappa(t) \qquad \text{and} \qquad
\bM \bM = \sum_{\varkappa} M_\varkappa\, M_\varkappa \:, \]
so that~\eqref{schstoch} can be written as
\[ \partial_t \psi(t) = -i \,\bM \cdot \bW_t \, \psi(t) \:. \]
Likewise, in the Dyson series~\eqref{dyson} we need to replace~$V_{\tau}$ by~$\bM \cdot \bW_\tau$.
When taking the statistical mean, we must take into account the contribution when the factor~$V_{\tau_1}$
in~\eqref{dyson} is paired with the factor~$V_{\tau_2}$. According to~\eqref{gaussW} this
gives rise to a factor~$\delta(\tau_1-\tau_2)$. As a consequence, the $\tau$-integral in~\eqref{dyson}
drops out. The fact that~$\tau_1=\tau_2$ corresponds to the upper limit of integration
of the second integral in~\eqref{dyson} is taken into account by a factor~$1/2$.
This gives the stochastic evolution equation which in the It\^o calculus is written as
\[ 
d\,|\psi \ra = \Big( - i \bM \cdot d\bW_t - \frac{1}{2}\: \bM \bM \,dt \Big) |\psi \ra \:. \]
Now one can proceed with the computation of composite expressions. One should keep in mind
to form the inner Gaussian pairings. For example,
\begin{align}
d\, |\psi \ra \la \psi| &= 
-i \big(  \bM \,  |\psi \ra \la \psi| -  |\psi \ra \la \psi|\, \bM \big) \cdot d\bW_t \label{t21} \\
&\quad\: -\frac{1}{2}\: \bM \bM\, |\psi \ra \la \psi|\,dt -\frac{1}{2}\: |\psi \ra \la \psi|\,\bM \bM\,dt
+ \sum_{\varkappa} M_\varkappa \,|\psi \ra \la \psi|\, M_\varkappa \:, \label{t2}
\end{align}
where the last term arises from a Gaussian pairing between bra and ket.
Taking the statistical mean, we obtain
\begin{align}
\frac{d}{dt} \bbra\: |\psi \ra \la \psi | \:\kket &=
-\frac{1}{2}\: \bM \bM\, |\psi \ra \la \psi| -\frac{1}{2}\: |\psi \ra \la \psi|\,\bM \bM
+ \sum_{\varkappa} M_\varkappa \,|\psi \ra \la \psi|\, M_\varkappa \notag \\
&=-\frac{1}{2} \sum_{\varkappa} 
\Big[ M_\varkappa, \big[ M_\varkappa,  |\psi \ra \la \psi| \big] \Big] \:. \label{dpsi2}
\end{align}
Taking the trace, we obtain
\[ \frac{d}{dt} \bbra\: \la \psi | \psi \ra \:\kket = 0 \:, \]
meaning that current conservation holds (or, equivalently, that the total probability is conserved)
in the statistical mean.

The evolution equation~\eqref{dpsi2} also gives rise to a corresponding
evolution equation for the statistical operator denoted by~$\sigma_t$,
\[ 
\sigma_t := \bbra \:\big| \psi(t) \big\ra \big\la \psi(t) \big| \:\kket \:. \]
Indeed, by linearity we obtain that
\begin{align*}
\frac{d \sigma_t}{dt} =- \frac{1}{2} \sum_{\varkappa} \Big[ M_\varkappa, \big[ M_\varkappa,  \sigma_t \big] \Big] \:.
\end{align*}
This is a deterministic evolution equation of Kossakowski-Lindblad form
(see~\cite{kossakowski,lindblad}).

\subsection{Why is there no Collapse?} \label{secnocollapse}
We next verify that the linear stochastic dynamics does {\em{not}} describe a collapse. To this end,
let~${\mathcal{O}}$ be an operator which commutes with the free Hamiltonian and with the stochastic potentials,
\beq \label{commute}
[{\mathcal{O}}, H_0] =0 \qquad \text{and} \qquad [{\mathcal{O}}, M_\varkappa] = 0 \quad \text{for all~$\varkappa$}\:.
\eeq
This assumption is motivated by physical situation in mind. Typically, the observable~${\mathcal{O}}$
corresponds to a position measurement, so that the eigenspaces of~${\mathcal{O}}$ are localized
in different spatial regions. Then the vanishing of the above commutators is a consequence of the
locality of the infinitesimal time evolution.

Denoting expectation values by
\[ \la {\mathcal{O}} \ra := \la \psi | {\mathcal{O}} | \psi \ra \:,\qquad \la {\mathcal{O}}^2 \ra := \la \psi | {\mathcal{O}\bl{^2}} | \psi \ra \:, \]
we thus obtain
\beq \label{expectO}
d\: \bbra\:  \la {\mathcal{O}} \ra\: \kket = \tr \Big( {\mathcal{O}}\:  d \:\bbra\: |\psi \ra \la \psi | \:\kket \Big) = 0 \qquad \text{and} \qquad
d \:\bbra\: \la {\mathcal{O}}^2 \ra\: \kket = 0 \:,
\eeq
simply because~\eqref{dpsi2} is formed of commutators which all vanish in view of~\eqref{commute}.
The first equation means that the statistical mean of the expectation value is time independent, as desired.

In order to detect collapse, we must consider the variance. A direct computation gives
\beq \label{nocollapse}
d \:\bbra \la {\mathcal{O}}^2 \ra - \la {\mathcal{O}} \ra^2 \kket =
\sum_\varkappa \la \psi | [M_\varkappa, {\mathcal{O}}] \, \psi \ra \; 
\la \psi | [M_\varkappa, {\mathcal{O}}] \, \psi \ra = 0 \:.
\eeq
Clearly, from~\eqref{nocollapse} we conclude that the variance is time independent, showing
that {\em{no}} collapse takes place.

Note that the term on the right arises from Gaussian pairings between the two factors
in~$\la {\mathcal{O}} \ra\,\la {\mathcal{O}} \ra$
(all the other pairings vanish in view of~\eqref{expectO}). The commutators in~\eqref{nocollapse}
come about because
of the second minus sign in~\eqref{t21}. Therefore, the basic shortcoming of our ansatz is that
the potential~$V$ in~\eqref{schstoch} is a symmetric operator, in agreement with the fact that
the scalar product~$\la \psi | \psi \ra$ is conserved. As a consequence, also all the operators~$W_\varkappa$
in~\eqref{schstoch} are symmetric. This in turn gives rise to the minus sign in~\eqref{t21}
and the commutators in~\eqref{nocollapse}.
As we shall see in the next section, the situation becomes more interesting if the non-relativistic
limit is taken more carefully by taking into account the nonlocality in time of the potential~$\B$.

\section{Microscopic Derivation of the Effective Collapse Model} \label{secmodel}
\subsection{A Symmetric Potential Nonlocal in Space and Time} \label{secnonloc}
We slightly generalize the setting in Section~\ref{secpurespatial} by allowing that the
nonlocal potential in~\eqref{Bstochastic} also depends on time.
We point out that, in contrast to the setting considered in Section~\ref{secnonrelativ}, now the
nonlocal potential~$\B$ is also nonlocal in time. This will be crucial for the following constructions.

\subsection{Hamiltonian Formulation} \label{sechamilton}
The Dirac equation can again be written in the Hamiltonian form~\eqref{dynHamilton}.
Moreover, the Hamiltonian~$H$ can again be decomposed into the free Hamiltonian and the
interaction~\eqref{H0V}. But now the potential~$V$ is nonlocal in time, i.e.\
\[ (V \psi)(t) = \int_{-\infty}^\infty V(t,t')\: \psi(t')\: dt' \:, \]
where the kernel~$V(t,t')$ is the spatial operator
\beq \label{Vdef2}
V(t,t') \::\: \H_{t'} \rightarrow \H_{t} \:,\:
(V(t,t') \psi)(\vec{x}) = \int_{\R^3} \Big(- \gamma^0 \, \B\big( (t,\vec{x}), (t', \vec{y}) \Big) \: \psi(\vec{y})\: d^3y \:.
\eeq
Here~$\H_t = L^2(\R^3, \C^4)$ is the Hilbert space of square integrable Dirac wave functions at time~$t$.
It is convenient to again work in the interaction picture by setting
\[ \check{\psi}(t) := e^{-i t H_0} \psi(t) \qquad \text{and} \qquad
\check{V}(t,t') := e^{-i t H_0}\: V(t,t')\: e^{i t' H_0} \:. \]
Then the Dirac equation in the Hamiltonian form takes the simple form
\beq \label{Hint}
i \partial_t \check{\psi} = \big( \check{V} \check{\psi} \big)(t) = \int_{-\infty}^\infty \check{V}(t,t')\: \check{\psi}(t')\: dt' \:.
\eeq
From now on, we shall always work in the interaction picture. With this in mind, the check will be omitted.

This equation can be solved with the usual Dyson series. However, one must take into account that the
Dyson series is no longer strictly retarded, because the potential~$V$ is nonlocal in time.
In more detail, given initial data at time~$t_0$,
\begin{align}
\psi(t) &= \psi(t_0) + \int_{t_0}^t \dot{\psi}(\tau)\: d\tau \notag \\
&= \psi(t_0) + \int_{t_0}^t \big(-i V \psi \big)(\tau)\: d\tau = \psi(t_0) + \int_{t_0}^t d\tau \int_{-\infty}^\infty d\zeta\: \big(-i V(\tau, \zeta) \big)\: \psi(\zeta) \:. \notag \\
\intertext{By applying this equation iteratively, we obtain}
\psi(t) &=  \psi(t_0)
+ \int_{t_0}^t d\tau \int_{-\infty}^\infty d\zeta\: \big(-i V(\tau, \zeta) \big)\: \psi(t_0) \notag \\
&\quad\: + \int_{t_0}^t d\tau_1 \int_{-\infty}^\infty d\zeta_1\: \big(-i V(\tau_1, \zeta_1) \big)
\int_{t_0}^{\zeta_1} d\tau_2 \int_{-\infty}^\infty d\zeta_2\: \big(-i V(\tau_2, \zeta_2) \big)\: \psi(t_0) \notag \\
&\quad\: + \cdots \:. \label{dysonnonloc}
\end{align}
We refer to this series as the {\em{nonlocal Dyson series}}.
As illustrated in Figure~\ref{dyson}, the retardation is violated on the time scale~$\ell_{\min}$.
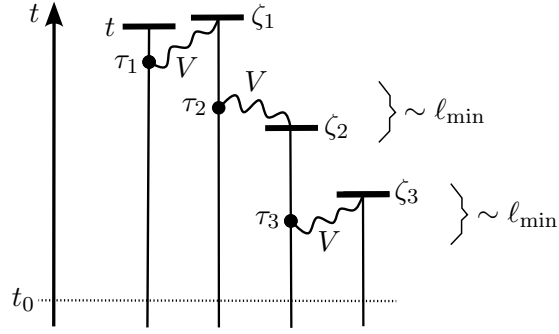
\begin{figure}
\psset{xunit=.5pt,yunit=.5pt,runit=.5pt}
\begin{pspicture}(323.68280366,245.65446989)
{
\newrgbcolor{curcolor}{0 0 0}
\pscustom[linewidth=4.4764723,linecolor=curcolor]
{
\newpath
\moveto(63.39203906,226.79784564)
\lineto(103.1518715,226.98942989)
}
}
{
\newrgbcolor{curcolor}{0 0 0}
\pscustom[linewidth=2.48314963,linecolor=curcolor]
{
\newpath
\moveto(11.48811591,1.09792249)
\lineto(11.72909858,230.01063682)
}
}
{
\newrgbcolor{curcolor}{0 0 0}
\pscustom[linestyle=none,fillstyle=solid,fillcolor=curcolor]
{
\newpath
\moveto(4.77262377,226.54154869)
\lineto(11.74556727,245.65447079)
\lineto(18.67825397,226.52690985)
\curveto(14.63189889,230.00758096)(8.84368032,230.01367438)(4.77262377,226.54154869)
\closepath
}
}
{
\newrgbcolor{curcolor}{0 0 0}
\pscustom[linewidth=1.79999626,linecolor=curcolor]
{
\newpath
\moveto(83.27194961,226.89363398)
\lineto(82.00555087,0.31994831)
}
}
{
\newrgbcolor{curcolor}{0 0 0}
\pscustom[linewidth=4.4764723,linecolor=curcolor]
{
\newpath
\moveto(114.80152063,233.54245036)
\lineto(154.56135307,233.73403461)
}
}
{
\newrgbcolor{curcolor}{0 0 0}
\pscustom[linewidth=1.72280315,linecolor=curcolor]
{
\newpath
\moveto(135.54283465,233.63823871)
\lineto(135.8280378,0.31492154)
}
}
{
\newrgbcolor{curcolor}{1 1 1}
\pscustom[linestyle=none,fillstyle=solid,fillcolor=curcolor]
{
\newpath
\moveto(186.43468395,144.03289095)
\curveto(186.43468395,142.66349748)(184.69022113,141.55338473)(182.53831717,141.55338473)
\curveto(180.38641322,141.55338473)(178.64195039,142.66349748)(178.64195039,144.03289095)
\curveto(178.64195039,145.40228442)(180.38641322,146.51239716)(182.53831717,146.51239716)
\curveto(184.69022113,146.51239716)(186.43468395,145.40228442)(186.43468395,144.03289095)
\closepath
}
}
{
\newrgbcolor{curcolor}{1 1 1}
\pscustom[linewidth=4.4764723,linecolor=curcolor]
{
\newpath
\moveto(186.43468395,144.03289095)
\curveto(186.43468395,142.66349748)(184.69022113,141.55338473)(182.53831717,141.55338473)
\curveto(180.38641322,141.55338473)(178.64195039,142.66349748)(178.64195039,144.03289095)
\curveto(178.64195039,145.40228442)(180.38641322,146.51239716)(182.53831717,146.51239716)
\curveto(184.69022113,146.51239716)(186.43468395,145.40228442)(186.43468395,144.03289095)
\closepath
}
}
{
\newrgbcolor{curcolor}{1 1 1}
\pscustom[linestyle=none,fillstyle=solid,fillcolor=curcolor]
{
\newpath
\moveto(214.06351544,144.38710632)
\curveto(214.06351544,138.90953244)(207.00636548,134.46908146)(198.30092982,134.46908146)
\curveto(189.59549417,134.46908146)(182.53834421,138.90953244)(182.53834421,144.38710632)
\curveto(182.53834421,149.86468019)(189.59549417,154.30513117)(198.30092982,154.30513117)
\curveto(207.00636548,154.30513117)(214.06351544,149.86468019)(214.06351544,144.38710632)
\closepath
}
}
{
\newrgbcolor{curcolor}{1 1 1}
\pscustom[linewidth=4.4764723,linecolor=curcolor]
{
\newpath
\moveto(214.06351544,144.38710632)
\curveto(214.06351544,138.90953244)(207.00636548,134.46908146)(198.30092982,134.46908146)
\curveto(189.59549417,134.46908146)(182.53834421,138.90953244)(182.53834421,144.38710632)
\curveto(182.53834421,149.86468019)(189.59549417,154.30513117)(198.30092982,154.30513117)
\curveto(207.00636548,154.30513117)(214.06351544,149.86468019)(214.06351544,144.38710632)
\closepath
}
}
{
\newrgbcolor{curcolor}{1 1 1}
\pscustom[linestyle=none,fillstyle=solid,fillcolor=curcolor]
{
\newpath
\moveto(230.35737139,116.34763243)
\curveto(230.35737139,107.5755768)(227.21090601,100.46442068)(223.32954426,100.46442068)
\curveto(219.44818251,100.46442068)(216.30171712,107.5755768)(216.30171712,116.34763243)
\curveto(216.30171712,125.11968805)(219.44818251,132.23084418)(223.32954426,132.23084418)
\curveto(227.21090601,132.23084418)(230.35737139,125.11968805)(230.35737139,116.34763243)
\closepath
}
}
{
\newrgbcolor{curcolor}{1 1 1}
\pscustom[linewidth=4.4764723,linecolor=curcolor]
{
\newpath
\moveto(230.35737139,116.34763243)
\curveto(230.35737139,107.5755768)(227.21090601,100.46442068)(223.32954426,100.46442068)
\curveto(219.44818251,100.46442068)(216.30171712,107.5755768)(216.30171712,116.34763243)
\curveto(216.30171712,125.11968805)(219.44818251,132.23084418)(223.32954426,132.23084418)
\curveto(227.21090601,132.23084418)(230.35737139,125.11968805)(230.35737139,116.34763243)
\closepath
}
}
{
\newrgbcolor{curcolor}{1 1 1}
\pscustom[linestyle=none,fillstyle=solid,fillcolor=curcolor]
{
\newpath
\moveto(278.5306527,118.17620726)
\curveto(278.5306527,96.9866003)(264.60023288,79.80902986)(247.41618919,79.80902986)
\curveto(230.2321455,79.80902986)(216.30172568,96.9866003)(216.30172568,118.17620726)
\curveto(216.30172568,139.36581423)(230.2321455,156.54338467)(247.41618919,156.54338467)
\curveto(264.60023288,156.54338467)(278.5306527,139.36581423)(278.5306527,118.17620726)
\closepath
}
}
{
\newrgbcolor{curcolor}{1 1 1}
\pscustom[linewidth=4.4764723,linecolor=curcolor]
{
\newpath
\moveto(278.5306527,118.17620726)
\curveto(278.5306527,96.9866003)(264.60023288,79.80902986)(247.41618919,79.80902986)
\curveto(230.2321455,79.80902986)(216.30172568,96.9866003)(216.30172568,118.17620726)
\curveto(216.30172568,139.36581423)(230.2321455,156.54338467)(247.41618919,156.54338467)
\curveto(264.60023288,156.54338467)(278.5306527,139.36581423)(278.5306527,118.17620726)
\closepath
}
}
{
\newrgbcolor{curcolor}{0.00784314 0.00784314 0.00784314}
\pscustom[linestyle=none,fillstyle=solid,fillcolor=curcolor]
{
\newpath
\moveto(87.29702946,200.14477608)
\curveto(87.29702946,197.92178856)(85.49494222,196.11970111)(83.27195495,196.11970111)
\curveto(81.04896768,196.11970111)(79.24688043,197.92178856)(79.24688043,200.14477608)
\curveto(79.24688043,202.3677636)(81.04896768,204.16985105)(83.27195495,204.16985105)
\curveto(85.49494222,204.16985105)(87.29702946,202.3677636)(87.29702946,200.14477608)
\closepath
}
}
{
\newrgbcolor{curcolor}{0 0 0}
\pscustom[linewidth=2.15457646,linecolor=curcolor,strokeopacity=0.02654868]
{
\newpath
\moveto(87.29702946,200.14477608)
\curveto(87.29702946,197.92178856)(85.49494222,196.11970111)(83.27195495,196.11970111)
\curveto(81.04896768,196.11970111)(79.24688043,197.92178856)(79.24688043,200.14477608)
\curveto(79.24688043,202.3677636)(81.04896768,204.16985105)(83.27195495,204.16985105)
\curveto(85.49494222,204.16985105)(87.29702946,202.3677636)(87.29702946,200.14477608)
\closepath
}
}
{
\newrgbcolor{curcolor}{0 0 0}
\pscustom[linewidth=1.79999768,linecolor=curcolor]
{
\newpath
\moveto(82.42870224,200.48985823)
\curveto(88.24293241,197.39247946)(94.05715578,194.29510749)(96.25581808,197.11516268)
\curveto(98.45448718,199.93522466)(97.03765644,208.67234642)(99.67087672,211.07138869)
\curveto(102.3040902,213.47043096)(108.9871234,209.53135291)(112.32870123,211.30249168)
\curveto(115.67027906,213.07365086)(115.67027906,220.55479556)(117.99618746,222.52433458)
\curveto(120.32210268,224.4938736)(124.97373581,220.95178655)(128.2888698,221.83311417)
\curveto(131.60400378,222.71443499)(133.58247534,228.01901401)(135.56098091,233.32370189)
}
}
{
\newrgbcolor{curcolor}{0.00784314 0.00784314 0.00784314}
\pscustom[linestyle=none,fillstyle=solid,fillcolor=curcolor]
{
\newpath
\moveto(139.71050852,165.69415256)
\curveto(139.71050852,163.47116504)(137.90842128,161.66907759)(135.68543401,161.66907759)
\curveto(133.46244674,161.66907759)(131.66035949,163.47116504)(131.66035949,165.69415256)
\curveto(131.66035949,167.91714008)(133.46244674,169.71922753)(135.68543401,169.71922753)
\curveto(137.90842128,169.71922753)(139.71050852,167.91714008)(139.71050852,165.69415256)
\closepath
}
}
{
\newrgbcolor{curcolor}{0 0 0}
\pscustom[linewidth=2.15457646,linecolor=curcolor,strokeopacity=0.02654868]
{
\newpath
\moveto(139.71050852,165.69415256)
\curveto(139.71050852,163.47116504)(137.90842128,161.66907759)(135.68543401,161.66907759)
\curveto(133.46244674,161.66907759)(131.66035949,163.47116504)(131.66035949,165.69415256)
\curveto(131.66035949,167.91714008)(133.46244674,169.71922753)(135.68543401,169.71922753)
\curveto(137.90842128,169.71922753)(139.71050852,167.91714008)(139.71050852,165.69415256)
\closepath
}
}
{
\newrgbcolor{curcolor}{0 0 0}
\pscustom[linewidth=4.4764723,linecolor=curcolor]
{
\newpath
\moveto(224.48543244,100.46439304)
\lineto(264.2452611,100.65597729)
}
}
{
\newrgbcolor{curcolor}{0.00784314 0.00784314 0.00784314}
\pscustom[linestyle=none,fillstyle=solid,fillcolor=curcolor]
{
\newpath
\moveto(194.01021902,80.48386001)
\curveto(194.01021902,78.26087249)(192.20813177,76.45878505)(189.9851445,76.45878505)
\curveto(187.76215723,76.45878505)(185.96006998,78.26087249)(185.96006998,80.48386001)
\curveto(185.96006998,82.70684753)(187.76215723,84.50893498)(189.9851445,84.50893498)
\curveto(192.20813177,84.50893498)(194.01021902,82.70684753)(194.01021902,80.48386001)
\closepath
}
}
{
\newrgbcolor{curcolor}{0 0 0}
\pscustom[linewidth=2.15457646,linecolor=curcolor,strokeopacity=0.02654868]
{
\newpath
\moveto(194.01021902,80.48386001)
\curveto(194.01021902,78.26087249)(192.20813177,76.45878505)(189.9851445,76.45878505)
\curveto(187.76215723,76.45878505)(185.96006998,78.26087249)(185.96006998,80.48386001)
\curveto(185.96006998,82.70684753)(187.76215723,84.50893498)(189.9851445,84.50893498)
\curveto(192.20813177,84.50893498)(194.01021902,82.70684753)(194.01021902,80.48386001)
\closepath
}
}
{
\newrgbcolor{curcolor}{0 0 0}
\pscustom[linewidth=1.75748033,linecolor=curcolor,linestyle=dashed,dash=0.93000001 1.86000001]
{
\newpath
\moveto(0.00195402,20.77383304)
\lineto(269.36286236,20.17538264)
}
}
{
\newrgbcolor{curcolor}{1 1 1}
\pscustom[linestyle=none,fillstyle=solid,fillcolor=curcolor]
{
\newpath
\moveto(186.43468395,144.03289095)
\curveto(186.43468395,142.66349748)(184.69022113,141.55338473)(182.53831717,141.55338473)
\curveto(180.38641322,141.55338473)(178.64195039,142.66349748)(178.64195039,144.03289095)
\curveto(178.64195039,145.40228442)(180.38641322,146.51239716)(182.53831717,146.51239716)
\curveto(184.69022113,146.51239716)(186.43468395,145.40228442)(186.43468395,144.03289095)
\closepath
}
}
{
\newrgbcolor{curcolor}{1 1 1}
\pscustom[linewidth=4.4764723,linecolor=curcolor]
{
\newpath
\moveto(186.43468395,144.03289095)
\curveto(186.43468395,142.66349748)(184.69022113,141.55338473)(182.53831717,141.55338473)
\curveto(180.38641322,141.55338473)(178.64195039,142.66349748)(178.64195039,144.03289095)
\curveto(178.64195039,145.40228442)(180.38641322,146.51239716)(182.53831717,146.51239716)
\curveto(184.69022113,146.51239716)(186.43468395,145.40228442)(186.43468395,144.03289095)
\closepath
}
}
{
\newrgbcolor{curcolor}{1 1 1}
\pscustom[linestyle=none,fillstyle=solid,fillcolor=curcolor]
{
\newpath
\moveto(214.06351544,144.38710632)
\curveto(214.06351544,138.90953244)(207.00636548,134.46908146)(198.30092982,134.46908146)
\curveto(189.59549417,134.46908146)(182.53834421,138.90953244)(182.53834421,144.38710632)
\curveto(182.53834421,149.86468019)(189.59549417,154.30513117)(198.30092982,154.30513117)
\curveto(207.00636548,154.30513117)(214.06351544,149.86468019)(214.06351544,144.38710632)
\closepath
}
}
{
\newrgbcolor{curcolor}{1 1 1}
\pscustom[linewidth=4.4764723,linecolor=curcolor]
{
\newpath
\moveto(214.06351544,144.38710632)
\curveto(214.06351544,138.90953244)(207.00636548,134.46908146)(198.30092982,134.46908146)
\curveto(189.59549417,134.46908146)(182.53834421,138.90953244)(182.53834421,144.38710632)
\curveto(182.53834421,149.86468019)(189.59549417,154.30513117)(198.30092982,154.30513117)
\curveto(207.00636548,154.30513117)(214.06351544,149.86468019)(214.06351544,144.38710632)
\closepath
}
}
{
\newrgbcolor{curcolor}{1 1 1}
\pscustom[linestyle=none,fillstyle=solid,fillcolor=curcolor]
{
\newpath
\moveto(230.35737139,116.34763243)
\curveto(230.35737139,107.5755768)(227.21090601,100.46442068)(223.32954426,100.46442068)
\curveto(219.44818251,100.46442068)(216.30171712,107.5755768)(216.30171712,116.34763243)
\curveto(216.30171712,125.11968805)(219.44818251,132.23084418)(223.32954426,132.23084418)
\curveto(227.21090601,132.23084418)(230.35737139,125.11968805)(230.35737139,116.34763243)
\closepath
}
}
{
\newrgbcolor{curcolor}{1 1 1}
\pscustom[linewidth=4.4764723,linecolor=curcolor]
{
\newpath
\moveto(230.35737139,116.34763243)
\curveto(230.35737139,107.5755768)(227.21090601,100.46442068)(223.32954426,100.46442068)
\curveto(219.44818251,100.46442068)(216.30171712,107.5755768)(216.30171712,116.34763243)
\curveto(216.30171712,125.11968805)(219.44818251,132.23084418)(223.32954426,132.23084418)
\curveto(227.21090601,132.23084418)(230.35737139,125.11968805)(230.35737139,116.34763243)
\closepath
}
}
{
\newrgbcolor{curcolor}{1 1 1}
\pscustom[linestyle=none,fillstyle=solid,fillcolor=curcolor]
{
\newpath
\moveto(309.0834582,166.32003445)
\curveto(309.0834582,145.13042748)(295.15303838,127.95285704)(277.96899469,127.95285704)
\curveto(260.78495101,127.95285704)(246.85453119,145.13042748)(246.85453119,166.32003445)
\curveto(246.85453119,187.50964141)(260.78495101,204.68721185)(277.96899469,204.68721185)
\curveto(295.15303838,204.68721185)(309.0834582,187.50964141)(309.0834582,166.32003445)
\closepath
}
}
{
\newrgbcolor{curcolor}{1 1 1}
\pscustom[linewidth=4.4764723,linecolor=curcolor]
{
\newpath
\moveto(309.0834582,166.32003445)
\curveto(309.0834582,145.13042748)(295.15303838,127.95285704)(277.96899469,127.95285704)
\curveto(260.78495101,127.95285704)(246.85453119,145.13042748)(246.85453119,166.32003445)
\curveto(246.85453119,187.50964141)(260.78495101,204.68721185)(277.96899469,204.68721185)
\curveto(295.15303838,204.68721185)(309.0834582,187.50964141)(309.0834582,166.32003445)
\closepath
}
}
{
\newrgbcolor{curcolor}{0.00784314 0.00784314 0.00784314}
\pscustom[linestyle=none,fillstyle=solid,fillcolor=curcolor]
{
\newpath
\moveto(87.29702946,200.14477608)
\curveto(87.29702946,197.92178856)(85.49494222,196.11970111)(83.27195495,196.11970111)
\curveto(81.04896768,196.11970111)(79.24688043,197.92178856)(79.24688043,200.14477608)
\curveto(79.24688043,202.3677636)(81.04896768,204.16985105)(83.27195495,204.16985105)
\curveto(85.49494222,204.16985105)(87.29702946,202.3677636)(87.29702946,200.14477608)
\closepath
}
}
{
\newrgbcolor{curcolor}{0 0 0}
\pscustom[linewidth=2.15457646,linecolor=curcolor,strokeopacity=0.02654868]
{
\newpath
\moveto(87.29702946,200.14477608)
\curveto(87.29702946,197.92178856)(85.49494222,196.11970111)(83.27195495,196.11970111)
\curveto(81.04896768,196.11970111)(79.24688043,197.92178856)(79.24688043,200.14477608)
\curveto(79.24688043,202.3677636)(81.04896768,204.16985105)(83.27195495,204.16985105)
\curveto(85.49494222,204.16985105)(87.29702946,202.3677636)(87.29702946,200.14477608)
\closepath
}
}
{
\newrgbcolor{curcolor}{0 0 0}
\pscustom[linewidth=4.4764723,linecolor=curcolor]
{
\newpath
\moveto(170.66163024,150.36962642)
\lineto(210.42146268,150.56121068)
}
}
{
\newrgbcolor{curcolor}{0 0 0}
\pscustom[linewidth=1.72280315,linecolor=curcolor]
{
\newpath
\moveto(189.61271433,150.77502611)
\lineto(190.3575685,0.00425949)
}
}
{
\newrgbcolor{curcolor}{0 0 0}
\pscustom[linewidth=1.79999768,linecolor=curcolor]
{
\newpath
\moveto(135.62599294,165.19261074)
\curveto(141.04093266,170.97626075)(146.45574312,176.7597679)(149.14972233,174.6652142)
\curveto(151.84370835,172.5706537)(151.81663181,162.59839291)(154.64286425,162.37173918)
\curveto(157.4691035,162.14505823)(163.14829795,171.66403877)(166.14250016,171.35572003)
\curveto(169.13670236,171.04740129)(169.44552454,160.91186491)(172.31191597,158.77727468)
\curveto(175.1783074,156.64268444)(180.60190753,162.50932608)(184.03291276,162.17273345)
\curveto(187.46391798,161.8361136)(188.90192693,155.29623908)(190.33996309,148.75619449)
}
}
{
\newrgbcolor{curcolor}{0 0 0}
\pscustom[linewidth=1.72119681,linecolor=curcolor]
{
\newpath
\moveto(244.29868346,101.48900784)
\lineto(244.57311496,1.2385776)
}
}
{
\newrgbcolor{curcolor}{0 0 0}
\pscustom[linewidth=4.4764723,linecolor=curcolor]
{
\newpath
\moveto(224.48543244,100.46439304)
\lineto(264.2452611,100.65597729)
}
}
{
\newrgbcolor{curcolor}{0.00784314 0.00784314 0.00784314}
\pscustom[linestyle=none,fillstyle=solid,fillcolor=curcolor]
{
\newpath
\moveto(194.01021902,80.48386001)
\curveto(194.01021902,78.26087249)(192.20813177,76.45878505)(189.9851445,76.45878505)
\curveto(187.76215723,76.45878505)(185.96006998,78.26087249)(185.96006998,80.48386001)
\curveto(185.96006998,82.70684753)(187.76215723,84.50893498)(189.9851445,84.50893498)
\curveto(192.20813177,84.50893498)(194.01021902,82.70684753)(194.01021902,80.48386001)
\closepath
}
}
{
\newrgbcolor{curcolor}{0 0 0}
\pscustom[linewidth=2.15457646,linecolor=curcolor,strokeopacity=0.02654868]
{
\newpath
\moveto(194.01021902,80.48386001)
\curveto(194.01021902,78.26087249)(192.20813177,76.45878505)(189.9851445,76.45878505)
\curveto(187.76215723,76.45878505)(185.96006998,78.26087249)(185.96006998,80.48386001)
\curveto(185.96006998,82.70684753)(187.76215723,84.50893498)(189.9851445,84.50893498)
\curveto(192.20813177,84.50893498)(194.01021902,82.70684753)(194.01021902,80.48386001)
\closepath
}
}
{
\newrgbcolor{curcolor}{0 0 0}
\pscustom[linewidth=1.79999768,linecolor=curcolor]
{
\newpath
\moveto(189.84392844,80.77567216)
\curveto(195.23703685,75.30554529)(200.630016,69.83555449)(203.46750765,71.43058012)
\curveto(206.30499931,73.02560576)(206.58674494,81.68535531)(209.64460422,83.00351997)
\curveto(212.7024703,84.32168463)(218.53621191,78.29798548)(221.45313713,79.0268137)
\curveto(224.37006236,79.75564192)(224.37006236,87.23678661)(226.69597077,89.20632564)
\curveto(229.02188598,91.17586466)(233.67351912,87.6337776)(236.9886531,88.51510523)
\curveto(240.30378709,89.39642605)(242.28225865,94.70100507)(244.26076422,100.00569294)
}
}
{
\newrgbcolor{curcolor}{0 0 0}
\pscustom[linewidth=1.12251965,linecolor=curcolor]
{
\newpath
\moveto(256.25738457,185.569508)
\lineto(264.64119307,174.77152249)
\lineto(264.64119307,167.08787902)
\lineto(268.75743118,162.14839178)
\lineto(264.36678047,158.02470422)
\lineto(264.36678047,148.1531754)
\lineto(256.25738457,139.64628658)
}
}
{
\newrgbcolor{curcolor}{0 0 0}
\pscustom[linewidth=1.12251965,linecolor=curcolor]
{
\newpath
\moveto(310.4129726,108.57956406)
\lineto(318.7967811,97.78157855)
\lineto(318.7967811,90.09793508)
\lineto(322.91301921,85.15844784)
\lineto(318.5223685,81.03476028)
\lineto(318.5223685,71.16323146)
\lineto(310.4129726,62.65634264)
}
\rput[bl](-7,230){$t$}
\rput[bl](50,220){$t$}
\rput[bl](57,192){$\tau_1$}
\rput[bl](160,225){$\zeta_1$}
\rput[bl](-20,13){$t_0$}
\rput[bl](108,160){$\tau_2$}
\rput[bl](215,142){$\zeta_2$}
\rput[bl](163,75){$\tau_3$}
\rput[bl](268,93){$\zeta_3$}
\rput[bl](210,58){$V$}
\rput[bl](155,178){$V$}
\rput[bl](103,189){$V$}
\rput[bl](275,154){$\sim \ell_{\min}$}
\rput[bl](330,76){$\sim \ell_{\min}$}
}
\end{pspicture}
\caption{The third-order contribution to the nonlocal Dyson series.}
\label{figdyson}
\end{figure}%
In what follows, we always assume that the potential~$\B$ is so small that this perturbation expansion
converges. Then, despite the nonlocality of the evolution equation in time, the Cauchy problem is well-posed.
We thus obtain a time evolution operator
\[ U^t_{t_0} \::\: \H_{t_0} \rightarrow \H_t \:, \qquad \psi(t_0) \mapsto \psi(t) \:. \]
This time evolution preserves the scalar product~$\la .|. \ra_t$ (which is also nonlocal in time).

\subsection{Transformation to the Standard $L^2$-Scalar Product} \label{sectransform}
The solution space of the Dirac equation~\eqref{dirnonloc} is endowed with the scalar
product~\eqref{c11}--\eqref{c3}, which is nonlocal in both space and time.
We choose~$\scrN$ as the Cauchy surface at time~$t_0$.
Taking the completion, we thus obtain the Hilbert space denoted by~$(\H_m, \la.|. \ra_t)$.
We now analyze how this scalar product can be related to the usual $L^2$-scalar product at time~$t_0$,
which we denote by
\[ ( \psi | \phi )_{t_0} := \int_{\R^3} \Sl \psi \,|\, \gamma^0 \phi \Sr(t_0, \vec{x})\: d^3x
= \int_{\R^3} \big( \psi^\dagger \phi \big)(t_0, \vec{x})\: d^3x \:, \]
where~$\psi^\dagger$ is the transposed complex conjugated spinor.
Using the notation~\eqref{Vdef2}, these two scalar products are related by
\beq \label{scalcomp}
\begin{split}
\la \psi | \phi \ra_{t_0} = ( \psi | \phi)_{t_0} &+i \int_{-\infty}^{t_0} d\tau \int_{t_0}^\infty d\tau' \:
\big( \psi(\tau) \,|\, V(\tau,\tau') \:\phi(\tau') \big)_\tau \\
&-i \int_{t_0}^\infty d\tau \int_{-\infty}^{t_0} d\tau' \:
\big( \psi(\tau) \,|\, V(\tau,\tau') \:\phi(\tau') \big)_\tau \:.
\end{split}
\eeq
Again assuming that the potential~$V$ is sufficiently small, these two scalar products are
bounded with respect to each other in the sense that there is a constant~$C$ such that
\[ \frac{1}{C}\: ( \psi | \psi)_{t_0} \leq \la \psi | \psi \ra_{t_0} \leq C\, ( \psi | \psi)_{t_0} \qquad \text{for all~$\psi \in \H_m$}\:. \]
Then we can represent one scalar product with respect to the other. In particular, there is a
bounded symmetric operator~$\Sig_{t_0} \in \Lin(\H_m)$ such that
\beq \label{Sigdef}
\la \psi | \phi \ra_{t_0} = \big( \psi \,|\, (\1+\Sig_{t_0}) \phi \big)_{t_0} \qquad \text{for all~$\psi, \phi \in \H_m$} \:.
\eeq
Moreover, the operator~$(\1+\Sig_{t_0})$ has a bounded inverse.

In what follows, it will sometimes be convenient to work on~$\H_m$ with the
simpler $L^2$-scalar product~$( .|. )_{t_0}$. This can be arranged by transforming the wave functions
at time~${t_0}$ according to
\beq \label{tildepsi}
\psi \mapsto \tilde{\psi} := \sqrt{\1+\Sig_{t_0}} \:\psi \:,
\eeq
so that
\[ \la \psi | \phi \ra_{t_0} = ( \tilde{\psi} | \tilde{\phi})_{t_0} \:. \]
In this way, the dependence of the scalar product on the nonlocal potential has been absorbed
into the wave functions. One should keep in mind, however, that the transformations from~$\psi$ to~$\tilde{\psi}$
and vice versa are nonlocal.

Performing the above construction at different times~$t_0$, we obtain a family of operators~$\Sig_t$
with~$t \in \R$. We point out that these operators in general depend on~$t$.
We denote the time evolution of the transformed wave function by
\[ 
\tilde{U}^t_{t_0} \::\: L^2(\R^3, \C^4) \rightarrow L^2(\R^3, \C^4)\:,\qquad
\tilde{\psi}(t_0) \mapsto \tilde{\psi}(t) \:. \]
This time evolution (which is again nonlocal in time) preserves the scalar product~$(.|.)_t$.
It is related to the time evolution~$U^t_{t_0}$ by
\[ \tilde{U}^t_{t_0} = \big( \1+\Sig_t \big)^\frac{1}{2}\: U^t_{t_0}\: \big( \1+\Sig_{t_0} \big)^{-\frac{1}{2}}\;. \]

\subsection{Markovian Gaussian Background Fields} \label{secmarkov}
Following the explanations in the introduction (shown in Figure~\ref{figcoll1}),
we now choose the nonlocal potential more concretely according to the ansatz~\eqref{Bstochastic}, i.e.\
\[ 
(V(t,t') \psi)(\vec{x}) = -\sum_{a=1}^N \gamma^0 B_a\Big( \frac{t+ t'}{2} , \frac{\vec{x}+\vec{y}}{2} \Big) 
\int_{\R^3} L_a\big( (t', \vec{y}) - (t,\vec{x}) \big)\: \psi(\vec{y})\: d^3y \:. \]
Similar to~\eqref{meanC} and~\eqref{covC} we again
assume that the stochastic potentials have mean zero and are Gaussian and Markovian, i.e.\
\begin{align}
\bbra \gamma^0 B_a (t, \vec{x}) \kket &= 0 \label{markov1} \\
\bbra \:\big(\gamma^0 B_a(t, \vec{x}) \big)^\alpha_\beta\: 
\big( \gamma^0 B_b(t',\vec{y}) \big)^\gamma_\delta \:\kket &= \delta(t-t')\: \delta_{ab}\:
C^{\;\:\,\alpha \gamma}_{a, \beta \delta}(t, \vec{x}, \vec{y}) \:. \label{markov2}
\end{align}
Proceeding exactly as in Lemma~\ref{lemmadiag}, we can write the potential as
\beq \label{Vrep2}
V(t,t') = \sum_{k,a} \: M_{k,a}(t,t')\: W_{k,a}\Big( \frac{t+t'}{2} \Big)
\eeq
where~$W_{k,a}(t)$ are normalized Gaussian random fields~\eqref{gaussW},
and~$M_{k,a}(t,t')$ are (for any fixed~$t$ and~$t'$) spatial operators with kernels
\[ 
\big(M_{k,a}(t,t') \big)^\alpha_\beta(\vec{x}, \vec{y}) := \sqrt{\lambda_k}\: 
\phi^{\alpha}_{k, \beta}\big( t, t', \vec{x}, \vec{y} \big) \: L_a\big( (t', \vec{y}) - (t,\vec{x}) \big)\:. \]
In order to keep the setting as simple as possible, we shall consider the case that the functions~$\phi^{\alpha}_{k, \beta}$
factor into a product of a function depending on time and a function depending on the spatial variables.
Moreover, for simplicity we assume that the covariance is static in the sense that it depends only on the time difference, i.e.\
\beq \label{Deltadef}
\big(M_{k,a}(t,t') \big)^\alpha_\beta(\vec{x}, \vec{y}) = \Delta_{k,a}(t'-t)\:
\big(N_{k,a}\big)^\alpha_\beta(\vec{x}, \vec{y})
\eeq
with a function~$\Delta \in C^\infty_0(\R)$ and new, purely spatial kernels~$N_{k,a}$.
In analogy to~\eqref{Masymm}, these kernels are symmetric in the sense that
\[ \overline{\Delta_{k,a}(\tau)} = \Delta_{k,a}(-\tau)\qquad \text{and} \qquad \overline{\big(N_{k,a}\big)^\alpha_\beta(\vec{x}, \vec{y})} = 
\big(N_{k,a}\big)^\beta_\alpha(\vec{y}, \vec{x}) \:. \]
The first relation means that~$\Delta_{k,a}(\tau) + \Delta_{k,a}(-\tau)$ is real.
Moreover, by flipping the sign of~$N_{k,a}$ we can arrange that this function is non-negative
at~$\tau=0$. For simplicity, we shall assume that this function does not change signs, i.e.\
\beq \label{Deltapos}
\Delta_{k,a}(\tau) + \Delta_{k,a}(-\tau) \geq 0 \qquad \text{for all~$\tau \in \R$}\:.
\eeq
Again, this assumption could be weakened considerably and is made for ease of presentation.

\subsection{Derivation of the Kossakowski-Lindblad Dynamics} \label{seclindblad}
As explained in Section~\ref{sectransform}, the transformed wave function~$\tilde{\psi}$
defined by~\eqref{tildepsi} has a unitary time evolution with respect to the standard scalar product.
With this in mind, we introduce the corresponding density operator~$\sigma_t$ as
the statistical mean of the projection operator to the state~$\tilde{\psi}(t)$, i.e.\
\beq \label{rhotildedef}
\sigma_t := \bbra \:\big| \tilde{\psi}(t) \big) \big( \tilde{\psi}(t) \big| \:\kket \:.
\eeq
In this section, we prove the following result.
\begin{Thm} \label{thmlindblad}
The statistical operator~\eqref{rhotildedef} satisfies a deterministic equation
of the Kossakowski-Lindblad form
\[ \frac{d \sigma_t}{dt} = -i [H, \sigma_t] - \frac{1}{2} \sint_\varkappa 
\big[ K_\varkappa, [K_\varkappa, \sigma_t] \big] \; \Big( 1 + \O \big( \ell_{\min}\, \|\B\| \big) \Big)\:. \]
Here~$\varkappa$ stands for the combination of parameters
\beq \label{kappadef}
\varkappa = (k, a, \zeta) \qquad \text{and} \qquad
\sint_\varkappa\;\cdots := \sum_{k,a} \int_{-\infty}^\infty d\zeta \;\cdots \:.
\eeq
The effective Hamiltonian takes the form
\begin{align}
H = H_0 &- i \sint_\varkappa
\int_{-\infty}^{-\zeta} \big[ M_{k,a}(t,t+2 \zeta), M_{k,a}(t+\zeta+\nu, t+\zeta-\nu) \big] \: d\nu \label{Hn1} \\
&-i \sint_\varkappa \int_{-\infty}^{-\zeta} \big[ M_{k,a}(t+2 \zeta, t), M_{k,a}(t+\zeta-\nu, t+\zeta+\nu) \big] \: d\nu \\
&+2i \sint_\varkappa \int_{-\infty}^\zeta \Big( M_{k,a}(t, t+2 \zeta) \:M_{k,a}(t+\zeta+\nu, t+\zeta-\nu) \\
&\qquad\qquad \;\;\:
- M_{k,a}(t+\zeta-\nu, t+\zeta+\nu) \:M_{k,a}(t+2 \zeta, t) \Big)\: d\nu \:. \label{Hn4}
\end{align}
Moreover, the kernels~$K_\varkappa$ are given by
\begin{align}
K_\varkappa &:= \sqrt{\frac{\alpha(\zeta)\, \beta(\zeta)}{2}}\: N_{k \alpha} \qquad \text{and} \label{Kkappa} \\
a(\zeta) &:= \Delta_{k,a}\big( (t+2 \zeta) - t \big) + \Delta_{k,a}\big( t - (t+2 \zeta) \big) \label{adef} \\
b(\zeta) &:= \int_{-\infty}^{-\zeta} \Big( \Delta_{k,a}\big( (t+\zeta-\nu) - (t+\zeta+\nu) \big) \notag \\
&\qquad\quad\: + \Delta_{k,a}\big( (t+\zeta+\nu) - (t+\zeta-\nu) \big) \Big) \:d\nu \label{bdef} 
\end{align}
(and~$N_{k,a}(\vec{x}, \vec{y})$ as in~\eqref{Deltadef}).
\end{Thm}

The remainder of this section is devoted to the derivation of this result.
We also make use of the detailed computations worked out in Appendix~\ref{applindblad}.
We proceed in several steps.
\subsubsection{Computation of~$\bbra \partial_t \psi \kket$} \label{sectpsi}
 In order to explain our methods, we first explain how to compute
the statistical mean of the time derivative of the untransformed wave function
\[ \bbra \,\partial_t \psi(t)\, \kket \:. \]
To this end, we differentiate the nonlocal Dyson series~\eqref{dysonnonloc} to obtain
\begin{align*}
\frac{d}{dt} \psi(t) &= 
\int_{-\infty}^\infty d\zeta\: \big(-i V(t, \zeta) \big)\: \psi(t_0) \notag \\
&\quad\: + \int_{-\infty}^\infty d\zeta_1\: \big(-i V(t, \zeta_1) \big)
\int_{t_0}^{\zeta_1} d\tau_2 \int_{-\infty}^\infty d\zeta_2\: \big(-i V(\tau_2, \zeta_2) \big)\: \psi(t_0) 
+ \cdots \:,
\end{align*}
where the dots stand for the higher orders in perturbation theory.
Writing the potential in the form~\eqref{Vrep2}, we can compute the statistical mean to each order
by taking Gaussian pairings with the help of~\eqref{gaussW}.
Assuming that~$\psi(t_0)$ is fixed, we can take it out of the statistical mean.
Clearly, the first order vanishes because the mean is zero. We thus obtain
\begin{align}
&\tfrac{d}{dt} \bbra \psi(t) \kket = - \int_{-\infty}^\infty d\zeta_1
\int_{t_0}^{\zeta_1} d\tau_2 \int_{-\infty}^\infty d\zeta_2\: 
\bbra V(t, \zeta_1)\: V(\tau_2, \zeta_2) \kket\: \psi(t_0) + \cdots \notag \\
&= -\int_{-\infty}^\infty d\zeta_1 \int_{-\infty}^{\zeta_1} d\tau_2 \int_{-\infty}^\infty d\zeta_2 
\:\delta \Big( \frac{t+\zeta_1}{2} - \frac{\tau_2+\zeta_2}{2} \Big) \notag \\
&\qquad \times \sum_{k,a}
M_{k,a}( t, \zeta_1)\: M_{k,a}( \tau_2, \zeta_2)\: \psi(t_0) + \cdots \notag \\
&= -2 \int_{-\infty}^\infty d\zeta_1 \int_{-\infty}^{\zeta_1} d\tau_2 
\sum_{k,a} M_{k,a}(t,\zeta_1)\: M_{k,a}(\tau_2, t+\zeta_1 - \tau_2) \: \psi(t_0) + \cdots \:, \label{resum}
\end{align}
were we omitted the spinorial indices and understand products as matrix products, i.e.\
\[ \big(M_{k,a}(t_1, t_2)\, M_{k,a}(t_3, t_4) \big)^\alpha_\beta = (M_{k,a})^\alpha_\gamma(t_1,t_2)\, (M_{k,a})^\gamma_\beta (t_3,t_4) \]
with the Einstein summation convention. Introducing the new integration variables
\[ \zeta = \frac{1}{2} \:\big( \zeta_1 - t \big)\:,\qquad \nu = \tau_2 - t-\zeta \:, \]
we obtain
\[ \frac{d}{dt} \bbra \psi(t) \kket = -4 \int_{-\infty}^\infty d\zeta \int_{-\infty}^\zeta d\nu
\sum_{k,a} M_{k,a}(t, t+2 \zeta)\:  M_{k,a}(t+\zeta+\nu, t+\zeta-\nu)\: \psi(t_0) + \cdots \:. \]
{Note that the Gaussian pairing took place at time~$t+\zeta$,
\Claudio{More details?}%
as is indicated by the dashed line in Figure~\ref{figgauss1}.
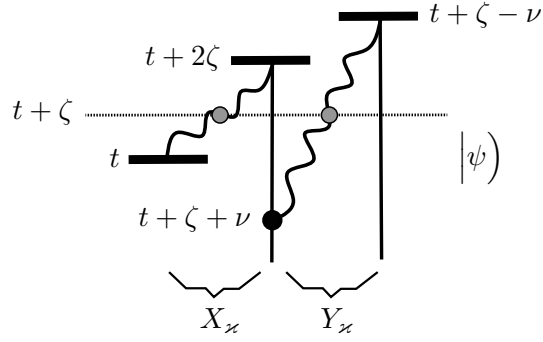
\begin{figure}
\psset{xunit=.5pt,yunit=.5pt,runit=.5pt}
\begin{pspicture}(286.30771907,239.81073815)
{
\newrgbcolor{curcolor}{0 0 0}
\pscustom[linewidth=6.71470867,linecolor=curcolor]
{
\newpath
\moveto(109.61665512,202.97440807)
\lineto(169.25640189,203.26176555)
}
}
{
\newrgbcolor{curcolor}{0 0 0}
\pscustom[linewidth=2.58420659,linecolor=curcolor]
{
\newpath
\moveto(140.72862614,203.11809059)
\lineto(140.32355528,51.00700146)
}
}
{
\newrgbcolor{curcolor}{1 1 1}
\pscustom[linestyle=none,fillstyle=solid,fillcolor=curcolor]
{
\newpath
\moveto(217.06636532,68.71007806)
\curveto(217.06636532,66.65598786)(214.44967097,64.99081874)(211.22181492,64.99081874)
\curveto(207.99395887,64.99081874)(205.37726452,66.65598786)(205.37726452,68.71007806)
\curveto(205.37726452,70.76416827)(207.99395887,72.42933739)(211.22181492,72.42933739)
\curveto(214.44967097,72.42933739)(217.06636532,70.76416827)(217.06636532,68.71007806)
\closepath
}
}
{
\newrgbcolor{curcolor}{1 1 1}
\pscustom[linewidth=6.71470867,linecolor=curcolor]
{
\newpath
\moveto(217.06636532,68.71007806)
\curveto(217.06636532,66.65598786)(214.44967097,64.99081874)(211.22181492,64.99081874)
\curveto(207.99395887,64.99081874)(205.37726452,66.65598786)(205.37726452,68.71007806)
\curveto(205.37726452,70.76416827)(207.99395887,72.42933739)(211.22181492,72.42933739)
\curveto(214.44967097,72.42933739)(217.06636532,70.76416827)(217.06636532,68.71007806)
\closepath
}
}
{
\newrgbcolor{curcolor}{1 1 1}
\pscustom[linestyle=none,fillstyle=solid,fillcolor=curcolor]
{
\newpath
\moveto(258.50962764,69.24140111)
\curveto(258.50962764,61.0250403)(247.92390229,54.36436382)(234.86574831,54.36436382)
\curveto(221.80759434,54.36436382)(211.22186899,61.0250403)(211.22186899,69.24140111)
\curveto(211.22186899,77.45776193)(221.80759434,84.1184384)(234.86574831,84.1184384)
\curveto(247.92390229,84.1184384)(258.50962764,77.45776193)(258.50962764,69.24140111)
\closepath
}
}
{
\newrgbcolor{curcolor}{1 1 1}
\pscustom[linewidth=6.71470867,linecolor=curcolor]
{
\newpath
\moveto(258.50962764,69.24140111)
\curveto(258.50962764,61.0250403)(247.92390229,54.36436382)(234.86574831,54.36436382)
\curveto(221.80759434,54.36436382)(211.22186899,61.0250403)(211.22186899,69.24140111)
\curveto(211.22186899,77.45776193)(221.80759434,84.1184384)(234.86574831,84.1184384)
\curveto(247.92390229,84.1184384)(258.50962764,77.45776193)(258.50962764,69.24140111)
\closepath
}
}
{
\newrgbcolor{curcolor}{1 1 1}
\pscustom[linestyle=none,fillstyle=solid,fillcolor=curcolor]
{
\newpath
\moveto(249.1403623,149.75472119)
\curveto(249.1403623,128.56511423)(235.20994248,111.38754379)(218.0258988,111.38754379)
\curveto(200.84185511,111.38754379)(186.91143529,128.56511423)(186.91143529,149.75472119)
\curveto(186.91143529,170.94432816)(200.84185511,188.1218986)(218.0258988,188.1218986)
\curveto(235.20994248,188.1218986)(249.1403623,170.94432816)(249.1403623,149.75472119)
\closepath
}
}
{
\newrgbcolor{curcolor}{1 1 1}
\pscustom[linewidth=4.4764723,linecolor=curcolor]
{
\newpath
\moveto(249.1403623,149.75472119)
\curveto(249.1403623,128.56511423)(235.20994248,111.38754379)(218.0258988,111.38754379)
\curveto(200.84185511,111.38754379)(186.91143529,128.56511423)(186.91143529,149.75472119)
\curveto(186.91143529,170.94432816)(200.84185511,188.1218986)(218.0258988,188.1218986)
\curveto(235.20994248,188.1218986)(249.1403623,170.94432816)(249.1403623,149.75472119)
\closepath
}
}
{
\newrgbcolor{curcolor}{0 0 0}
\pscustom[linewidth=4.4764723,linecolor=curcolor]
{
\newpath
\moveto(195.09514205,132.04290697)
\lineto(234.85497071,132.23449122)
}
}
{
\newrgbcolor{curcolor}{0 0 0}
\pscustom[linewidth=6.71470867,linecolor=curcolor]
{
\newpath
\moveto(32.50243276,128.43591862)
\lineto(92.14217953,128.7233139)
}
}
{
\newrgbcolor{curcolor}{1 1 1}
\pscustom[linestyle=none,fillstyle=solid,fillcolor=curcolor]
{
\newpath
\moveto(217.06636532,68.71007806)
\curveto(217.06636532,66.65598786)(214.44967097,64.99081874)(211.22181492,64.99081874)
\curveto(207.99395887,64.99081874)(205.37726452,66.65598786)(205.37726452,68.71007806)
\curveto(205.37726452,70.76416827)(207.99395887,72.42933739)(211.22181492,72.42933739)
\curveto(214.44967097,72.42933739)(217.06636532,70.76416827)(217.06636532,68.71007806)
\closepath
}
}
{
\newrgbcolor{curcolor}{1 1 1}
\pscustom[linewidth=6.71470867,linecolor=curcolor]
{
\newpath
\moveto(217.06636532,68.71007806)
\curveto(217.06636532,66.65598786)(214.44967097,64.99081874)(211.22181492,64.99081874)
\curveto(207.99395887,64.99081874)(205.37726452,66.65598786)(205.37726452,68.71007806)
\curveto(205.37726452,70.76416827)(207.99395887,72.42933739)(211.22181492,72.42933739)
\curveto(214.44967097,72.42933739)(217.06636532,70.76416827)(217.06636532,68.71007806)
\closepath
}
}
{
\newrgbcolor{curcolor}{1 1 1}
\pscustom[linestyle=none,fillstyle=solid,fillcolor=curcolor]
{
\newpath
\moveto(258.50962764,69.24140111)
\curveto(258.50962764,61.0250403)(247.92390229,54.36436382)(234.86574831,54.36436382)
\curveto(221.80759434,54.36436382)(211.22186899,61.0250403)(211.22186899,69.24140111)
\curveto(211.22186899,77.45776193)(221.80759434,84.1184384)(234.86574831,84.1184384)
\curveto(247.92390229,84.1184384)(258.50962764,77.45776193)(258.50962764,69.24140111)
\closepath
}
}
{
\newrgbcolor{curcolor}{1 1 1}
\pscustom[linewidth=6.71470867,linecolor=curcolor]
{
\newpath
\moveto(258.50962764,69.24140111)
\curveto(258.50962764,61.0250403)(247.92390229,54.36436382)(234.86574831,54.36436382)
\curveto(221.80759434,54.36436382)(211.22186899,61.0250403)(211.22186899,69.24140111)
\curveto(211.22186899,77.45776193)(221.80759434,84.1184384)(234.86574831,84.1184384)
\curveto(247.92390229,84.1184384)(258.50962764,77.45776193)(258.50962764,69.24140111)
\closepath
}
}
{
\newrgbcolor{curcolor}{1 1 1}
\pscustom[linestyle=none,fillstyle=solid,fillcolor=curcolor]
{
\newpath
\moveto(249.1403623,149.75472119)
\curveto(249.1403623,128.56511423)(235.20994248,111.38754379)(218.0258988,111.38754379)
\curveto(200.84185511,111.38754379)(186.91143529,128.56511423)(186.91143529,149.75472119)
\curveto(186.91143529,170.94432816)(200.84185511,188.1218986)(218.0258988,188.1218986)
\curveto(235.20994248,188.1218986)(249.1403623,170.94432816)(249.1403623,149.75472119)
\closepath
}
}
{
\newrgbcolor{curcolor}{1 1 1}
\pscustom[linewidth=4.4764723,linecolor=curcolor]
{
\newpath
\moveto(249.1403623,149.75472119)
\curveto(249.1403623,128.56511423)(235.20994248,111.38754379)(218.0258988,111.38754379)
\curveto(200.84185511,111.38754379)(186.91143529,128.56511423)(186.91143529,149.75472119)
\curveto(186.91143529,170.94432816)(200.84185511,188.1218986)(218.0258988,188.1218986)
\curveto(235.20994248,188.1218986)(249.1403623,170.94432816)(249.1403623,149.75472119)
\closepath
}
}
{
\newrgbcolor{curcolor}{0 0 0}
\pscustom[linewidth=6.71470867,linecolor=curcolor]
{
\newpath
\moveto(190.72149165,236.16604524)
\lineto(250.36123843,236.45342162)
}
}
{
\newrgbcolor{curcolor}{0 0 0}
\pscustom[linewidth=2.58420659,linecolor=curcolor]
{
\newpath
\moveto(221.83345134,236.30335169)
\lineto(222.9507326,53.21206854)
}
}
{
\newrgbcolor{curcolor}{0 0 0}
\pscustom[linewidth=2.69999651,linecolor=curcolor]
{
\newpath
\moveto(142.02072113,82.78870363)
\curveto(149.33732485,86.60887283)(156.65381631,90.428991)(156.57185197,97.33479333)
\curveto(156.49021417,104.24060587)(149.00952869,114.23188826)(152.34612661,118.91142791)
\curveto(155.68273474,123.59095736)(169.83616706,122.95839711)(171.80766879,128.1505915)
\curveto(173.77917052,133.34277569)(163.56875187,144.3596635)(165.15756624,148.0983888)
\curveto(166.74637039,151.83711409)(180.13433613,148.29738092)(183.25882885,153.31015585)
\curveto(186.38332157,158.32290017)(179.24434129,171.88808114)(182.19432302,177.66114886)
\curveto(185.14431496,183.43421658)(198.18267704,181.41486489)(200.36579414,185.37371629)
\curveto(202.54892145,189.3325779)(193.87625272,199.26945891)(194.81076057,203.80669368)
\curveto(195.74530923,208.34393865)(206.28696302,207.48153739)(212.50512,211.88699856)
\curveto(218.72327698,216.29246993)(220.61749833,225.96543636)(222.5117605,235.63860687)
}
}
{
\newrgbcolor{curcolor}{0 0 0}
\pscustom[linewidth=2.69999651,linecolor=curcolor]
{
\newpath
\moveto(61.33443477,129.68848769)
\curveto(62.53681663,138.62116259)(63.73918828,147.55380687)(69.25233146,148.81547818)
\curveto(74.76548485,150.07714949)(84.58939956,143.66782741)(88.11147855,146.55836662)
\curveto(91.63355754,149.44890584)(88.85387225,161.63904102)(92.97193776,163.91981733)
\curveto(97.09002369,166.20059365)(108.10536038,158.57199069)(112.38459931,160.78224215)
\curveto(116.66382803,162.99249362)(114.20642835,175.04129336)(116.52955427,178.90651642)
\curveto(118.85266998,182.77173947)(125.95603578,178.45281437)(130.90094929,180.92352102)
\curveto(135.84587301,183.39418684)(138.63179339,192.65427012)(141.41777499,201.91453708)
}
}
{
\newrgbcolor{curcolor}{0.00784314 0.00784314 0.00784314}
\pscustom[linestyle=none,fillstyle=solid,fillcolor=curcolor]
{
\newpath
\moveto(146.75098942,82.78871153)
\curveto(146.75098942,79.45423012)(144.04785856,76.75109885)(140.71337765,76.75109885)
\curveto(137.37889674,76.75109885)(134.67576588,79.45423012)(134.67576588,82.78871153)
\curveto(134.67576588,86.12319293)(137.37889674,88.8263242)(140.71337765,88.8263242)
\curveto(144.04785856,88.8263242)(146.75098942,86.12319293)(146.75098942,82.78871153)
\closepath
}
}
{
\newrgbcolor{curcolor}{0 0 0}
\pscustom[linewidth=3.23186649,linecolor=curcolor,strokeopacity=0.02654868]
{
\newpath
\moveto(146.75098942,82.78871153)
\curveto(146.75098942,79.45423012)(144.04785856,76.75109885)(140.71337765,76.75109885)
\curveto(137.37889674,76.75109885)(134.67576588,79.45423012)(134.67576588,82.78871153)
\curveto(134.67576588,86.12319293)(137.37889674,88.8263242)(140.71337765,88.8263242)
\curveto(144.04785856,88.8263242)(146.75098942,86.12319293)(146.75098942,82.78871153)
\closepath
}
}
{
\newrgbcolor{curcolor}{0 0 0}
\pscustom[linewidth=1.68377948,linecolor=curcolor]
{
\newpath
\moveto(131.84683465,43.72610823)
\lineto(115.64985827,31.15039547)
\lineto(104.12439307,31.15039547)
\lineto(96.7151622,24.97603831)
\lineto(90.52962898,31.56201248)
\lineto(75.72233575,31.56201248)
\lineto(62.96200441,43.72610823)
}
}
{
\newrgbcolor{curcolor}{0 0 0}
\pscustom[linewidth=2.06929139,linecolor=curcolor,linestyle=dashed,dash=1.09500003 1.09500003]
{
\newpath
\moveto(0.00072945,161.88737563)
\lineto(272.40868913,162.07793941)
}
}
{
\newrgbcolor{curcolor}{0 0 0}
\pscustom[linewidth=1.68377948,linecolor=curcolor]
{
\newpath
\moveto(220.02506457,43.72610823)
\lineto(203.82808819,31.15039547)
\lineto(192.30262299,31.15039547)
\lineto(184.89339213,24.97603831)
\lineto(178.7078589,31.56201248)
\lineto(163.90056567,31.56201248)
\lineto(151.14023433,43.72610823)
}
}
{
\newrgbcolor{curcolor}{0.58823532 0.58823532 0.58823532}
\pscustom[linestyle=none,fillstyle=solid,fillcolor=curcolor]
{
\newpath
\moveto(107.39112807,161.98146632)
\curveto(107.39112807,158.64698492)(104.6879972,155.94385365)(101.35351629,155.94385365)
\curveto(98.01903539,155.94385365)(95.31590452,158.64698492)(95.31590452,161.98146632)
\curveto(95.31590452,165.31594772)(98.01903539,168.01907899)(101.35351629,168.01907899)
\curveto(104.6879972,168.01907899)(107.39112807,165.31594772)(107.39112807,161.98146632)
\closepath
}
}
{
\newrgbcolor{curcolor}{0 0 0}
\pscustom[linewidth=1.13385831,linecolor=curcolor,strokeopacity=0.998285]
{
\newpath
\moveto(107.39112807,161.98146632)
\curveto(107.39112807,158.64698492)(104.6879972,155.94385365)(101.35351629,155.94385365)
\curveto(98.01903539,155.94385365)(95.31590452,158.64698492)(95.31590452,161.98146632)
\curveto(95.31590452,165.31594772)(98.01903539,168.01907899)(101.35351629,168.01907899)
\curveto(104.6879972,168.01907899)(107.39112807,165.31594772)(107.39112807,161.98146632)
\closepath
}
}
{
\newrgbcolor{curcolor}{0.58823532 0.58823532 0.58823532}
\pscustom[linestyle=none,fillstyle=solid,fillcolor=curcolor]
{
\newpath
\moveto(189.69158636,161.98146632)
\curveto(189.69158636,158.64698492)(186.98845549,155.94385365)(183.65397458,155.94385365)
\curveto(180.31949368,155.94385365)(177.61636281,158.64698492)(177.61636281,161.98146632)
\curveto(177.61636281,165.31594772)(180.31949368,168.01907899)(183.65397458,168.01907899)
\curveto(186.98845549,168.01907899)(189.69158636,165.31594772)(189.69158636,161.98146632)
\closepath
}
}
{
\newrgbcolor{curcolor}{0 0 0}
\pscustom[linewidth=1.13385831,linecolor=curcolor,strokeopacity=0.998285]
{
\newpath
\moveto(189.69158636,161.98146632)
\curveto(189.69158636,158.64698492)(186.98845549,155.94385365)(183.65397458,155.94385365)
\curveto(180.31949368,155.94385365)(177.61636281,158.64698492)(177.61636281,161.98146632)
\curveto(177.61636281,165.31594772)(180.31949368,168.01907899)(183.65397458,168.01907899)
\curveto(186.98845549,168.01907899)(189.69158636,165.31594772)(189.69158636,161.98146632)
\closepath
}
\rput[bl](280,110){$\Big| \psi \Big)$}
\rput[bl](18,123){$t$}
\rput[bl](-55,155){$t+\zeta$}
\rput[bl](45,195){$t+2 \zeta$}
\rput[bl](40,74){$t+\zeta+\nu$}
\rput[bl](259,228){$t+\zeta-\nu$}
\rput[bl](86,-2){$X_\varkappa$}
\rput[bl](177,-2){$Y_\varkappa$}
}
\end{pspicture}
\caption{Gaussian pairing in ket.}
\label{figgauss1}
\end{figure}%
The fact that~$\zeta$ is non-zero reflects the nonlocality of the Dirac equation in time.
It is useful to combine the sums over~$k$ and~$a$ with the integration over~$\zeta$.
Introducing the short notation~\eqref{kappadef}, we obtain the short formula
\[ \frac{d}{dt} \bbra \psi(t) \kket = -\sint_\varkappa X_\varkappa(t)\, Y_\varkappa(t) \, \psi(t_0) + \cdots \]
with spatial operators~$X_\varkappa(t)$ and~$Y_\varkappa(t)$ given by
\begin{align}
X_\varkappa(t) &:= 2 M_{k,a}(t, t+2 \zeta) \label{Xdef} \\
Y_\varkappa(t) &:= 2 \int_{-\infty}^\zeta M_{k,a}(t+ \zeta+\nu, t+\zeta-\nu) \: d\nu \:. \label{Ydef}
\end{align}
These operators are also shown in Figure~\ref{figgauss1}.

Clearly, proceeding to higher order in perturbation theory, we have more factors~$V$,
and taking the statistical means gives the sum of all Gaussian pairings of these potentials.
But the situation is simplified considerably by noting
that, to leading order in~$\ell_{\min} \|V\|$, it suffices to take into account 
all adjacent pairings, meaning that each pairing connects two neighboring factors~$V$
(an example of non-adjacent pairings is shown in
Figure~\ref{figfourth} on page~\pageref{figfourth}). In order not to distract from the main construction steps,
the precise scaling argument will be given in Section~\ref{secerror} below. Here we simply use this
result, which means that, to every order in perturbation theory, the first factor~$V$ is paired with
the second factor~$V$, exactly as shown in Figure~\ref{figgauss1}.
All the remaining factors~$V$ of the Dyson series give precisely~$\psi(t)$.
Therefore, the dynamical equation for the statistical mean of~$\psi(t)$ is obtained
simply by replacing~$\psi(t_0)$ in~\eqref{resum} by the statistical mean of~$\psi(t)$. We thus obtain
\[ \frac{d}{dt} \bbra \psi(t) \kket = -\sint_\varkappa X_\varkappa(t)\, Y_\varkappa(t) \, \bbra \psi(t) \kket \:, \]
which holds up to the error term specified in Theorem~\ref{thmlindblad}.
For notational simplicity, we shall omit these error terms in the following computations,
noting that they will be treated systematically in Section~\ref{secerror}.

\subsubsection{Computation of~$\bbra |\partial_t \psi )( \psi| \kket$} \label{sectpsipsi}
Clearly, we do not want to compute the statistical mean of~$\psi$, but the statistical mean of
a ket/bra combination like~\eqref{rhotildedef}. In preparation, we now want to compute
the statistical mean of~$|\partial_t \psi)(\psi|$.
The general method is to express both the ket~$|\psi(t))$ and the bra~$(\psi(t)|$ as a nonlocal Dyson series.
Taking the statistical mean, we again obtain Gaussian parings, but now also between
a factor~$V$ contained in bra and another factor contained in ket.
Similar as explained after~\eqref{Ydef}, to leading order in~$\ell_{\min} \|V\|$ it suffices to take into account 
all adjacent pairings. But now we must consider the Gaussian paring between the first factor~$V$ in bra
and the first factor in ket. More precisely, we obtain
\begin{align}
&\bbra \Big( \frac{d}{dt} |\psi(t)) \Big)  ( \psi(t) | \kket + 
\sint_\varkappa X_\varkappa(t)\, Y_\varkappa(t) \, \bbra \big| \psi(t) \big) \big( \psi(t) \big| \kket \notag \\
&= -i \int_{-\infty}^\infty d\zeta_1 \: \bbra \big| V(t, \zeta_1)\, \psi(t) \big) \big( \psi(t) \big| \kket
+ \sint_\varkappa X_\varkappa(t)\, Y_\varkappa(t) \, \bbra \big| \psi(t) \big) \big( \psi(t) \big| \kket \notag \\
&\overset{(*)}{=} \int_{-\infty}^\infty d\zeta_1 \int_{-\infty}^t d\tau_1 \int_{-\infty}^\infty d\zeta_2\:
\bbra \big| V(t, \zeta_1)\, \psi(t) \big) \big( \psi(t) \big| V(\tau_1, \zeta_2)^\dagger \kket \notag \\
&= \sum_{k,a} \int_{-\infty}^\infty d\zeta_1 \int_{-\infty}^t d\tau_1 \int_{-\infty}^\infty d\zeta_2 \notag \\
&\qquad \times \delta \Big( \frac{t+\zeta_1}{2} - \frac{\tau_1+\zeta_2}{2} \Big)\:
M_{k,a} (t, \zeta_1)\:|\psi)(\psi|\: M_{k,a} (\tau_1, \zeta_2)^\dagger \notag \\
&= 2 \sum_{k,a} \int_{-\infty}^\infty d\zeta_1 \int_{-\infty}^t d\tau_1\:
M_{k,a} (t, \zeta_1)\:|\psi)(\psi|\: M_{k,a} (\tau_1, t+\zeta_1-\tau_1)^\dagger \:. \label{b1}
\end{align}
We note for clarity that in~$(*)$, the pairing of~$V(t, \zeta_1)$ with the ket~$|\psi(t))$
as already computed in~\eqref{resum} cancels with the $\varkappa$-sum.
Therefore, it remains to take into account the pairing of~$V(t, \zeta_1)$ with the bra~$(\psi(t)|$.

Introducing the integration variables
\[ \zeta = \frac{1}{2}\:(\zeta_1-t) \:,\qquad \nu = \tau_1 - t - \zeta\:, \]
we obtain
\begin{align*}
\eqref{b1} &= 4 \sint_\varkappa \int_{-\infty}^{-\zeta} d\nu\: 
M_{k,a} (t, t+2 \zeta)\:|\psi)(\psi|\: M_{k,a} (t+\zeta+\nu, t+\zeta-\nu)^\dagger \\
&= \sint_\varkappa X_\varkappa \:|\psi)(\psi|\: Z_\varkappa^\dagger
\end{align*}
with
\beq \label{Zdef}
Z_\varkappa := 2 \int_{-\infty}^{-\zeta} M_{k,a} (t+\zeta+\nu, t+\zeta-\nu) \: d\nu\:.
\eeq
This is illustrated in Figure~\ref{figgauss2}.

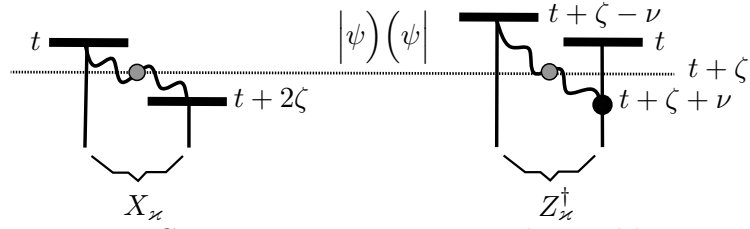
\begin{figure}
\psset{xunit=.5pt,yunit=.5pt,runit=.5pt}
\begin{pspicture}(499.57215797,127.87734144)
{
\newrgbcolor{curcolor}{0 0 0}
\pscustom[linewidth=2.69999651,linecolor=curcolor]
{
\newpath
\moveto(135.52357984,59.1205514)
\curveto(134.4830143,59.22882352)(133.44244876,59.3369936)(132.70097348,63.2880588)
\curveto(131.9594982,67.23908318)(131.5171234,75.03290062)(127.68861694,75.27921206)
\curveto(123.86012069,75.5255541)(116.64567647,68.22448198)(112.9923035,71.1582484)
\curveto(109.33893052,74.09201483)(109.24695534,87.26046673)(104.14748107,87.42764053)
\curveto(99.0480068,87.59479392)(88.94107427,74.76071001)(84.19393852,75.31646951)
\curveto(79.44681298,75.8722188)(80.05981077,89.81736248)(77.15119899,92.72607631)
\curveto(74.24258721,95.63479014)(67.81274343,87.50667612)(63.68812649,87.69166736)
\curveto(59.56349934,87.87667902)(57.74458885,96.37490818)(55.92563754,104.87333124)
}
}
{
\newrgbcolor{curcolor}{0 0 0}
\pscustom[linewidth=6.71470867,linecolor=curcolor]
{
\newpath
\moveto(397.31140157,124.23266743)
\lineto(337.67166614,124.52002491)
}
}
{
\newrgbcolor{curcolor}{0 0 0}
\pscustom[linewidth=2.58420659,linecolor=curcolor]
{
\newpath
\moveto(366.19942677,124.37634995)
\lineto(365.85586772,26.5658731)
}
}
{
\newrgbcolor{curcolor}{0 0 0}
\pscustom[linewidth=6.71470867,linecolor=curcolor]
{
\newpath
\moveto(475.8088252,105.4842227)
\lineto(416.16908976,105.77161798)
}
}
{
\newrgbcolor{curcolor}{0 0 0}
\pscustom[linewidth=6.71470867,linecolor=curcolor]
{
\newpath
\moveto(88.58757921,105.11851184)
\lineto(28.94784378,105.40590711)
}
}
{
\newrgbcolor{curcolor}{0 0 0}
\pscustom[linewidth=2.58420659,linecolor=curcolor]
{
\newpath
\moveto(57.52356661,105.95135341)
\lineto(55.89746268,26.21863278)
}
}
{
\newrgbcolor{curcolor}{0 0 0}
\pscustom[linewidth=2.69999651,linecolor=curcolor]
{
\newpath
\moveto(445.59358791,50.94674705)
\curveto(444.39120605,59.87942195)(443.18883439,68.81206623)(437.67569121,70.07373754)
\curveto(432.16253783,71.33540884)(422.33862312,64.92608677)(418.81654413,67.81662598)
\curveto(415.29446513,70.70716519)(418.07415043,82.89730037)(413.95608491,85.17807669)
\curveto(409.83799899,87.458853)(398.8226623,79.83025004)(394.54342337,82.04050151)
\curveto(390.26419465,84.25075297)(392.72159433,96.29955272)(390.39846841,100.16477577)
\curveto(388.07535269,104.02999883)(380.9719869,99.71107373)(376.02707339,102.18178037)
\curveto(371.08214967,104.6524462)(368.29622929,113.91252947)(365.51024769,123.17279644)
}
}
{
\newrgbcolor{curcolor}{0 0 0}
\pscustom[linewidth=1.68377948,linecolor=curcolor]
{
\newpath
\moveto(371.09981102,20.27045766)
\lineto(387.2967874,7.69474491)
\lineto(398.82226772,7.69474491)
\lineto(406.23150236,1.52038774)
\lineto(412.41703937,8.10636192)
\lineto(427.22432126,8.10636192)
\lineto(439.98464882,20.27045766)
}
}
{
\newrgbcolor{curcolor}{0 0 0}
\pscustom[linewidth=2.06929139,linecolor=curcolor,linestyle=dashed,dash=1.09500003 1.09500003]
{
\newpath
\moveto(0.00327307,83.0458668)
\lineto(499.56890079,81.46683688)
}
}
{
\newrgbcolor{curcolor}{0 0 0}
\pscustom[linewidth=1.68377948,linecolor=curcolor]
{
\newpath
\moveto(60.81016819,19.9047468)
\lineto(77.00714457,7.32903404)
\lineto(88.53258709,7.32903404)
\lineto(95.94182173,1.15467688)
\lineto(102.12735874,7.74065105)
\lineto(116.93460283,7.74065105)
\lineto(129.6948926,19.9047468)
}
}
{
\newrgbcolor{curcolor}{0.58823532 0.58823532 0.58823532}
\pscustom[linestyle=none,fillstyle=solid,fillcolor=curcolor]
{
\newpath
\moveto(399.53694178,83.23970671)
\curveto(399.53694178,79.9052253)(402.24007265,77.20209403)(405.57455356,77.20209403)
\curveto(408.90903446,77.20209403)(411.61216533,79.9052253)(411.61216533,83.23970671)
\curveto(411.61216533,86.57418811)(408.90903446,89.27731938)(405.57455356,89.27731938)
\curveto(402.24007265,89.27731938)(399.53694178,86.57418811)(399.53694178,83.23970671)
\closepath
}
}
{
\newrgbcolor{curcolor}{0 0 0}
\pscustom[linewidth=1.13385831,linecolor=curcolor,strokeopacity=0.998285]
{
\newpath
\moveto(399.53694178,83.23970671)
\curveto(399.53694178,79.9052253)(402.24007265,77.20209403)(405.57455356,77.20209403)
\curveto(408.90903446,77.20209403)(411.61216533,79.9052253)(411.61216533,83.23970671)
\curveto(411.61216533,86.57418811)(408.90903446,89.27731938)(405.57455356,89.27731938)
\curveto(402.24007265,89.27731938)(399.53694178,86.57418811)(399.53694178,83.23970671)
\closepath
}
}
{
\newrgbcolor{curcolor}{0.58823532 0.58823532 0.58823532}
\pscustom[linestyle=none,fillstyle=solid,fillcolor=curcolor]
{
\newpath
\moveto(89.09681583,82.82671016)
\curveto(89.09681583,79.49222876)(91.7999467,76.78909749)(95.1344276,76.78909749)
\curveto(98.46890851,76.78909749)(101.17203938,79.49222876)(101.17203938,82.82671016)
\curveto(101.17203938,86.16119156)(98.46890851,88.86432283)(95.1344276,88.86432283)
\curveto(91.7999467,88.86432283)(89.09681583,86.16119156)(89.09681583,82.82671016)
\closepath
}
}
{
\newrgbcolor{curcolor}{0 0 0}
\pscustom[linewidth=1.13385831,linecolor=curcolor,strokeopacity=0.998285]
{
\newpath
\moveto(89.09681583,82.82671016)
\curveto(89.09681583,79.49222876)(91.7999467,76.78909749)(95.1344276,76.78909749)
\curveto(98.46890851,76.78909749)(101.17203938,79.49222876)(101.17203938,82.82671016)
\curveto(101.17203938,86.16119156)(98.46890851,88.86432283)(95.1344276,88.86432283)
\curveto(91.7999467,88.86432283)(89.09681583,86.16119156)(89.09681583,82.82671016)
\closepath
}
}
{
\newrgbcolor{curcolor}{0 0 0}
\pscustom[linewidth=2.58420659,linecolor=curcolor]
{
\newpath
\moveto(445.2343748,105.3858794)
\lineto(445.51935118,26.57054837)
}
}
{
\newrgbcolor{curcolor}{0.00784314 0.00784314 0.00784314}
\pscustom[linestyle=none,fillstyle=solid,fillcolor=curcolor]
{
\newpath
\moveto(439.48190441,58.42017853)
\curveto(439.48190441,55.08569713)(442.18503528,52.38256586)(445.51951618,52.38256586)
\curveto(448.85399709,52.38256586)(451.55712796,55.08569713)(451.55712796,58.42017853)
\curveto(451.55712796,61.75465994)(448.85399709,64.45779121)(445.51951618,64.45779121)
\curveto(442.18503528,64.45779121)(439.48190441,61.75465994)(439.48190441,58.42017853)
\closepath
}
}
{
\newrgbcolor{curcolor}{0 0 0}
\pscustom[linewidth=3.23186649,linecolor=curcolor,strokeopacity=0.02654868]
{
\newpath
\moveto(439.48190441,58.42017853)
\curveto(439.48190441,55.08569713)(442.18503528,52.38256586)(445.51951618,52.38256586)
\curveto(448.85399709,52.38256586)(451.55712796,55.08569713)(451.55712796,58.42017853)
\curveto(451.55712796,61.75465994)(448.85399709,64.45779121)(445.51951618,64.45779121)
\curveto(442.18503528,64.45779121)(439.48190441,61.75465994)(439.48190441,58.42017853)
\closepath
}
}
{
\newrgbcolor{curcolor}{0 0 0}
\pscustom[linewidth=2.58420659,linecolor=curcolor]
{
\newpath
\moveto(134.21577449,60.92082995)
\lineto(134.20065638,26.21915436)
}
}
{
\newrgbcolor{curcolor}{0 0 0}
\pscustom[linewidth=6.71470867,linecolor=curcolor]
{
\newpath
\moveto(162.74357669,60.77943404)
\lineto(103.10383748,61.06682932)
}
\rput[bl](245,90){$\Big| \psi\Big) \Big( \psi \Big|$}
\rput[bl](15,100){$t$}
\rput[bl](510,75){$t+\zeta$}
\rput[bl](485,100){$t$}
\rput[bl](168,53){$t+2 \zeta$}
\rput[bl](405,117){$t+ \zeta - \nu$}
\rput[bl](458,52){$t+\zeta+\nu$}
\rput[bl](397,-28){$Z_\varkappa^\dagger$}
\rput[bl](85,-28){$X_\varkappa$}
}
\end{pspicture}
\caption{Gaussian pairing connecting bra and ket.}
\label{figgauss2}
\end{figure}%

\subsubsection{Computation of~$\partial_t \sigma_t$}
In order to complete the computation of~$\partial_t \sigma_t$, we must analyze the
transformed wave functions~$\tilde{\psi}$ defined in~\eqref{tildepsi}.
We now outline step by step how this can be done, referring for the detailed computations to
Appendix~\ref{applindblad}.
The first step is to compute the operator~$\Sig_t$ defined in~\eqref{Sigdef}. First, from~\eqref{scalcomp}
\[ \big( \psi \,|\, \Sig_t\, \phi \big)_t =
i \bigg( \int_{-\infty}^t \!\!\!\!\!d\tau \int_t^\infty \!\!\!\!\!d\tau' \!-\! \int_t^\infty \!\!\!\!\!d\tau \int_{-\infty}^t \!\!\!\!\!d\tau' \bigg) \: \big( \psi(\tau) \,|\, V(\tau,\tau') \:\psi(\tau') \big)_\tau \:. \]
Employing the transformation
\begin{align*}
\big( \psi(\tau) \,|\, V(\tau,\tau') \:\psi(\tau') \big)_\tau
&= \big( U^\tau_t \,\psi(t) \,|\, V(\tau,\tau') \:U^{\tau'}_t\, \psi(t) \big)_\tau \\
&= \big( \psi(t) \,|\, (U^\tau_t)^\dagger \,V(\tau,\tau') \:U^{\tau'}_t\, \psi(t) \big)_\tau \:,
\end{align*}
we conclude that
\beq \label{Sform}
\Sig_t = i \bigg( \int_{-\infty}^t \!\!\!\!\!d\tau \int_t^\infty \!\!\!\!\!d\tau' \!-\! \int_t^\infty \!\!\!\!\!d\tau \int_{-\infty}^t \!\!\!\!\!d\tau' \bigg) \: (U^\tau_t)^\dagger \,V(\tau,\tau') \:U^{\tau'}_t \:.
\eeq
The next step is to compute~$\bbra \partial_t \tilde{\psi} \kket$. To this end, we begin with the formula
\begin{align*}
\frac{d \tilde{\psi}}{dt}  &= \frac{d}{dt} \Big( \sqrt{\1 + \Sig_t}\: \psi(t) \Big)
= \frac{d}{dt} \Big( \psi(t) + \frac{\Sig_t}{2}\: \psi(t) - \frac{\Sig_t^2}{8} \: \psi(t) \Big) \:,
\end{align*}
again up to the error terms.
Now the time derivative can be computed with the product rule. To first order,
\begin{align*}
\frac{d}{dt} \psi(t) &= 
\int_{-\infty}^\infty d\zeta\: \big(-i V(t, \zeta) \big)\: \psi(t) \\
\frac{d}{dt} \Sig_t \psi(t) &= i \int_{-\infty}^\infty V(t,\tau') \:\psi(\tau')\: d\tau' - i \int_{-\infty}^\infty \,V(\tau,t)
\:\psi(t)\: d\tau \\
\frac{d}{dt} \tilde{\psi}(t) &= \frac{d}{dt} \tilde{\psi}(t) + \frac{1}{2}\: \frac{d}{dt}  \Sig_t \psi(t)
= -\frac{i}{2} \int_{-\infty}^\infty \big( V(t, \zeta) + V(\zeta,t) \big)\:d\zeta\; \psi(t) \:.
\end{align*}
The higher orders can be computed similarly, as is worked out systematically in Appendix~\ref{applindblad}.
Taking the statistical mean, the Gaussian parings can be computed similar as explained in Section~\ref{sectpsi}.
Finally, in order to compute the statistical mean of~$|\partial_t \tilde{\psi} )( \tilde{\psi}|$,
similar as explained in Section~\ref{sectpsipsi}, one must
take into account Gaussian parings between bra and ket.

Collecting all the terms, one ends up with
the following deterministic evolution equation for the statistical operator~\eqref{rhotildedef},
\[ \frac{d \sigma_t}{dt} = \sint_\varkappa \Big( C_\varkappa \,\sigma_t +  \sigma_t\, C_\varkappa^\dagger
+ A_\varkappa \,\sigma_t\, B_\varkappa + B_\varkappa \,\sigma_t\, A_\varkappa \Big) \:, \]
where~$A_\varkappa$, $B_\varkappa$ and~$C_\varkappa$ are operators involving~$M_{k,a}$.
Before writing this equation in more detail, it is useful to simplify it by
combining parts of it with the Hamiltonian. Indeed, transforming the Hamiltonian according to
\[ H_0 \rightarrow H_0 - \frac{i}{2} \sint_\varkappa \big( C_\varkappa - C_\varkappa^\dagger \big) \:, \]
we can interpret the last summand as a symmetric deterministic potential in the Dirac equation.
A direct computation yields (for details see Appendix~\ref{applindblad})
\begin{align*}
\frac{1}{2}\:\big( C_\varkappa - C_\varkappa^\dagger \big)
&= \frac{1}{4}\: \big[ X_\varkappa, Z_\varkappa \big] + \frac{1}{4}\: \big[ X_\varkappa^\dagger, Z_\varkappa^\dagger \big] 
-\frac{1}{2}\: X_\varkappa\, Y_\varkappa + \frac{1}{2} Y_\varkappa^\dagger\, X_\varkappa^\dagger \:,
\end{align*}
with~$X_\varkappa$, $Y_\varkappa$ and~$Z_\varkappa$ given by~\eqref{Xdef}, \eqref{Ydef} and~\eqref{Zdef}.
The transformed Hamiltonian can be written in the form~\eqref{Hn1}--\eqref{Hn4}.
Modifying the interaction picture by including this potential, the dynamical equation for~$\sigma_t$ changes to
\[ \frac{d \sigma_t}{dt} = \sint_\varkappa \Big( \frac{1}{2} \,\big( C_\varkappa + C_\varkappa^\dagger \big) \,\sigma_t
+ \frac{1}{2} \,\sigma_t\, \big(C_\varkappa + C_\varkappa^\dagger \big)
+ A_\varkappa \,\sigma_t\, B_\varkappa + B_\varkappa \,\sigma_t\, A_\varkappa \Big) \:. \]
A direct computation shows that the operator~$C_\varkappa + C_\varkappa^\dagger$ can be
expressed in terms of~$A_\varkappa$ and~$B_\varkappa$ (for details see Appendix~\ref{applindblad}).
We thus obtain
\begin{align*}
\frac{d \sigma_t}{dt} &= \sint_\varkappa \Big( -\frac{1}{2} \,\big( A_\varkappa B_\varkappa+ B_\varkappa A_\varkappa \big) \,\sigma_t
- \frac{1}{2} \,\sigma_t\, \big(A_\varkappa B_\varkappa+ B_\varkappa A_\varkappa \big)
+ A_\varkappa \,\sigma_t\, B_\varkappa + B_\varkappa \,\sigma_t\, A_\varkappa \Big)
\end{align*}
with operators~$A_\varkappa$ and~$B_\varkappa$ given by (see also~\eqref{Xdef} and~\eqref{Zdef})
\begin{align*}
A_\varkappa(t) &= \frac{1}{2} \:\big( X_\varkappa(t) + X_\varkappa(t)^* \big)
= M_{k,a}(t, t+2 \zeta) + M_{k,a}(t+2 \zeta, t) \\ 
B_\varkappa(t) &= \frac{1}{2} \:\big( Z_\varkappa(t) + Z_\varkappa(t)^* \big) \notag \\
&=\int_{-\infty}^{-\zeta} \Big( M_{k,a}(t+\zeta+\nu, t+\zeta-\nu)
+ M_{k,a}(t+\zeta-\nu, t+\zeta+\nu) \Big) \:d\nu \:. 
\end{align*}
This can be simplified further with the help of~\eqref{Deltadef}. Indeed,
\[ A_\varkappa = a(\zeta)\: N_{k,a} \qquad \text{and} \qquad
B_\varkappa = b(\zeta)\: N_{k,a} \]
with the functions~$a(\zeta)$ and~$b(\zeta)$ as in~\eqref{adef} and~\eqref{bdef}.
Note that these functions are non-negative in view of~\eqref{Deltapos}.
We conclude that
\[ \frac{d \sigma_t}{dt} = -\sint_\varkappa a(\zeta)\, b(\zeta)\:
\big[ N_{k,a}, [N_{k,a}, \sigma_t] \big] \:. \]
Rescaling the operators~$N_{k,a}$ according to~\eqref{Kkappa} gives
the desired Kossakowski-Lindblad equation.

\subsubsection{Scaling of the Error Terms} \label{secerror}
In order to complete the proof of Theorem~\ref{thmlindblad}, it remains to analyze the error terms.
As already mentioned after~\eqref{Ydef}, when taking statistical means we must sum over all
possible Gaussian pairings. In the previous calculations we took into account only adjacent pairings.
A typical example of a non-adjacent pairing in the nonlocal Dyson series is shown in Figure~\ref{figfourth}.
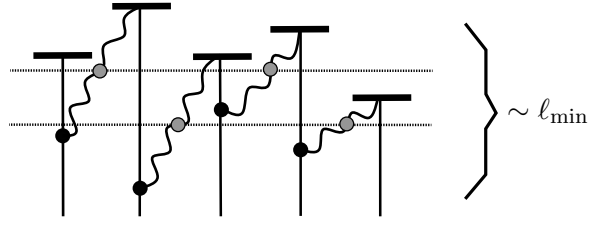
\begin{figure}
\psset{xunit=.5pt,yunit=.5pt,runit=.5pt}
\begin{pspicture}(368.05741918,182.65686035)
{
\newrgbcolor{curcolor}{0 0 0}
\pscustom[linewidth=4.97385821,linecolor=curcolor]
{
\newpath
\moveto(18.3551811,142.85740712)
\lineto(62.53277102,143.07027012)
}
}
{
\newrgbcolor{curcolor}{0 0 0}
\pscustom[linewidth=1.99999742,linecolor=curcolor]
{
\newpath
\moveto(40.44397228,142.96383862)
\lineto(40.71927307,21.88548649)
}
}
{
\newrgbcolor{curcolor}{0 0 0}
\pscustom[linewidth=4.97385821,linecolor=curcolor]
{
\newpath
\moveto(77.70792189,179.95709846)
\lineto(121.88551181,180.16996145)
}
}
{
\newrgbcolor{curcolor}{0 0 0}
\pscustom[linewidth=1.91422485,linecolor=curcolor]
{
\newpath
\moveto(98.5227326,180.0623583)
\lineto(98.40050268,21.88321122)
}
}
{
\newrgbcolor{curcolor}{1 1 1}
\pscustom[linestyle=none,fillstyle=solid,fillcolor=curcolor]
{
\newpath
\moveto(155.06923169,50.89635597)
\curveto(155.06923169,49.37480764)(153.13093971,48.14134901)(150.73993537,48.14134901)
\curveto(148.34893103,48.14134901)(146.41063905,49.37480764)(146.41063905,50.89635597)
\curveto(146.41063905,52.41790429)(148.34893103,53.65136292)(150.73993537,53.65136292)
\curveto(153.13093971,53.65136292)(155.06923169,52.41790429)(155.06923169,50.89635597)
\closepath
}
}
{
\newrgbcolor{curcolor}{1 1 1}
\pscustom[linewidth=4.97385821,linecolor=curcolor]
{
\newpath
\moveto(155.06923169,50.89635597)
\curveto(155.06923169,49.37480764)(153.13093971,48.14134901)(150.73993537,48.14134901)
\curveto(148.34893103,48.14134901)(146.41063905,49.37480764)(146.41063905,50.89635597)
\curveto(146.41063905,52.41790429)(148.34893103,53.65136292)(150.73993537,53.65136292)
\curveto(153.13093971,53.65136292)(155.06923169,52.41790429)(155.06923169,50.89635597)
\closepath
}
}
{
\newrgbcolor{curcolor}{1 1 1}
\pscustom[linestyle=none,fillstyle=solid,fillcolor=curcolor]
{
\newpath
\moveto(185.76791422,51.2899318)
\curveto(185.76791422,45.20373849)(177.92663648,40.26990398)(168.2539302,40.26990398)
\curveto(158.58122392,40.26990398)(150.73994618,45.20373849)(150.73994618,51.2899318)
\curveto(150.73994618,57.37612511)(158.58122392,62.30995962)(168.2539302,62.30995962)
\curveto(177.92663648,62.30995962)(185.76791422,57.37612511)(185.76791422,51.2899318)
\closepath
}
}
{
\newrgbcolor{curcolor}{1 1 1}
\pscustom[linewidth=4.97385821,linecolor=curcolor]
{
\newpath
\moveto(185.76791422,51.2899318)
\curveto(185.76791422,45.20373849)(177.92663648,40.26990398)(168.2539302,40.26990398)
\curveto(158.58122392,40.26990398)(150.73994618,45.20373849)(150.73994618,51.2899318)
\curveto(150.73994618,57.37612511)(158.58122392,62.30995962)(168.2539302,62.30995962)
\curveto(177.92663648,62.30995962)(185.76791422,57.37612511)(185.76791422,51.2899318)
\closepath
}
}
{
\newrgbcolor{curcolor}{0.00784314 0.00784314 0.00784314}
\pscustom[linestyle=none,fillstyle=solid,fillcolor=curcolor]
{
\newpath
\moveto(44.91628967,82.48587003)
\curveto(44.91628967,80.01588387)(42.91397046,78.01356446)(40.44398455,78.01356446)
\curveto(37.97399864,78.01356446)(35.97167943,80.01588387)(35.97167943,82.48587003)
\curveto(35.97167943,84.95585619)(37.97399864,86.9581756)(40.44398455,86.9581756)
\curveto(42.91397046,86.9581756)(44.91628967,84.95585619)(44.91628967,82.48587003)
\closepath
}
}
{
\newrgbcolor{curcolor}{0 0 0}
\pscustom[linewidth=2.39397537,linecolor=curcolor,strokeopacity=0.02654868]
{
\newpath
\moveto(44.91628967,82.48587003)
\curveto(44.91628967,80.01588387)(42.91397046,78.01356446)(40.44398455,78.01356446)
\curveto(37.97399864,78.01356446)(35.97167943,80.01588387)(35.97167943,82.48587003)
\curveto(35.97167943,84.95585619)(37.97399864,86.9581756)(40.44398455,86.9581756)
\curveto(42.91397046,86.9581756)(44.91628967,84.95585619)(44.91628967,82.48587003)
\closepath
}
}
{
\newrgbcolor{curcolor}{0 0 0}
\pscustom[linewidth=1.99999742,linecolor=curcolor]
{
\newpath
\moveto(40.44397606,82.48586445)
\curveto(45.81383055,82.48586445)(51.18366992,82.48586445)(50.78248819,86.88215043)
\curveto(50.38132913,91.27844397)(44.2090885,100.07101594)(46.12683591,103.66093972)
\curveto(48.04458331,107.25085594)(58.0522885,105.63813153)(59.98015748,108.57559547)
\curveto(61.90802646,111.51306697)(55.75605165,119.00073437)(57.36204094,123.41060397)
\curveto(58.96803024,127.82047358)(68.3318022,129.15253027)(70.79633386,133.77068964)
\curveto(73.26086551,138.38884145)(68.82605858,146.29256681)(71.37875906,150.30227515)
\curveto(73.93147465,154.3119835)(83.4713726,154.4271835)(86.34086551,157.37513185)
\curveto(89.21035843,160.32308019)(85.40909102,166.10340649)(86.53989543,170.35702287)
\curveto(87.67072252,174.61063925)(93.73363654,177.33706193)(99.79667906,180.06354508)
}
}
{
\newrgbcolor{curcolor}{0 0 0}
\pscustom[linewidth=4.97385821,linecolor=curcolor]
{
\newpath
\moveto(138.52418646,141.96964145)
\lineto(182.70177638,142.18250445)
}
}
{
\newrgbcolor{curcolor}{0 0 0}
\pscustom[linewidth=1.91422485,linecolor=curcolor]
{
\newpath
\moveto(158.82826205,142.07134098)
\lineto(159.4358589,21.88247043)
}
}
{
\newrgbcolor{curcolor}{0 0 0}
\pscustom[linewidth=1.99999742,linecolor=curcolor]
{
\newpath
\moveto(98.68118173,43.0077013)
\curveto(105.14418142,44.89747264)(111.60712063,46.78722886)(111.86139213,51.47234634)
\curveto(112.11567874,56.15745626)(106.1612674,63.63788208)(108.05786457,66.41492019)
\curveto(109.95446929,69.1919583)(119.70204472,67.26555578)(121.34736378,71.15452334)
\curveto(122.99268283,75.0434909)(116.53576819,84.74773027)(118.94814236,88.29922382)
\curveto(121.36051654,91.85072492)(132.64181669,89.24936681)(134.88502677,92.9766235)
\curveto(137.12824441,96.70388019)(130.33319811,106.75926791)(131.46925606,110.85823358)
\curveto(132.60532157,114.95719925)(141.67234016,113.09923641)(144.62155087,115.55299641)
\curveto(147.57076913,118.00676397)(144.40180913,124.77195956)(145.75803213,130.13578193)
\curveto(147.11427024,135.49960429)(152.99566866,139.46146382)(158.87719559,143.42340649)
}
}
{
\newrgbcolor{curcolor}{0.00784314 0.00784314 0.00784314}
\pscustom[linestyle=none,fillstyle=solid,fillcolor=curcolor]
{
\newpath
\moveto(103.15348141,43.007711)
\curveto(103.15348141,40.53772484)(101.15116221,38.53540544)(98.6811763,38.53540544)
\curveto(96.21119038,38.53540544)(94.20887118,40.53772484)(94.20887118,43.007711)
\curveto(94.20887118,45.47769717)(96.21119038,47.48001657)(98.6811763,47.48001657)
\curveto(101.15116221,47.48001657)(103.15348141,45.47769717)(103.15348141,43.007711)
\closepath
}
}
{
\newrgbcolor{curcolor}{0 0 0}
\pscustom[linewidth=2.39397537,linecolor=curcolor,strokeopacity=0.02654868]
{
\newpath
\moveto(103.15348141,43.007711)
\curveto(103.15348141,40.53772484)(101.15116221,38.53540544)(98.6811763,38.53540544)
\curveto(96.21119038,38.53540544)(94.20887118,40.53772484)(94.20887118,43.007711)
\curveto(94.20887118,45.47769717)(96.21119038,47.48001657)(98.6811763,47.48001657)
\curveto(101.15116221,47.48001657)(103.15348141,45.47769717)(103.15348141,43.007711)
\closepath
}
}
{
\newrgbcolor{curcolor}{0 0 0}
\pscustom[linewidth=1.91244088,linecolor=curcolor]
{
\newpath
\moveto(219.4824,162.68747704)
\lineto(219.5504315,22.16669846)
}
}
{
\newrgbcolor{curcolor}{0 0 0}
\pscustom[linewidth=4.97385821,linecolor=curcolor]
{
\newpath
\moveto(196.74819402,162.58366098)
\lineto(240.92578394,162.79652397)
}
}
{
\newrgbcolor{curcolor}{0 0 0}
\pscustom[linewidth=1.99999742,linecolor=curcolor]
{
\newpath
\moveto(220.98095244,71.56627641)
\curveto(226.4742085,70.17980208)(231.9673285,68.79336555)(233.77252913,72.66183295)
\curveto(235.5777222,76.53030035)(233.69489764,85.65316523)(236.69322709,87.36734728)
\curveto(239.69157165,89.08152933)(247.57075276,83.38681689)(251.41077543,85.5441183)
\curveto(255.25079055,87.70141972)(255.05117102,97.71045531)(257.88506079,99.64936807)
\curveto(260.71895811,101.58828082)(266.58602457,95.4568213)(270.36938457,95.88701468)
\curveto(274.15274457,96.31720051)(275.85195969,103.30916775)(277.5512126,110.30125594)
}
}
{
\newrgbcolor{curcolor}{0 0 0}
\pscustom[linewidth=2.41133863,linecolor=curcolor]
{
\newpath
\moveto(343.42459843,166.95502476)
\lineto(359.00505827,147.16094728)
\lineto(359.00505827,113.84358413)
\lineto(366.65467087,99.64875956)
\lineto(358.85754331,87.79834571)
\lineto(358.85754331,53.26812838)
\lineto(343.42459843,34.98349972)
}
}
{
\newrgbcolor{curcolor}{0 0 0}
\pscustom[linewidth=1.91422485,linecolor=curcolor]
{
\newpath
\moveto(279.38652094,111.446895)
\lineto(279.44623748,21.85904492)
}
}
{
\newrgbcolor{curcolor}{0 0 0}
\pscustom[linewidth=1.87464564,linecolor=curcolor,linestyle=dashed,dash=0.99199998 0.99199998]
{
\newpath
\moveto(0.86814614,131.65319421)
\lineto(318.83960315,131.42196271)
}
}
{
\newrgbcolor{curcolor}{0 0 0}
\pscustom[linewidth=1.87464564,linecolor=curcolor,linestyle=dashed,dash=0.99199998 0.99199998]
{
\newpath
\moveto(0.00068409,90.87292334)
\lineto(317.97212598,90.64169185)
}
}
{
\newrgbcolor{curcolor}{0.58823532 0.58823532 0.58823532}
\pscustom[linestyle=none,fillstyle=solid,fillcolor=curcolor]
{
\newpath
\moveto(73.58822399,131.23062918)
\curveto(73.58822399,128.36833471)(71.26787566,126.04798658)(68.40558094,126.04798658)
\curveto(65.54328622,126.04798658)(63.22293789,128.36833471)(63.22293789,131.23062918)
\curveto(63.22293789,134.09292365)(65.54328622,136.41327178)(68.40558094,136.41327178)
\curveto(71.26787566,136.41327178)(73.58822399,134.09292365)(73.58822399,131.23062918)
\closepath
}
}
{
\newrgbcolor{curcolor}{0 0 0}
\pscustom[linewidth=0.97329635,linecolor=curcolor,strokeopacity=0.998285]
{
\newpath
\moveto(73.58822399,131.23062918)
\curveto(73.58822399,128.36833471)(71.26787566,126.04798658)(68.40558094,126.04798658)
\curveto(65.54328622,126.04798658)(63.22293789,128.36833471)(63.22293789,131.23062918)
\curveto(63.22293789,134.09292365)(65.54328622,136.41327178)(68.40558094,136.41327178)
\curveto(71.26787566,136.41327178)(73.58822399,134.09292365)(73.58822399,131.23062918)
\closepath
}
}
{
\newrgbcolor{curcolor}{0.00784314 0.00784314 0.00784314}
\pscustom[linestyle=none,fillstyle=solid,fillcolor=curcolor]
{
\newpath
\moveto(164.32616381,102.13981271)
\curveto(164.32616381,99.66982655)(162.3238446,97.66750714)(159.85385869,97.66750714)
\curveto(157.38387278,97.66750714)(155.38155357,99.66982655)(155.38155357,102.13981271)
\curveto(155.38155357,104.60979887)(157.38387278,106.61211828)(159.85385869,106.61211828)
\curveto(162.3238446,106.61211828)(164.32616381,104.60979887)(164.32616381,102.13981271)
\closepath
}
}
{
\newrgbcolor{curcolor}{0 0 0}
\pscustom[linewidth=2.39397537,linecolor=curcolor,strokeopacity=0.02654868]
{
\newpath
\moveto(164.32616381,102.13981271)
\curveto(164.32616381,99.66982655)(162.3238446,97.66750714)(159.85385869,97.66750714)
\curveto(157.38387278,97.66750714)(155.38155357,99.66982655)(155.38155357,102.13981271)
\curveto(155.38155357,104.60979887)(157.38387278,106.61211828)(159.85385869,106.61211828)
\curveto(162.3238446,106.61211828)(164.32616381,104.60979887)(164.32616381,102.13981271)
\closepath
}
}
{
\newrgbcolor{curcolor}{0 0 0}
\pscustom[linewidth=1.99999742,linecolor=curcolor]
{
\newpath
\moveto(161.05649008,102.17918728)
\curveto(167.25466583,99.11210067)(173.45269795,96.04508964)(175.94666457,100.14278539)
\curveto(178.44063118,104.2404887)(177.23033575,115.50234697)(180.64296945,118.42833626)
\curveto(184.05560315,121.35433311)(192.09073512,115.94411326)(195.14525102,120.1906846)
\curveto(198.19977449,124.43725594)(196.27328882,138.34007421)(198.77724094,142.01631484)
\curveto(201.28119307,145.69254791)(208.21526551,139.14173468)(212.29706457,140.88282697)
\curveto(216.37887118,142.62391925)(217.60792063,152.65690193)(218.83699276,162.69009626)
}
}
{
\newrgbcolor{curcolor}{0.58823532 0.58823532 0.58823532}
\pscustom[linestyle=none,fillstyle=solid,fillcolor=curcolor]
{
\newpath
\moveto(201.91883076,132.59051168)
\curveto(201.91883076,129.72821721)(199.59848243,127.40786908)(196.73618772,127.40786908)
\curveto(193.873893,127.40786908)(191.55354467,129.72821721)(191.55354467,132.59051168)
\curveto(191.55354467,135.45280615)(193.873893,137.77315428)(196.73618772,137.77315428)
\curveto(199.59848243,137.77315428)(201.91883076,135.45280615)(201.91883076,132.59051168)
\closepath
}
}
{
\newrgbcolor{curcolor}{0 0 0}
\pscustom[linewidth=0.97329635,linecolor=curcolor,strokeopacity=0.998285]
{
\newpath
\moveto(201.91883076,132.59051168)
\curveto(201.91883076,129.72821721)(199.59848243,127.40786908)(196.73618772,127.40786908)
\curveto(193.873893,127.40786908)(191.55354467,129.72821721)(191.55354467,132.59051168)
\curveto(191.55354467,135.45280615)(193.873893,137.77315428)(196.73618772,137.77315428)
\curveto(199.59848243,137.77315428)(201.91883076,135.45280615)(201.91883076,132.59051168)
\closepath
}
}
{
\newrgbcolor{curcolor}{0.58823532 0.58823532 0.58823532}
\pscustom[linestyle=none,fillstyle=solid,fillcolor=curcolor]
{
\newpath
\moveto(132.45581038,90.80123724)
\curveto(132.45581038,87.93894277)(130.13546205,85.61859464)(127.27316733,85.61859464)
\curveto(124.41087261,85.61859464)(122.09052428,87.93894277)(122.09052428,90.80123724)
\curveto(122.09052428,93.66353171)(124.41087261,95.98387984)(127.27316733,95.98387984)
\curveto(130.13546205,95.98387984)(132.45581038,93.66353171)(132.45581038,90.80123724)
\closepath
}
}
{
\newrgbcolor{curcolor}{0 0 0}
\pscustom[linewidth=0.97329635,linecolor=curcolor,strokeopacity=0.998285]
{
\newpath
\moveto(132.45581038,90.80123724)
\curveto(132.45581038,87.93894277)(130.13546205,85.61859464)(127.27316733,85.61859464)
\curveto(124.41087261,85.61859464)(122.09052428,87.93894277)(122.09052428,90.80123724)
\curveto(122.09052428,93.66353171)(124.41087261,95.98387984)(127.27316733,95.98387984)
\curveto(130.13546205,95.98387984)(132.45581038,93.66353171)(132.45581038,90.80123724)
\closepath
}
}
{
\newrgbcolor{curcolor}{0.00784314 0.00784314 0.00784314}
\pscustom[linestyle=none,fillstyle=solid,fillcolor=curcolor]
{
\newpath
\moveto(224.02285753,71.93693329)
\curveto(224.02285753,69.46694713)(222.02053832,67.46462772)(219.55055241,67.46462772)
\curveto(217.0805665,67.46462772)(215.07824729,69.46694713)(215.07824729,71.93693329)
\curveto(215.07824729,74.40691945)(217.0805665,76.40923886)(219.55055241,76.40923886)
\curveto(222.02053832,76.40923886)(224.02285753,74.40691945)(224.02285753,71.93693329)
\closepath
}
}
{
\newrgbcolor{curcolor}{0 0 0}
\pscustom[linewidth=2.39397537,linecolor=curcolor,strokeopacity=0.02654868]
{
\newpath
\moveto(224.02285753,71.93693329)
\curveto(224.02285753,69.46694713)(222.02053832,67.46462772)(219.55055241,67.46462772)
\curveto(217.0805665,67.46462772)(215.07824729,69.46694713)(215.07824729,71.93693329)
\curveto(215.07824729,74.40691945)(217.0805665,76.40923886)(219.55055241,76.40923886)
\curveto(222.02053832,76.40923886)(224.02285753,74.40691945)(224.02285753,71.93693329)
\closepath
}
}
{
\newrgbcolor{curcolor}{0 0 0}
\pscustom[linewidth=4.97385821,linecolor=curcolor]
{
\newpath
\moveto(258.63046677,111.04101153)
\lineto(302.80805669,111.25387452)
}
}
{
\newrgbcolor{curcolor}{0.58823532 0.58823532 0.58823532}
\pscustom[linestyle=none,fillstyle=solid,fillcolor=curcolor]
{
\newpath
\moveto(259.42398271,91.40015072)
\curveto(259.42398271,88.53785625)(257.10363438,86.21750812)(254.24133966,86.21750812)
\curveto(251.37904494,86.21750812)(249.05869661,88.53785625)(249.05869661,91.40015072)
\curveto(249.05869661,94.26244519)(251.37904494,96.58279332)(254.24133966,96.58279332)
\curveto(257.10363438,96.58279332)(259.42398271,94.26244519)(259.42398271,91.40015072)
\closepath
}
}
{
\newrgbcolor{curcolor}{0 0 0}
\pscustom[linewidth=0.97329635,linecolor=curcolor,strokeopacity=0.998285]
{
\newpath
\moveto(259.42398271,91.40015072)
\curveto(259.42398271,88.53785625)(257.10363438,86.21750812)(254.24133966,86.21750812)
\curveto(251.37904494,86.21750812)(249.05869661,88.53785625)(249.05869661,91.40015072)
\curveto(249.05869661,94.26244519)(251.37904494,96.58279332)(254.24133966,96.58279332)
\curveto(257.10363438,96.58279332)(259.42398271,94.26244519)(259.42398271,91.40015072)
\closepath
}
\rput[bl](375,92){$\sim \ell_{\min}$}
}
\end{pspicture}
\caption{Non-adjacent Gaussian pairing.}
\label{figfourth}
\end{figure}%
Since each pairing takes place at fixed time, and the Dyson series is causal
except for the nonlocality on the scale~$\ell_{\min}$, one sees that, in this example, all the potentials
are evaluated in a time strip of width~$\sim \ell_{\min}$.
Therefore, all the paired potentials together can be regarded as one nonlocal potential
evaluated at time~$t$; similar as is the case for the paired potential in Figure~\ref{figgauss1}.
Therefore, the scaling of the error term can be obtained simply by comparing
the terms in Figure~\ref{figfourth} with those in Figure~\ref{figgauss1}.
Clearly, in Figure~\ref{figfourth} we have two more factors~$\B$. Moreover, we have two
more black dots, each of which corresponds to a time integral, giving a factor~$\ell_{\min}$.
Therefore, the non-adjacent pairing gives an additional scaling factor~$\ell_{\min}^2\, \|\B\|^2$.

This argument applies similarly to higher order non-adjacent pairings, as well as to
pairings between bra and ket (as shown in Figure~\ref{figgauss2}).
We thus conclude that the non-adjacent pairings can indeed be neglected up to an error of the order
\[ 1 + \O \Big( \ell_{\min}^2\, \|\B\|^2 \Big) \:. \]
This concludes the derivation of Theorem~\ref{thmlindblad}.

\subsection{Reduction of the State Vector} \label{seccollapse}
Assume that a state~$\psi(t)$ given at time~$t_0$ is evaluated at a later time~$t_1$.
The question is whether the time evolution has led to a collapse of the wave function.
In order to keep the setting clean and simple, we assume that the operator~$\Sig_t$ vanishes
near the initial and final times, as shown in Figure~\ref{figscatter},
\begin{figure}
\psset{xunit=.5pt,yunit=.5pt,runit=.5pt}
\begin{pspicture}(404.98344842,244.77890856)
{
\newrgbcolor{curcolor}{0.80000001 0.80000001 0.80000001}
\pscustom[linestyle=none,fillstyle=solid,fillcolor=curcolor]
{
\newpath
\moveto(1.00606564,171.6087706)
\lineto(403.98184894,172.0567958)
\lineto(403.98188674,61.71779738)
\lineto(1.49995162,61.3536021)
\closepath
}
}
{
\newrgbcolor{curcolor}{0.80000001 0.80000001 0.80000001}
\pscustom[linewidth=2.0031495,linecolor=curcolor]
{
\newpath
\moveto(1.00606564,171.6087706)
\lineto(403.98184894,172.0567958)
\lineto(403.98188674,61.71779738)
\lineto(1.49995162,61.3536021)
\closepath
}
}
{
\newrgbcolor{curcolor}{0 0 0}
\pscustom[linewidth=2.75905519,linecolor=curcolor]
{
\newpath
\moveto(25.24584945,0.0134421)
\lineto(23.03101606,227.39768525)
}
}
{
\newrgbcolor{curcolor}{0 0 0}
\pscustom[linestyle=none,fillstyle=solid,fillcolor=curcolor]
{
\newpath
\moveto(15.34365056,223.45994606)
\lineto(22.86171448,244.77890843)
\lineto(30.79362672,223.61043636)
\curveto(26.25619859,227.4291001)(19.82514601,227.36645851)(15.34365056,223.45994606)
\closepath
}
}
{
\newrgbcolor{curcolor}{0 0 0}
\pscustom[linewidth=2.0031495,linecolor=curcolor]
{
\newpath
\moveto(1.25281512,19.98260242)
\lineto(403.73064567,20.80736336)
}
}
{
\newrgbcolor{curcolor}{0 0 0}
\pscustom[linewidth=2.0031495,linecolor=curcolor]
{
\newpath
\moveto(1.25281512,212.60324619)
\lineto(403.73060787,213.42800714)
}
\rput[bl](-23,10){$t_0$}
\rput[bl](-20,205){$t_1$}
\rput[bl](50,30){$\Sig_t=0$}
\rput[bl](50,180){$\Sig_t=0$}
\rput[bl](50,110){$\Sig_t \neq 0$}
\rput[bl](200,30){$\tilde{\psi} = \psi = \psi^\res$}
\rput[bl](200,180){$\tilde{\psi} = \psi = \psi^\res$}
\rput[bl](200,110){$\tilde{\psi} \neq \psi \neq \psi^\res$}
}
\end{pspicture}
\caption{Situation for the wave function collapse.}
\label{figscatter}
\end{figure}
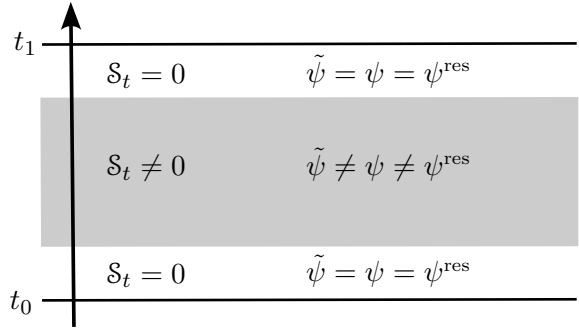%
so that at these times we do not distinguish
between the transformed and untransformed wave functions, i.e.\
\[ \psi(t_0) = \tilde{\psi}(t_0) \qquad \text{and} \qquad \psi(t_1) = \tilde{\psi}(t_1) \:. \]
Consequently, we can consider equivalently the curves~$\psi(t)$ or~$\tilde{\psi}(t)$.
In the study of the collapse, one usually makes the assumption that the observable~${\mathcal{O}}$
used for the measurement commutes with the operators which appear in the stochastic Schr\"odinger equation. Having a position measurement in mind, this is justified by the locality and causality of the time evolution.
In our setting, we must keep in mind that the transformed time evolution~$\tilde{\psi}(t)$ is not causal,
because the operator~$\sqrt{1+\Sig_t}$ is nonlocal. For this reason, in order to get into the position to
follow the standard argumentation, we must consider the untransformed curve~$\psi(t)$.
Doing so, we are facing the difficulty that the norm~$(\psi(t)|\psi(t))$ is no longer conserved.
However, fluctuations of the scalar product do not comply with the standard argumentation. For this reason,
we rescale the wave function such as to preserve the norm. We thus consider the curve
\[ 
\psi^\res(t) := c(t)\: \psi(t) \qquad \text{with} \qquad c(t) := \frac{1}{\sqrt{\bbra (\psi(t) | \psi(t)) \kket}} \:. \]
We point out for clarity that
\beq \label{psinit}
\psi^\res(t_0) = \tilde{\psi}(t_0) \qquad \text{and} \qquad \psi^\res(t_1) = \tilde{\psi}(t_1) \:.
\eeq
Therefore, considering the curve~$\psi^\res(t)$ does not change the initial or end state.
Choosing the curve~$\psi^\res(t)$ over~$\tilde{\psi}(t)$ is a matter of convenience, having the advantage
that the corresponding dynamical equations can be studied similar as the CSL model.

Let us work out the corresponding dynamical equations.
We begin by computing the change of the $L^2$-norm,
\begin{align*}
\frac{d}{dt} \bbra (\psi | \psi) \kket &= \bbra (\partial_t \psi | \psi) +  (\psi | \partial_t \psi) \kket \\
&= \sint_\varkappa \big( \psi \:\big|\: (-X_\varkappa Y_\varkappa - Y_\varkappa^\dagger X_\varkappa^\dagger
+ Z_\varkappa^\dagger X_\varkappa + X_\varkappa^\dagger Z_\varkappa) \psi \big) \:.
\end{align*}
Consequently, the statistical operator of the rescaled wave function has the following dynamics,
\begin{align}
&\frac{d}{dt} \bbra |\psi^\res(t))(\psi^\res(t)| \kket = \frac{d}{dt} \bbra \frac{| \psi)(\psi |}{\bbra (\psi | \psi) \kket} \kket \notag \\
&= \frac{1}{\bbra (\psi | \psi) \kket} \sint_\varkappa \Big( - X_\varkappa Y_\varkappa |\psi)(\psi| - |\psi)(\psi| Y_\varkappa^\dagger X_\varkappa^\dagger + X_\varkappa |\psi)(\psi| Z_\varkappa^\dagger + Z_\varkappa |\psi)(\psi| X_\varkappa^\dagger \Big) \notag \\
&\quad\: -\frac{1}{\bbra (\psi | \psi) \kket^2} \sint_\varkappa \big( \psi \:\big|\: (-X_\varkappa Y_\varkappa - Y_\varkappa^\dagger X_\varkappa^\dagger
+ Z_\varkappa^\dagger X_\varkappa + X_\varkappa^\dagger Z_\varkappa) \psi \big) \: |\psi)(\psi| \:.
\label{resred}
\end{align}

Now we need to assume that the observable~${\mathcal{O}}$ commutes with all the operators and that
\beq \label{composite}
\sint_\varkappa (-X_\varkappa Y_\varkappa - Y_\varkappa^\dagger X_\varkappa^\dagger
+ Z_\varkappa^\dagger X_\varkappa + X_\varkappa^\dagger Z_\varkappa) \sim \1 \:.
\eeq
These assumptions can be justified as follows.
The assumption that the observable commutes with the operators describing the dynamics
is usually justified by considering position measurements and using the locality of the interaction.
This argument also applies in our setting, provided that the length scale for the position measurement
(like for example the distance of neighboring photo-diodes in the measurement device) is much larger than
the length scale~$\ell_{\min}$. With the same argumentation, one obtains that the sum of the operators
in~\eqref{composite} can be treated by a local operator (again on the length scale of the measurement
device which is assumed to be much larger than~$\ell_{\min}$.
Assuming that the averages over the fluctuating fields as taken in~\eqref{composite}
are spatially homogeneous, this local operator is indeed a multiple of the identity.

Using these assumptions in~\eqref{resred}, we conclude that
\beq \label{dtO}
\frac{d}{dt} \bbra (\psi^\res \,|\, {\mathcal{O}} \, \psi^\res) \kket = 0 \:.
\eeq
Integrating this equation from~$t_0$ to~$t_1$ and using~\eqref{psinit}, we conclude that
\[ \bbra (\tilde{\psi}(t_1) \,|\, {\mathcal{O}} \, \tilde{\psi}(t_1)) \kket = \bbra (\tilde{\psi}(t_0) \,|\, {\mathcal{O}} \, \tilde{\psi}(t_0)) \kket \:. \]
This means that, as desired, the statistical mean of the expectation value remains preserved in the
reduction process. 

We next consider how the variance changes. We first note that~\eqref{dtO} also holds for the
square of the observable,
\[ \frac{d}{dt} \bbra (\psi^\res \,|\, {\mathcal{O}}^2 \, \psi^\res) \kket = 0 \:. \]
However, in the evolution equation of the variance, we must take into account Gaussian pairings between
the two expectation values. A straightforward computation similar to that in Section~\ref{sectpsipsi} gives
\begin{align}
&\frac{d}{dt} \bbra (\psi^\res \,|\, {\mathcal{O}}^2 \, \psi^\res) - (\psi^\res \,|\, {\mathcal{O}} \, \psi^\res)^2 \kket \notag \\
&= - 4 \re \;\bbra (\psi^\res \,|\, {\mathcal{O}} \, \partial_t \psi^\res)\: (\partial_t \psi^\res \,|\, {\mathcal{O}} \, \psi^\res) \kket \label{coll1} \\
&= - 4 \sint_\varkappa \Big( (\psi \,|\, {\mathcal{O}}\, (X_\varkappa - X_\varkappa^\dagger)\, \psi) \: (\psi | \psi) - 
(\psi \,|\, {\mathcal{O}} \psi) \: (\psi \,|\, (X_\varkappa - X_\varkappa^\dagger)\, \psi) \Big) \notag \\
&\qquad \qquad \times \Big( (\psi \,|\, {\mathcal{O}}\, (Z_\varkappa - Z_\varkappa^\dagger)\, \psi) \: (\psi | \psi) - 
(\psi \,|\, {\mathcal{O}} \psi) \: (\psi \,|\, (Z_\varkappa - Z_\varkappa^\dagger)\, \psi) \Big) \:. \label{coll2}
\end{align}
It is clear from~\eqref{coll1} that this expression is smaller or equal to zero.
Moreover, one sees from~\eqref{coll2} that the expression is in general non-zero
unless~$\psi$ is an eigenstate of~${\mathcal{O}}$.
This shows that a collapse takes place and proves the Born rule.

\subsection{Description with Fock Spaces} \label{secfock}
For the sake of simplicity, so far we considered the one-particle Dirac or Schr\"odinger equations.
But causal fermion systems can also be described in the language of quantum field theory
in terms of states on bosonic and fermionic Fock spaces (see~\cite{fockbosonic, fockfermionic, fockentangle}).
In this setting, the stochastic dynamics and nonlinear dynamics
is formulated in the standard way with creation and annihilation operators, giving agreement with
the standard formulation of CSL models.
We note that, as worked out in detail~\cite{fockdynamics}, the formulation of the dynamics with a nonlocal
potential in the Dirac equation~\eqref{dirnonloc} can be related directly to a quantum dynamics in Fock spaces.
In order to analyze the collapse of a many-particle wave function, it suffices to consider a $q$-particle
Hartree-Fock state
\beq \label{HF}
\Psi = \psi_1 \wedge \cdots \wedge \psi_q \:.
\eeq
The dynamics of each one-particle wave function is described by the nonlocal Dirac equation.
Following the standard argument in collapse theory, the many-particle wave function collapses once
one of the one-particle wave functions does. In this way, collapse phenomena are predominant for
mesoscopic and macroscopic systems. This explains why our macroscopic world behaves classically.

\subsection{Parameters of the Effective Collapse Model} \label{secparameters}
The model involves the following free parameters:
\bitem
\itemD The length scale~$\ell_{\min}$ (see~\eqref{ellmindef}). Equivalently, the number~$N$ of fields
(see~\eqref{Nscaleintro}).
\itemD The strength of the stochastic field as described by the covariance.
\eitem
The same parameters come into play in the quantum field theory limit of causal fermion systems
as obtained in~\cite{fockdynamics}. We hope that exploring these connections further will give
constraints for the above parameters.

\subsection{How Does the Nonlinearity Enter?} \label{secnonlin}
Clearly, the nonlinearity of the dynamics is crucial for the collapse of the wave function.
Despite this fact, in our above derivation of the effective collapse model in Section~\ref{secmodel},
we did not need to specify the structure and strength of the nonlinear coupling.
One way of understanding how this comes about is to note that we made essential use of the
conservation law for the commutator inner product. This conservation law holds only because the
nonlinearity of the dynamics as described by the causal action principle guarantees
that this conservation law is respected. In our effective description with the nonlocal Dirac equation, this
enters by the assumption that the potential is symmetric~\eqref{Bsymm}. But one should keep in mind
this symmetric potential comes about on the level of the causal action principle by a complicated nonlinear
interaction of all one-particle wave functions.

Nevertheless, it is an important question to describe the nonlinearity more specifically.
We see two alternative methods for doing so. One is to work with the nonlinear coupling as described
by the nonlocal Maxwell equations~\eqref{maxnonloc}. This method has the advantage that it can be
used for a quantitative analysis. However, one should keep in mind that adding the Maxwell action~\eqref{SAdef}
is only an approximate description, which does not take into account effects due to the non-smoothness
and possibly discreteness of spacetime. A systematic study of all these effects goes beyond the scope of this
article. An alternative, more conceptual way of understanding the nonlocality is to note that, in the
causal fermion system description, the one-particle wave functions also describe the bosonic fields.
This statement is made precise by the construction of the quantum state in~\cite{fockbosonic},
which acts on an observable algebra involving both bosonic and fermionic field operators.
In order to explain the idea in the simplest possible setting, we return to the Hartree-Fock state~\eqref{HF}.
Here the bosonic potentials are encoded in the wave functions~$\psi_1, \ldots, \psi_q$ in the sense that
the potential~$\B$ in the Dirac equation~\eqref{dirnonloc} is determined uniquely by the condition that all the
wave functions satisfy this Dirac equation.
Having this picture in mind, the potential~$\B$ depends on the Hartree-Fock state. Consequently,
the Dirac equation by itself describes a nonlinear dynamics on the Fock state.

\section{Outlook: The Relativistic Model} \label{secrelativity}
Since causal fermion systems incorporate the principle of causality and the equivalence principle
(for details see for example~\cite[Section~5.9]{intro}), it should also give rise to a relativistic collapse model.
However, this relativistic model will be considerably more involved. We plan to work out the
details separately in~\cite{relcoll}. Here we merely make a few remarks.
The main simplification when taking the non-relativistic limit is that it becomes possible with
equal-time relations like in the Markov property~\eqref{markov2}.
In the relativistic setting, these equal-time relations must be replaced by relations stating that
the stochastic field propagates with the speed of light. In other words, these relations should involve
the bosonic causal propagator.
When doing so, the treatment becomes relativistic in the sense that the Dirac equation
and the equations describing the dynamics of the bosonic fields are covariant.
However, the stochastic background field itself will not be covariant.
This is already obvious from the ansatz~\eqref{Bstochastic}, where the kernels~$L_a(y-x)$
break Lorentz invariance.
Our concept is that these stochastic background fields originate from the early universe
and/or are generated by the matter on earth and of the surrounding stars and galaxies.
With this in mind, it is obvious and natural that Lorentz invariance is broken by these fields.

Working out the relativistic collapse model should also give
a better and more detailed understanding for the locality assumptions used in
our derivation of the collapse mechanism. In particular, in the relativistic setting
the identity operator in~\eqref{composite} should be replaced by an operator which is causal
with strong decay properties in lightlike directions,
up to nonlocal corrections on the scale~$\ell_{\min}$ which break Lorentz invariance.

\appendix
\section{Detailed Computation of the Gaussian Pairings} \label{applindblad}
In this appendix we provide details of the computations in Section~\ref{seclindblad}.
\addtocontents{toc}{\setcounter{tocdepth}{1}}%
\subsection{Computation of~$\bbra |\partial_t \Sig_t \psi) \kket$}
We first derive the evolution equation for~$\tilde{\psi}$. Differentiating~\eqref{tildepsi} 
and using~\eqref{Hint} gives
\begin{align}
&\frac{d}{dt} \tilde{\psi} = \frac{d}{dt} \big( \sqrt{\1+\Sig_{t}} \:\psi(t) \big)
= \Big( \partial_t \sqrt{\1+\Sig_{t}} \Big)\:\psi(t) + \sqrt{\1+\Sig_{t}} \:\partial_t \psi(t) \notag \\
&= \Big( \partial_t \sqrt{\1+\Sig_{t}} \Big) \big( \1+\Sig_{t} \big)^{-\frac{1}{2}}\: \tilde{\psi}(t)
-i  \sqrt{\1+\Sig_{t}}\: \int_{-\infty}^\infty V(t,\tau)\: \big( \1+\Sig_\tau \big)^{-\frac{1}{2}}\: \tilde{\psi}(\tau)\: d\tau \:.
\label{Sigexpand}
\end{align}
The operator~$\Sig_t$ involves a factor of~$V$ (see~\eqref{Sigdef}). Moreover, because of the
factor~$V(t,\tau)$, we know that~$|t-\tau| \lesssim \ell_{\min}$.
Therefore, the scaling argument explained in Section~\ref{secerror} again applies,
making it possible to expand in powers of~$\Sig_t$ and~$\Sig_\tau$ up to second order in
these operators. We begin with the linear terms, whereas the terms quadratic in these operators
will be considered in Sections~\ref{secSS1} and~\ref{secSS2}.
Our starting point is the formula~\eqref{Sform} for~$\Sig_t$. Multiplying by~$\psi(t)$ and differentiating
with respect to~$t$ gives
\begin{align*}
\Sig_t \psi(t) &=  i \bigg( \int_{-\infty}^t \!\!\!\!\!d\tau \int_t^\infty \!\!\!\!\!d\tau' \!-\! \int_t^\infty \!\!\!\!\!d\tau \int_{-\infty}^t \!\!\!\!\!d\tau' \bigg) \: (U^\tau_t)^\dagger \,V(\tau,\tau') \:\psi(\tau') \\
\frac{d}{dt} \Sig_t \psi(t) &=
i \bigg( \int_{-\infty}^t \!\!\!\!\!d\tau \int_t^\infty \!\!\!\!\!d\tau' \!-\! \int_t^\infty \!\!\!\!\!d\tau \int_{-\infty}^t \!\!\!\!\!d\tau' \bigg) 
\int_{-\infty}^\infty d\zeta \:(-i) V(t, \zeta)^\dagger \,(U^\tau_\zeta)^\dagger\, V(\tau,\tau') \:\psi(\tau') \\
&\quad\: + i \int_{-\infty}^\infty V(t,\tau') \:\psi(\tau')\: d\tau' - i \int_{-\infty}^\infty (U^\tau_t)^\dagger \,V(\tau,t)
\:\psi(t)\: d\tau \:.
\end{align*}
Taking the statistical mean and again restricting attention to adjacent parings, we obtain
\begin{align}
&\bbra \frac{d}{dt} \Sig_t \psi(t) \kket \notag \\
&= \bigg( \int_{-\infty}^t \!\!\!\!\!d\tau \int_t^\infty \!\!\!\!\!d\tau' \!-\! \int_t^\infty \!\!\!\!\!d\tau \int_{-\infty}^t \!\!\!\!\!d\tau' \bigg) \int_{-\infty}^\infty d\zeta_2 \: \bbra V(t, \zeta_2)^\dagger \,V(\tau,\tau') \kket \:\tilde{\psi}(t) \label{a1} \\
&\quad\: + \int_{-\infty}^\infty d\zeta_1 \int_{-\infty}^{\zeta_1} d\tau_2 \int_{\infty}^\infty
d\zeta_2\: \bbra V(t,\zeta_1) \:V(\tau_2, \zeta_2) \kket \:\tilde{\psi}(t) \label{a2} \\
&\quad\: - \int_{-\infty}^\infty d\zeta_1 \int_{-\infty}^t d\tau_2 \int_{\infty}^\infty
d\zeta_2\: \bbra V(\zeta_1, t) \:V(\tau_2, \zeta_2) \kket \:\tilde{\psi}(t) \label{a22} \\
&\quad\: + \int_{-\infty}^\infty d\zeta_1 \int_t^{\zeta_1} d\tau_1 \int_{-\infty}^\infty d\zeta_2\:
\bbra V(\tau_1, \zeta_2)^\dagger\,V(\zeta_1,t) \kket\:\tilde{\psi}(t) \:, \label{a3}
\end{align}
again up to error terms as specified in Theorem~\ref{thmlindblad}.
Here we could replace the factors~$\psi(\tau')$ and~$\psi(t)$ on the right by
a factor~$\tilde{\psi}(t)$ using that we are considering the linear term of the expansion
of~\eqref{Sigexpand} in powers of~$\Sig_t$ and~$\Sig_\tau$.

The term~\eqref{a1} can be computed as follows,
\begin{align}
\eqref{a1}
&= \sum_{k,a}
\bigg( \int_{-\infty}^t \!\!\!\!\!d\tau \int_t^\infty \!\!\!\!\!d\tau' \!-\! \int_t^\infty \!\!\!\!\!d\tau \int_{-\infty}^t \!\!\!\!\!d\tau' \bigg) \int_{-\infty}^\infty d\zeta_2 \notag \\
&\qquad \times \delta \Big( \frac{t+\zeta_2}{2} - \frac{\tau+\tau'}{2}\Big) \;
M_{k,a}(t, \zeta_2)^\dagger \,M_{k,a}(\tau,\tau') \:\tilde{\psi}(t) \notag \\
&= 2 \sum_{k,a}
\bigg( \int_{-\infty}^t \!\!\!\!\!d\tau \int_t^\infty \!\!\!\!\!d\tau' \!-\! \int_t^\infty \!\!\!\!\!d\tau \int_{-\infty}^t \!\!\!\!\!d\tau' \bigg)  \
M_{k,a}(t, \tau+\tau'-t)^\dagger \,M_{k,a}(\tau,\tau') \:\tilde{\psi}(t) \:. \notag
\end{align}
Introducing the integration variables
\[ \zeta = \frac{\tau+\tau'}{2} -t \:,\qquad \nu = \frac{\tau-\tau'}{2}\:, \]
we obtain
\begin{align*}
\eqref{a1} &= -4 \sint_\varkappa 
\int_{\R \setminus [-|\zeta|, |\zeta|]} \!\!\!\!\!\!\!\!\!\!\!\!\epsilon(\nu)\:
M_{k,a}(t, t+2 \zeta)^\dagger \,M_{k,a}(t+\zeta+\nu, t+\zeta-\nu) \:\tilde{\psi}(t) \: d\nu \:.
\end{align*}

The terms~\eqref{a2} and~\eqref{a22} can be computed as follows,
\begin{align*}
&\eqref{a2}+\eqref{a22} \\
&= \int_{-\infty}^\infty d\zeta_1 \int_{-\infty}^\infty d\tau_2 \int_{\infty}^\infty
d\zeta_2 \\
&\qquad \times \bbra \big( V(t,\zeta_1) \: \Theta(\zeta_1-\tau_2)- V(\zeta_1, t) 
 \: \Theta(t-\tau_2\big)\:V(\tau_2, \zeta_2) \kket \:\tilde{\psi}(t) \\
&\quad\:+ \int_{-\infty}^\infty d\zeta_1 \int_{-\infty}^{\zeta_1} d\tau_2 \int_{\infty}^\infty
d\zeta_2\: \bbra \big( V(t,\zeta_1) - V(\zeta_1, t) \big)\:V(\tau_2, \zeta_2) \kket \:\tilde{\psi}(t) \\
&= \sum_{k,a} \int_{-\infty}^\infty d\zeta_1 \int_{-\infty}^\infty d\tau_2 \int_{\infty}^\infty
d\zeta_2 \: \delta \Big( \frac{t+\zeta_1}{2} - \frac{\tau_2+\zeta_2}{2} \Big) \\
&\qquad \times 
\: \big( M_{k,a}(t,\zeta_1)\: \Theta(\zeta_1-\tau_2) - M_{k,a}(\zeta_1, t) 
\: \Theta(t-\tau_2)\big)\:M_{k,a}(\tau_2, \zeta_2)\:\tilde{\psi}(t) \\
&= 2 \sum_{k,a} \int_{-\infty}^\infty d\zeta_1 \int_{-\infty}^\infty d\tau_2 \;
\big( M_{k,a}(t,\zeta_1) \: \Theta(\zeta_1-\tau_2)- M_{k,a}(\zeta_1, t) \: \Theta(t-\tau_2)\big) \\
&\qquad \times \:M_{k,a}(\tau_2, 
t+\zeta_1-\tau_2)\:\tilde{\psi}(t) \:.
\end{align*}
Introducing the integration variables
\[ \zeta = \frac{1}{2} \:\big( \zeta_1 - t \big)\:,\qquad \nu = \tau_2 - t- \zeta \:, \]
we obtain
\[ \eqref{a2}+\eqref{a22}
= 4 \sint_\varkappa \int_{-\zeta}^\zeta M_{k,a}(t, t+2 \zeta) \:
M_{k,a}(t+\zeta+\nu, t +\zeta- \nu)\: d\nu \:. \]

The term~\eqref{a3} can be computed as follows,
\begin{align*}
\eqref{a3}
&= \sum_{k,a}
\int_{-\infty}^\infty d\zeta_1 \int_t^{\zeta_1} d\tau_1 \int_{-\infty}^\infty d\zeta_2 \\
&\qquad \times
\delta \Big( \frac{(\tau_1+\zeta_2}{2} - \frac{\zeta_1+t}{2} \Big)\:
M_{k,a}(\tau_1, \zeta_2)^\dagger\,M_{k,a}(\zeta_1,t) \:\tilde{\psi}(t) \\
&= 2 \sum_{k,a}
\int_{-\infty}^\infty d\zeta_1 \int_t^{\zeta_1} d\tau_1\:
M_{k,a}(\tau_1, \zeta_1+t-\tau_1)^\dagger\,M_{k,a}(\zeta_1,t) \:\tilde{\psi}(t) \:.
\end{align*}
Introducing the integration variables
\[ \zeta = \frac{\zeta_1-t}{2} \:,\qquad \nu = t_1-t-\zeta\:, \]
we obtain
\begin{align*}
\eqref{a3} &= 4 \sint_\varkappa
\int_{-\zeta}^\zeta d\nu\:
M_{k,a}(t+\zeta+\nu ,t+\zeta-\nu)^\dagger\,M_{k,a}(t+2 \zeta,t) \:\tilde{\psi}(t) \:.
\end{align*}

\subsection{Computation of~$\bbra |\partial_t \Sig_t \psi)(\psi| \kket$}
In this and the next section we compute the contributions linear in~$\Sig_t$ with
a Gaussian pairing between bra and ket.
\begin{align}
&\bbra \Big( \frac{d}{dt} | \Sig_t \psi(t)) \Big)  ( \psi(t) | \kket - \bbra \Big( \frac{d}{dt} |\Sig_t \psi) \Big) \kket \; ( \psi(t)| \notag \\
&= i \int_{-\infty}^\infty d\tau_1 \;\bbra \big( V(t,\tau_1) - V(\tau_1,t) \big) \:|\psi(t))(\psi(t)| \kket \notag \\
&= -\int_{-\infty}^\infty d\tau_1 \int_{-\infty}^t d\tau_2 \int_{-\infty}^\infty d\zeta_2
\;\bbra \big( V(t,\tau_1) - V(\tau_1,t) \big) \:|\psi)(\psi|\: V(\tau_2, \zeta_2)^\dagger \kket \notag \\
&= - \sum_{k,a}\int_{-\infty}^\infty d\tau_1 \int_{-\infty}^t d\tau_2 \int_{-\infty}^\infty d\zeta_2 \notag \\
&\qquad \times \delta \Big( \frac{t+\tau_1}{2} - \frac{\tau_2+\zeta_2}{2} \Big) \:
\big( M_{k,a}(t,\tau_1) - M_{k,a}(\tau_1,t) \big) \:|\psi)(\psi|\: M_{k,a}(\tau_2, \zeta_2)^\dagger
\kket \notag \\
&= -2 \sum_{k,a}\int_{-\infty}^\infty d\tau_1 \int_{-\infty}^t d\tau_2 \notag \\
&\qquad \times
\big( M_{k,a}(t,\tau_1) - M_{k,a}(\tau_1,t) \big) \: M_{k,a}(\tau_2, t+\tau_1-\tau_2)^\dagger \kket \:. \label{b2}
\end{align}
Introducing the integration variables
\[ \zeta = \frac{1}{2}\:(\tau_1-t) \:,\qquad \nu = \tau_2 - t + \zeta\:, \]
we obtain
\begin{align*}
\eqref{b2} &= -4 \sint_\varkappa \int_{-\infty}^{-\zeta} d\nu \:
\big( M_{k,a} (t, t+2 \zeta)- M_{k,a} (t+2 \zeta, t)  \big) \\
&\qquad \qquad \times\: |\psi)(\psi|\: M_{k,a} (t+\zeta+\nu, t+\zeta-\nu)^\dagger
\end{align*}

\subsection{Computation of~$\bbra |\partial_t \psi)(\Sig_t \psi| \kket$}
\begin{align}
&\bbra \Big( \frac{d}{dt} | \psi(t)) \Big)  ( \Sig_t \psi(t) | \kket - \bbra \Big( \frac{d}{dt} |\psi) \Big) \kket \;
( \Sig_t \psi(t)| \notag \\
&= -i \int_{-\infty}^\infty d\zeta_1 \: \bbra \big| V(t, \zeta_1)\, \psi(t) \big) \big( \Sig_t \psi(t) \big| \kket \notag \\
&= -\int_{-\infty}^\infty d\zeta_1 \bigg( \int_{-\infty}^t \!\!\!\!\!d\tau \int_t^\infty \!\!\!\!\!d\tau' \!-\! \int_t^\infty \!\!\!\!\!d\tau \int_{-\infty}^t \!\!\!\!\!d\tau' \bigg) \: \bbra V(t, \zeta_1)\: | \psi)(\psi|\:V(\tau,\tau')^\dagger \kket \notag \\
&= - \sum_{k,a} \int_{-\infty}^\infty d\zeta_1 \bigg( \int_{-\infty}^t \!\!\!\!\!d\tau \int_t^\infty \!\!\!\!\!d\tau' \!-\! \int_t^\infty \!\!\!\!\!d\tau \int_{-\infty}^t \!\!\!\!\!d\tau' \bigg) \notag \\
&\qquad \times \delta \Big( \frac{t+\zeta_1}{2} - \frac{\tau+\tau'}{2} \Big)\:
M_{k,a}(t, \zeta_1)\: | \psi)(\psi|\:M_{k,a}(\tau,\tau')^\dagger \notag \\
&= -2\sum_{k,a} \bigg( \int_{-\infty}^t \!\!\!\!\!d\tau \int_t^\infty \!\!\!\!\!d\tau' \!-\! \int_t^\infty \!\!\!\!\!d\tau \int_{-\infty}^t \!\!\!\!\!d\tau' \bigg) \: M_{k,a}(t, \tau+\tau'-t)\: | \psi)(\psi|\:M_{k,a}(\tau,\tau')^\dagger \:. \label{b3}
\end{align}
Introducing the integration variables
\[ \zeta = \frac{1}{2}\:(\tau+\tau')-t \:,\qquad \nu = \frac{1}{2}\: (\tau-\tau')\:, \]
we obtain
\begin{align*}
\eqref{b3} &= 4 \sint_\varkappa \int_{\R \setminus [-|\zeta|, |\zeta|]} \epsilon(\nu)\:
M_{k,a} (t, t+2 \zeta) \\
&\qquad\qquad \times\:|\psi)(\psi|\: M_{k,a} (t+\zeta+\nu, t+\zeta-\nu)^\dagger
\end{align*}

\subsection{Computation of~$\bbra |\partial_t \Sig_t \psi)(\Sig_t \psi| \kket$} \label{secSS1}
In the remaining two sections we consider the contributions involving two factors~$\Sig_t$.
\begin{align}
&\bbra \Big( \frac{d}{dt} | \Sig_t \psi(t)) \Big)  ( \Sig_t \psi(t) | \kket - \bbra \Big( \frac{d}{dt} |\Sig_t \psi) \Big) \kket \; ( \Sig_t \psi(t)| \notag \\
&= i \int_{-\infty}^\infty d\tau_1 \;\bbra \big( V(t,\tau_1) - V(\tau_1,t) \big) \:|\psi(t))( \Sig_t \psi(t)| \kket \notag \\
&= \int_{-\infty}^\infty d\tau_1 \:\bigg( \int_{-\infty}^t \!\!\!\!\!d\tau \int_t^\infty \!\!\!\!\!d\tau' \!-\! \int_t^\infty \!\!\!\!\!d\tau \int_{-\infty}^t \!\!\!\!\!d\tau' \bigg) \notag \\
&\qquad \times \bbra \big( V(t,\tau_1) - V(\tau_1,t) \big) \:|\psi)(\psi|\: V(\tau,\tau')^\dagger \kket \notag \\
&= \sum_{k,a} \int_{-\infty}^\infty d\tau_1 \:\bigg( \int_{-\infty}^t \!\!\!\!\!d\tau \int_t^\infty \!\!\!\!\!d\tau' \!-\! \int_t^\infty \!\!\!\!\!d\tau \int_{-\infty}^t \!\!\!\!\!d\tau' \bigg) \notag \\
&\qquad \times \delta \Big( \frac{t+\tau_1}{2} - \frac{\tau+\tau'}{2} \Big) \: 
\big( M_{k,a}(t,\tau_1) - M_{k,a}(\tau_1,t) \big) \:|\psi)(\psi|\: M_{k,a}(\tau,\tau')^\dagger \notag \\
&= 2 \sum_{k,a} \bigg( \int_{-\infty}^t \!\!\!\!\!d\tau \int_t^\infty \!\!\!\!\!d\tau' \!-\! \int_t^\infty \!\!\!\!\!d\tau \int_{-\infty}^t \!\!\!\!\!d\tau' \bigg) \notag \\
&\qquad \times  \big( M_{k,a}(t,\tau+\tau'-t) - M_{k,a}(\tau+\tau'-t,t) \big)
\:|\psi)(\psi|\: M_{k,a}(\tau,\tau')^\dagger \:. \label{d1}
\end{align}
Introducing the integration variables
\[ \zeta = \frac{1}{2}\:(\tau+\tau') \:,\qquad \nu = \frac{\tau-\tau'}{2} \:, \]
we obtain
\begin{align*}
\eqref{d1} &= 4 \sint_\varkappa \int_{\R \setminus [-|\zeta|, |\zeta|]} \epsilon(\nu)\:
\big( M_{k,a} (t, t+2 \zeta)- M_{k,a} (t+2 \zeta, t)  \big)  \\
&\qquad\qquad \times\:|\psi)(\psi|\: M_{k,a} (t+\zeta-\nu, t+ \zeta+\nu)^\dagger\: d\nu
\end{align*}

\subsection{Computation of~$\bbra \partial_t \Sig_t^2 \kket$}  \label{secSS2}
\[ \frac{d}{dt} \Sig_t^2 = \dot{\Sig}_t \,\Sig_t + \Sig_t \,\dot{\Sig}_t
= \dot{\Sig}_t \,\Sig_t + \big( \dot{\Sig}_t\, \Sig_t \big)^\dagger \:. \]
Using~\eqref{Sform},
\begin{align}
&\bbra \Sig_t \dot{\Sig}_t \kket =
\bbra - \bigg( \int_{-\infty}^t \!\!\!\!\!d\tau_1 \int_t^\infty \!\!\!\!\!d\tau_2 \!-\! \int_t^\infty \!\!\!\!\!d\tau_1 \int_{-\infty}^t \!\!\!\!\!d\tau_2 \bigg) \: V(\tau_1,\tau_2) \notag \\
&\qquad \times
\bigg( \frac{d}{dt} \bigg( \int_{-\infty}^t \!\!\!\!\!d\tau \int_t^\infty \!\!\!\!\!d\tau' \!-\! \int_t^\infty \!\!\!\!\!d\tau \int_{-\infty}^t \!\!\!\!\!d\tau' \bigg) \: V(\tau,\tau') \bigg) \ket \notag \\
&=- \bigg( \int_{-\infty}^t \!\!\!\!\!d\tau_1 \int_t^\infty \!\!\!\!\!d\tau_2 \!-\! \int_t^\infty \!\!\!\!\!d\tau_1 \int_{-\infty}^t \!\!\!\!\!d\tau_2 \bigg) \int_{-\infty}^\infty d\tau'\: \bbra V(\tau_1,\tau_2)\,V(t,\tau') \kket \notag \\
&\quad \:+ \bigg( \int_{-\infty}^t \!\!\!\!\!d\tau_1 \int_t^\infty \!\!\!\!\!d\tau_2 \!-\! \int_t^\infty \!\!\!\!\!d\tau_1 \int_{-\infty}^t \!\!\!\!\!d\tau_2 \bigg) \int_{-\infty}^\infty d\tau\:\bbra V(\tau_1,\tau_2) \,V(\tau,t) \kket \notag \\
&=-\bigg( \int_{-\infty}^t \!\!\!\!\!d\tau_1 \int_t^\infty \!\!\!\!\!d\tau_2 \!-\! \int_t^\infty \!\!\!\!\!d\tau_1 \int_{-\infty}^t \!\!\!\!\!d\tau_2 \bigg) \int_{-\infty}^\infty d\tau\: \bbra V(\tau_1,\tau_2)\,\big( V(t,\tau) - V(\tau, t) \big) \kket \notag \\ 
&=-\sum_{k,a} \bigg( \int_{-\infty}^t \!\!\!\!\!d\tau_1 \int_t^\infty \!\!\!\!\!d\tau_2 \!-\! \int_t^\infty \!\!\!\!\!d\tau_1 \int_{-\infty}^t \!\!\!\!\!d\tau_2 \bigg)  \int_{-\infty}^\infty d\tau \notag \\
&\qquad \times \delta \Big(\frac{t+\tau}{2} - \frac{\tau_1+\tau_2}{2} \Big) \: 
M_{k,a}(\tau_1,\tau_2) \, \big( M_{k,a}(t,\tau) - M_{k,a}(\tau, t) \big) \notag \\
&=-2 \sum_{k,a}
\bigg( \int_{-\infty}^t \!\!\!\!\!d\tau_1 \int_t^\infty \!\!\!\!\!d\tau_2 \!-\! \int_t^\infty \!\!\!\!\!d\tau_1 \int_{-\infty}^t \!\!\!\!\!d\tau_2 \bigg) \notag \\
&\qquad \times \: 
M_{k,a}(\tau_1,\tau_2)\,\big( M_{k,a}(t, \tau_1+\tau_2-t) - M_{k,a}(\tau_1+\tau_2-t, t) \big) \:. \label{c1}
\end{align}
Introducing the integration variables
\[ \zeta = \frac{\tau_1+\tau_2}{2} - t \:,\qquad \nu = \frac{\tau-\tau'}{2}\:, \]
we obtain
\begin{align*}
\eqref{c1} &= 4 \sint_\varkappa \int_{\R \setminus [-|\zeta|, |\zeta|]} \epsilon(\nu)\:
M_{k,a}(t+\zeta+ \nu ,t+\zeta-\nu) \\
&\qquad\qquad \times \:\big( M_{k,a}(t, t+2 \zeta) - M_{k,a}(t+2 \zeta, t) \big) \,d\nu
\end{align*}

\addtocontents{toc}{\setcounter{tocdepth}{2}}

\Thanks{{{\em{Acknowledgments:}} We are grateful to Catalina Curceanu, Patrick Fischer, Jos{\'e} M.\ Isidro,  Antonino Marcian{\`o}, Simone Murro, Kristian Piscicchia and Tejinder P.\ Singh for helpful discussions
and comments on the manuscript. We would like to thank the referees for useful suggestions.


\end{document}